%% file: ms.tex
\documentclass[]{elsarticle}
\usepackage[utf8]{inputenc}

\usepackage{lineno,hyperref}
\usepackage{url}
\usepackage{breakurl}

\usepackage{amsmath,amssymb,amsfonts}
\usepackage{algorithmic}
\usepackage{graphicx}
\usepackage{textcomp}
\usepackage{xcolor}
\usepackage{soul}
\usepackage{array}
\usepackage{tabularx}
\usepackage{pdflscape}
\usepackage{float}\modulolinenumbers[5]
\usepackage{listings}

\makeatletter
\def\ps@pprintTitle{%
	\let\@oddhead\@empty
	\let\@evenhead\@empty
	\def\@oddfoot{\footnotesize\itshape
		\hfill\today}%
	\let\@evenfoot\@oddfoot
}
\makeatother

\begin{document}

\begin{frontmatter}

\title{HPIM-DM: a fast and reliable dense-mode\\multicast routing protocol (extended version)}

\author{Pedro Oliveira}
\ead{pedro.francisco.oliveira@tecnico.ulisboa.pt}

\author{Alexandre Silva}
\ead{alexandre.s.silva@tecnico.ulisboa.pt}

\author{Rui Valadas}
\ead{rui.valadas@tecnico.ulisboa.pt}

\address{Instituto de Telecomunicações and Instituto Superior Técnico, University of Lisbon, Portugal}




\begin{abstract}
This paper proposes the HPIM-DM (Hard-state Protocol Independent Multicast - Dense Mode) multicast routing protocol. HPIM-DM is a hard-state version of PIM-DM that keeps its main characteristics but has faster convergence and better resilience to replay attacks. Like PIM-DM, HPIM-DM is meant for dense networks and supports its operation on a unicast routing protocol and reverse path forwarding checks. However, routers maintain sense of the multicast trees at all times, allowing fast reconfiguration in the presence of network failures or unicast route changes. This is achieved by (i) keeping information on all upstream neighbors from which multicast data can be received, (ii) ensuring the reliable transmission and sequencing of control messages, and (iii) synchronizing the routing information immediately when a new router joins the network. The protocol was fully implemented in Python, and the implementation is publicly available. Finally, the correctness of the protocol was extensively validated using model checking, logical reasoning and tests performed over the protocol implementation.%RV 25/1/2020
\end{abstract}


\end{frontmatter}

%\linenumbers

\input{cc_introduction}
\input{cc_pimdm_v3}
\input{cc_hpimdm_v20}

\input{cc_correctness_v4}
\input{cc_implementation_v4}
\input{cc_tests_v1}
\input{cc_conclusions}

\bibliography{ms}

\end{document}

%% file: cc_introduction.tex
\section{Introduction}\label{sec:intro}

Multicast routing protocols allow an efficient support of group communications (e.g. videoconferencing, Internet TV distribution, collaborative work) by providing a distribution overlay, usually comprised of one or more trees, for delivering multicast traffic from sources to groups of interested receivers \cite{livro_primer_multicast,livro_cisco,livro_multicast_communication,livro_adhocwireless}.%RV 31/1/2020

Over the years several multicast routing protocols have been proposed for both fixed and wireless networks. Examples are DVMRP \cite{rfc1075}, MOSPF \cite{rfc1584}, CBT \cite{cbt}, and PIM \cite{rfc3973,RFC7761}. The PIM protocols are currently the preferred ones for intra-domain multicast routing in fixed networks. They are implemented by all vendors and keep being updated by the PIM working group of IETF \cite{ietf_pim_documents}. The PIM protocols rely on a unicast routing protocol to build unidirectional multicast distribution trees using Reverse Path Forwarding (RPF) checks. There are two main versions of PIM: PIM-DM (Dense Mode) and PIM-SM (Sparse Mode). PIM-DM \cite{rfc3973} is meant for dense multicast networks and builds one distribution tree per multicast source and group using a flood-and-prune algorithm; it belongs to the class of source-based multicast routing protocols. PIM-SM \cite{RFC7761} is meant for sparse multicast networks and builds a single distribution tree per group rooted at a special router, called Rendezvous Point (RP), to which all sources of the corresponding group must connect. Both protocols are soft-state protocols.%RV 31/1/2020

Despite their popularity, the PIM protocols suffer from slow convergence, which can cause excessive loss of multicast data and network congestion, limiting their applicability in high-speed networks. Moreover, they are prone to replay attacks.%RV 31/1/2020

This paper proposes HPIM-DM (Hard-state Protocol Independent Multicast - Dense Mode), which is an hard-state version of PIM-DM. It keeps the main characteristics of PIM-DM, such as reliance on unicast routing protocol and RPF checking, but achieves much faster convergence due to its novel multicast routing concepts. In particular, each router maintains information on all its upstream neighbors (i.e. neighbors from which multicast data can be received), and not just on its current parent in the multicast distribution tree. Moreover, a new router joining the network obtains information on the active trees very fast by synchronizing with each neighbor, and can start receiving multicast data immediately, if interested. In HPIM-DM, all control messages are reliably transmitted and sequenced. The protocol is fully optimized to minimize the amount of stored state information. In addition to its convergence properties, HPIM-DM is much more resilient to security attacks (e.g. replay attacks) than PIM-DM. We have implemented HPIM-DM in Python, and the implementation is publicly available. Finally, the correctness of the protocol was extensively verified using model checking, logical reasoning and tests performed over the implementation.%RV 31/1/2020

This paper is organized as follows. Section \ref{sec:sourcerouting} introduces the principles of source-based multicast routing. Section \ref{sec:pimdm} reviews the operation of PIM-DM and section \ref{sec:pimdmissues} highlights its main issues. The specification of HPIM-DM is presented in Section \ref{sec:hpimdm} and section \ref{sec:correctness} discusses its correctness. Section \ref{sec:implementation} gives an overview of the implementation and section \ref{sec:tests} describes the tests performed on the implementation. Finally, in section \ref{sec:conclusions}, we present the conclusions and directions for future work.%RV 31/1/2020

%% file: cc_pimdm_v3.tex
\section{Source-based multicast routing}\label{sec:sourcerouting}

\medskip
\noindent \textbf{IP multicast routing and source-based trees} IP multicast routing protocols establish one or more logical routing structures (usually trees) over the physical network to deliver multicast packets from sources of multicast traffic to groups of interested receivers; the sources and receivers are hosts, i.e. IP endpoints. In source-based IP multicast routing, each source delivers traffic to a group of interested receivers through \textit{a single} tree, sometimes referred to as \textit{source-based} and/or \textit{distribution} tree. The tree routing structure allows saving resources (i.e. packet transmissions) in relation to the possibility of having dedicated unicast routes connecting the source to each individual interested receiver. Indeed, a packet sent by the source is transmitted only once in each link that belongs to the tree, and a given router may transmit a received packet through more than one interface.%RV 25/1/2020

\medskip
\noindent \textbf{IP multicast model and addresses} In the IP multicast model \cite{rfc1112}, communication groups are identified by IP multicast addresses. When a source wants to send a packet to a group of receivers, it uses as IP destination address for the packet the IP multicast address that identifies the group. Moreover, receivers are preprogrammed to process only IP multicast addresses of the groups they are interested in. Both IPv4 and IPv6 have address blocks reserved for IP multicasting. In IPv4, the block is 224.0.0.0/4 \cite{multicast_addresses_ipv4} and, in IPv6, is FF00::/8 \cite{multicast_addresses_ipv6}.%RV 25/1/2020

\medskip
\noindent \textbf{Overlaid and non-permanent source-based trees} In source-based IP multicast routing, there is a different distribution tree for each source-group pair. Thus, a multicast routing domain may have several source-based trees overlaid on the physical network. Each tree is completely identified by the source unicast IP address and the group multicast IP address. These addresses are usually represented by S and G, respectively, and the tree is usually referred to as the (S,G) tree.%RV 25/1/2020

A tree is only maintained as long as the multicast source is active, i.e. keeps sending multicast packets. This contrasts with unicast routing protocols, where the paths are maintained independently of the presence of unicast data.%RV 25/1/2020

\medskip
\noindent \textbf{How hosts express their interest} In the IP multicast model \cite{rfc1112}, a host interested in receiving packets of a multicast group signals the routers directly attached to its subnet using the IGMP or MLD protocols, in case of IPv4 or IPv6, respectively. Depending on the protocol version, the interest can be expressed per group (IGMPv2 and MLDv1), or per source and per group, i.e. per tree (IGMPv3 and MLDv2). Expressing interest per group impacts all trees of the corresponding group; expressing interest per tree impacts only the corresponding tree.%RV 25/1/2020

\medskip
\noindent \textbf{Broadcast versus multicast trees} In order to understand the operation of multicast routing protocols, it is important to distinguish between the broadcast and multicast trees. The \textit{broadcast} tree spans all routers of a multicast routing domain. The \textit{multicast} tree is the tree required to deliverer multicast packets to the \textit{interested} receivers (and only to those), and is a subset of the broadcast tree. While forming a multicast tree is the ultimate goal of a multicast routing protocol, in many cases, this implies building first a broadcast tree.%RV 25/1/2020

\medskip
\noindent \textbf{Link types} Regarding the various link technologies (layer-2) used to connect routers, they can be abstracted in two types: (i) point-to-point links, connecting only two routers, and (ii) shared links, connecting two or more routers. Shared links may abstract relatively complex layer-2 networks and often offer a broadcast capability. As it will be discussed later, shared links impose specific requirements on multicast routing protocols.%RV 25/1/2020

\medskip
\noindent \textbf{Originators, parents and children} In the broadcast and multicast trees it is important to distinguish among the roles of the various routers. The routers directly connected to the source's subnet are called \textit{originator} routers. Note that there can be more than one originator for a given tree. A broadcast or multicast tree can be seen as directed tree rooted at the source's subnet. In this tree, the \textit{parent} of a router \textit{r} is the router connected to \textit{r} on the path from the root to $r$, from which multicast data must be received. The \textit{child} of a router \textit{r} is a router for which \textit{r} is the parent, and to which multicast data must be sent. The \textit{leaf} routers of the tree are the routers that have no children.%RV 25/1/2020

\medskip
\noindent \textbf{Tree configuration} A tree can be defined in terms of the router interfaces that belong to it. While there are several possible criteria to define a distribution tree, in this work we concentrate in the criteria used in PIM protocols. In this case, the tree configuration relies on a unicast routing protocol, and the router interfaces that belong to the tree are the ones in the shortest paths from the multicast receivers to the source's subnet. This type of tree is sometimes called \textit{reverse shortest path} tree, because the tree is defined in a direction opposite to the flow of data.%RV 25/1/2020

\medskip
\noindent \textbf{Interface role and RPC} At each router, the shortest path from the router to the source's subnet is determined through the unicast routing protocol. The cost of a path from a router interface to the source's subnet is calculated as the sum of the unicast costs assigned to the interfaces that transmit a packet in that path in the upstream direction (would this transmission take place), i.e. from the router to the source's subnet; the costs of receiving interfaces are not considered. Among the various router interfaces, the one that provides the shortest path to the source's subnet is called the \textit{root} interface and the corresponding path cost is called Root Path Cost (RPC); the remaining interfaces are called \textit{non-root} interfaces.%RV 25/1/2020

\medskip
\noindent \textbf{Forwarding states} Once a tree is formed, packets are received on root interfaces and forwarded on non-root interfaces. However, some non-root interfaces may not be allowed to transmit multicast packets (i.e. may not be part of the distribution tree). This happens for two reasons: (i) to avoid duplicate transmissions at links or (ii) because the interface has no downstream routers or hosts interested in receiving multicast traffic. Thus, a non-root interface can be in one of two states: FORWARDING, when allowed to transmit multicast data, or PRUNED, when not allowed to transmit multicast data. Under stable conditions, a router receives multicast packets on its root interface, and retransmits them through its FORWARDING non-root interfaces.%RV 25/1/2020

\medskip
\noindent \textbf{Pruning to avoid duplicate transmissions} As mentioned above, to avoid duplicate transmissions at links, some non-root interfaces must be pruned. Specifically, at shared links only one non-root interface connected to the link must be allowed to forward multicast packets. The selection of this interface is performed by the \textit{Assert Protocol}; the selected interface is named the \textit{Assert Winner} (AW) and the remaining interfaces the \textit{Assert Losers} (ALs). The AW is usually the non-root interface that, among the one attached to the shared link, offers the lowest RPC; the interface identifiers are usually used as a tiebreaker in case of identical RPCs. At point-to-point link interfaces both interfaces must be pruned if they are both non-root interfaces, since non-root interfaces have no interest in receiving multicast packets. In the broadcast tree, all non-root interfaces that are not placed in PRUNED state to avoid duplicate transmissions, stay in FORWARDING state. Moreover, the broadcast tree is completely determined by the set of root interfaces and non-root interfaces in FORWARDING state.%RV 25/1/2020

\medskip
\noindent \textbf{Multicast routing table} The multicast routing table gathers, at a router, the information required to forward multicast packets. It comprises an entry per tree, and each entry includes information on the root interface and on the non-root interfaces in FORWARDING state of the corresponding tree.%RV 25/1/2020

\medskip
\noindent \textbf{RPF checks} Verifying whether an incoming packet entered a router through a root or a non-root interface is sometimes called the \textit{RPF check}, since it determines if the packet arrived on the interface providing the reverse shortest path to the source (i.e. the root interface). Packets can only be accepted if received on a root interface, i.e. if they passed the RPF check.%RV 25/1/2020

\medskip
\noindent \textbf{Reverse Path Forwarding} The presence of a unicast routing protocol (and, therefore, of root and non-root interfaces) allows implementing a method for controlled flooding on a network, i.e. a method to deliver the packets transmitted by one router to all other network routers, without indefinite circulation of packets. The method is simply, at each router, (i) to accept packets received on the root interface and forward them on all non-root interfaces, and (ii) to discard packets received on non-root interfaces. This method is known as \textit{Reverse Path Forwarding} (RPF) \cite{reverse_path_forwarding} and is used by PIM-DM to build the broadcast tree. It ensures that the packets sent by the source router are delivered to all other routers. However, in some cases, the same packet can be transmitted on a link more than once.%RV 25/1/2020

\medskip
\noindent \textbf{Reconfiguration of the broadcast tree and Hello protocol} Once the broadcast tree is formed, several events may dictate its reconfiguration, such as the removal/addition of routers or links, and changes in the unicast paths. Multicast routing protocols must be equipped with the mechanisms required to reconfigure the broadcast tree dynamically, when these events happen. One such mechanism is the \textit{Hello protocol} whereby routers periodically announce their presence to their neighbors; this protocol allows detecting router or link removals (e.g. failures) and router or link additions to the network.%RV 25/1/2020

\medskip
\noindent \textbf{From the broadcast to the multicast tree} As noted above, the broadcast tree must be pruned from the non-root interfaces that should not transmit traffic due to lack of interest or nonexistence of multicast receivers in some zones of the multicast routing domain. If a non-root interface has no downstream routers or directly attached hosts interested in receiving multicast traffic for some tree, it should be placed in PRUNED state for that tree. Recall that a host interested in receiving multicast traffic, signals the routers directly attached to its subnet using the IGMP or MLD protocols, in case of IPv4 or IPv6, respectively.%RV 25/1/2020

A router that has no non-root interfaces interested in transmitting multicast traffic (i.e. with all non-root interfaces in PRUNED state) becomes itself not interested in receiving multicast traffic. In this case, the router signals its upstream routers on the tree about this lack of interest using specific control messages, such that the tree is pruned and its path to the originator is removed. This is how the broadcast tree leads to the multicast tree. Likewise, if a router becomes interested again, it signals its upstream routers such that its path to the originator is reinstalled.%RV 25/1/2020

\begin{figure}[t!]
	\centering
	\includegraphics[scale=0.45]{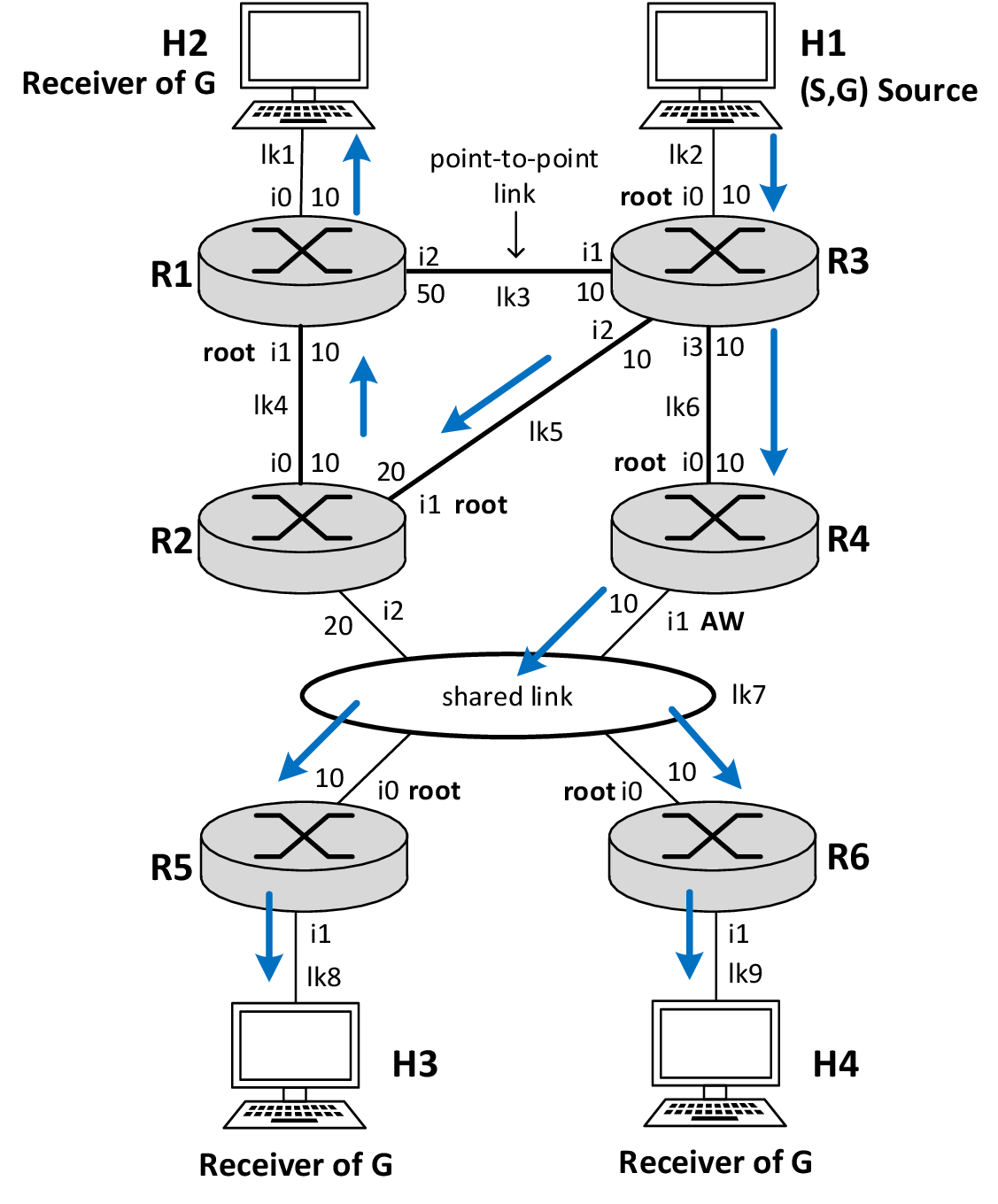}
	\caption{Broadcast tree example.} 
	\label{fig:broadtree}
\end{figure}

\medskip
\noindent \textbf{Examples} Figure \ref{fig:broadtree} gives an example of a broadcast tree. The network has six routers (R1 to R6), nine links (lk1 to lk9), and four hosts (H1 to H4). Host H1 is a multicast source, and the other hosts are multicast receivers for group G. R3 is an originator router. There is one shared link connecting routers R2, R4, R5, and R6; all other links are point-to-point links. The figure indicates, near each interface, the interface identifier and the unicast interface cost. It also indicates the root interface of each router and the AW of the shared link. The arrows represent the paths followed by the multicast packets from the source to the receivers. Recall that packets always arrive at routers through their root interfaces, and that the root interface is the interface that provides the shortest path cost to the source's subnet. For example, router R2 has three interfaces leading to the source's subnet. Interface i0 offers a cost of 10+50+10=70, interface i1 offers a cost of 20+10=30, and interface i2 offers a cost of 20+10+10=40. Thus, the root interface is i1, providing an RPC of 30. At the shared link the AW is interface i1-R4. This is because, among the non-root interfaces attached to the link, i.e. i2-R2 and i1-R4, i1-R4 offers an RPC of 10+10=20 and i2-R2 offers a higher RPC, i.e. 20+10=30 as noted before. Remember that, for the path cost calculations, only the interfaces that would transmit packets towards the source's subnet, i.e. in the reverse path, must be considered. For example, the cost of the path R2$\rightarrow$R3$\rightarrow$H1 excludes the cost of interface i2-R3 since, for the purpose of path cost calculations, this interface is a receiver of packets and not a transmitter. In this tree, the non-root interfaces placed in PRUNED state are i2-R1 and i1-R3 (two non-root interfaces connected by a point-to-point link) and i2-R2 (an AL at a shared link). All other non-root interfaces are in FORWARDING state. Moreover, R3 is parent of R2 and R4, R2 is parent of R1, and R4 is parent of R5 and R6.%RV 25/1/2020

%RV 14/6/2019

\begin{figure}[t!]
	\centering
	\includegraphics[scale=0.45]{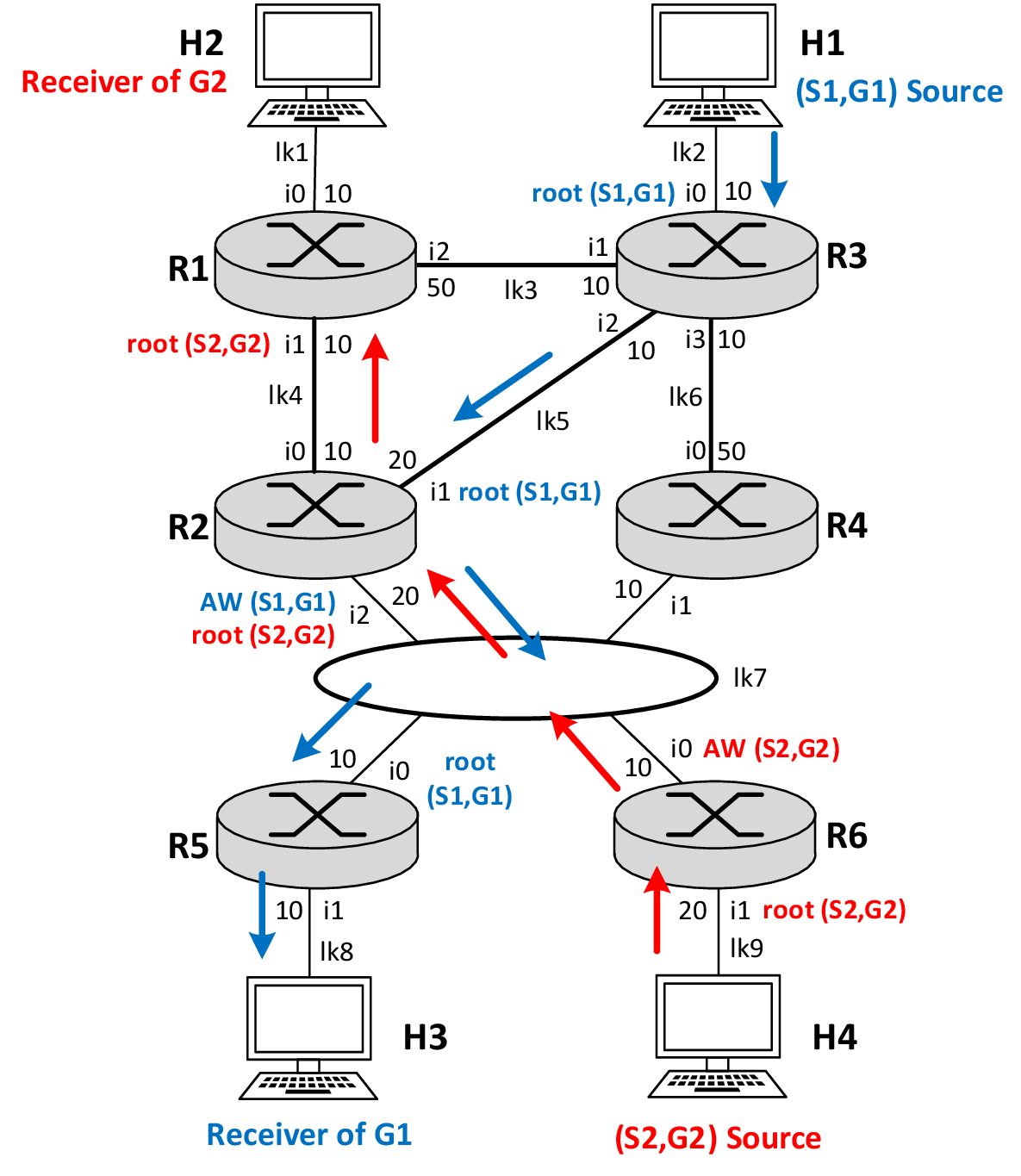}
	\caption{Two multicast trees overlaid on the same physical network.} 
	\label{fig:broad2trees}
\end{figure}

Figure \ref{fig:broad2trees} illustrates a network with two multicast trees. The physical network is the same of Figure \ref{fig:broadtree}, but now H1 and H4 are sources for groups G1 and G2, respectively, H3 is receiver for G1 and H2 is receiver for G2. We also increased the cost of interface i0-R4 to 50. There are now two multicast trees, (S1,G1) and (S2,G2), the first having R3 as originator and the second having R6 as originator. The interface roles (root versus non-root) and AW are specific of each tree. For example, in router R2 interface i1 is root for (S1,G1), but the root for (S2,G2) is interface i2; moreover, i2 is also the AW for (S1,G1). Note also that some routers are not part of the multicast trees due to lack of interest. Specifically, R4 is not part of the two trees, R1 and R6 are not part of (S1,G1), and R3 and R5 are not part of (S2,G2).%RV 25/1/2020

%RV 14/6/2019

\begin{figure}[t!]
	\centering
	\includegraphics[scale=0.65]{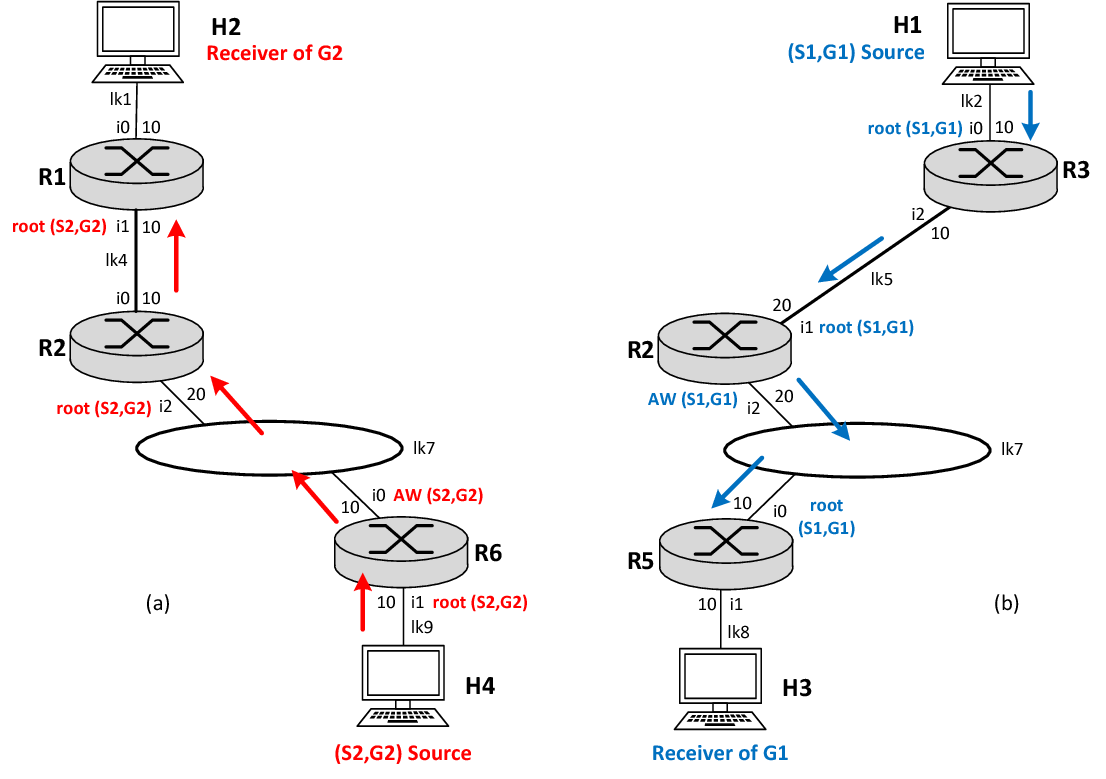}
	\caption{Separate views of the multicast trees of Figure \ref{fig:broad2trees}; (a) (S2,G2) tree and (b) (S1,G1) tree.} 
	\label{fig:separate_trees}
\end{figure}%RV 14/6/2019

\section{PIM-DM operation}\label{sec:pimdm}

PIM-DM \cite{rfc3973} is a source-based, data-driven, dense-mode, and soft-state multicast routing protocol, that relies on a unicast routing protocol and on RPF checks. PIM protocols are called \textit{independent} since they do not rely on any particular unicast routing protocol.%RV 31/1/2020

\medskip
\noindent \textbf{Hello protocol} In PIM-DM, routers interested in participating in the multicast routing protocol establish and maintain neighborhood relationships with their neighboring routers through the periodic exchange of Hello messages. A router considers an unknown router its neighbor as soon as it receives the first Hello message from it. Thus, neighborhood relationships need not be bidirectional, as required in other routing protocols (e.g. OSPF); PIM-DM Hello messages do not even include the addresses of the link neighbors. The periodicity of Hello transmissions is called \textit{Hello Period}, and has a default value of 30 seconds. Moreover, a router considers a neighbor dead if it ceases receiving Hello messages from it for more than a time period called \textit{Hold Time}, which has a default value of 3.5 $\times$ Hello period, i.e. 105 seconds. The Hold Time value is carried in Hello messages. The timer that regulates the periodic transmission of Hello messages (according to Hello Time) is called \textit{Hello Timer} (HT) and the one that regulates the liveness of neighborhood relationships (according to Hold Time) is called \textit{Neighbor Liveness Timer} (NLT).%RV 26/1/2020

Hello messages include a field to detect neighbor reboots, called \textit{Generation ID}. This field is a random number generated when an interface is started, and included in all Hello messages it sends. Thus, if a router receives two Hello messages from the same neighbor with different Generation IDs, it learns that the neighbor has rebooted. In this case, an Hello message is sent immediately to the rebooted neighbor, without waiting for the next time scheduled for the transmission of a periodic Hello. This feature allows rebooted neighbors to speedup the establishment of neighborhood relationships.%RV 26/1/2020

\medskip
\noindent \textbf{Broadcast tree construction} In PIM-DM networks, one broadcast tree is built for each (S,G) pair. The construction of the broadcast tree is triggered by the source, when it starts transmitting multicast data; that is why the protocol is said to be \textit{data-driven}. When a source S starts transmitting multicast data for group G, the data packets are flooded from the originator routers to all other routers, using the RPF technique. Specifically, when a packet is received at a router interface for the first time, the router checks, using the unicast routing table, if the packet was received through a root or a non-root interface. If received through a root interface, the router builds a multicast routing table for the corresponding (S,G) tree, and forwards the packet through all its non-root interfaces; otherwise, the packet is discarded. Except for the originators, the root interface is the one with lowest RPC; the root interface of an originator is the interface attached to the source's subnet, irrespective of the RPC. Initially, the non-root interfaces are all placed in FORWARDING state. In this way, the initial packet is delivered to all network routers, without indefinite circulation.%RV 26/1/2020

\begin{figure}[t!]
	\centering
	\includegraphics[scale=0.45]{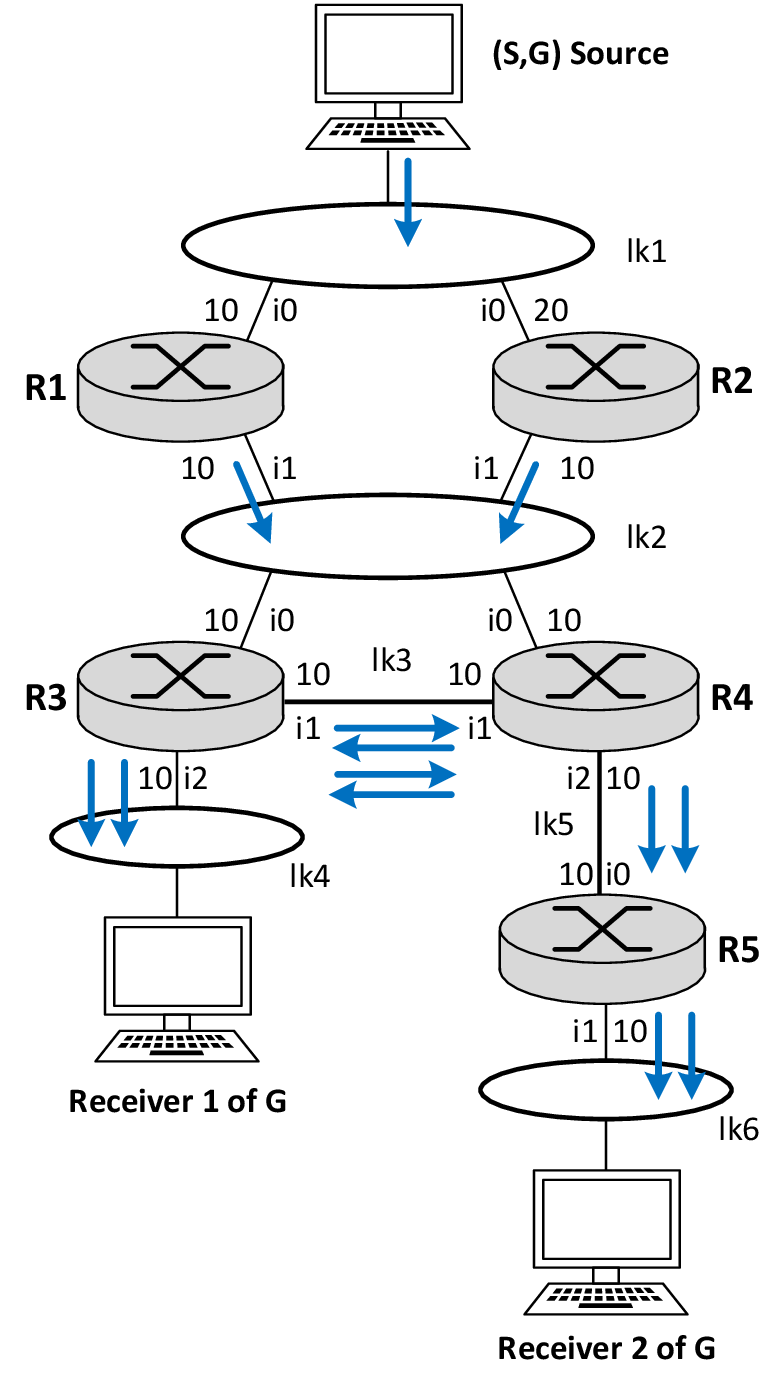}
	\caption{Initial flood of multicast data packets in PIM-DM.} 
	\label{fig:pimdm_initialflood}
\end{figure}

\medskip
\noindent \textbf{Example} Figure \ref{fig:pimdm_initialflood} illustrates the initial flood of data packets in PIM-DM. In the network, there is one multicast source S and two multicast receivers for group G and, therefore, an (S,G) tree must be built. The network includes five routers and six links; links lk3 and lk5 are point-to-point links, and the remaining ones are shared links. We assume that there is a unicast routing protocol running and that the two multicast receivers already joined the multicast group G (Receiver 1 signaled R3 and Receiver 2 signaled R5 using the IGMP protocol). The figure includes, near each router interface, the interface identifier and the interface costs associated with the unicast routing protocol. All interface costs are 10, except the cost of interface i0-R2 which is 20. The RPC is 10 for R1, 20 for R2, R3, and R4, and 30 for R5, and the root interface is, for all routers, the i0 interface. In this network there are two originators relative to the (S,G) tree: R1 and R2.%RV 26/1/2020

%RV 22/10/2018

When the source sends the first multicast data packet, routers R1 and R2 receive it through their interfaces i0-R1 and i0-R2, determine that their non-root interfaces are i1-R1 and i1-R2, install an entry in the multicast routing table for this tree, and transmit the packet to the shared link. Recall that non-root interfaces are placed initially in the FORWARDING state. The packet transmitted by one interface is received by all other interfaces attached to the shared link. Taking the case of the packet transmitted by R1, this packet is discarded when received at interface i1-R2 since the interface is non-root. However, when received at interfaces i0-R3 and i0-R4, the routers verify that the packets were received through root interfaces, determine the non-root interfaces (which are the i1 and i2 interfaces of R3 and R4), and transmit the packets through these interfaces. The non-root interfaces of R3 and R4 transmit two replicas of the initial packets, one for each packet received at their root interfaces. Thus, Receiver 1 ends up receiving two replicas of the initial data packet. The packets transmitted by either side of the point-to-point link lk3 are discarded by the interface of the other side, since that interface is non-root. Finally, the packets transmitted through interface i2-R4 are received by the root interface of R5, i.e. i0-R5, and are delivered to Receiver 2 through the non-root interface i1-R5. Receiver 2 also receives two replicas of the initial data packet. To conclude, the initial data packet was delivered to all routers and links without the indefinite circulation of packets (i.e. without uncontrolled flooding).%RV 26/1/2020

As illustrated in this example, the RPF technique ensures that there is no indefinite circulation of packets. However, it does not prevent packets from being transmitted more than once on shared links, and on point-to-point links attaching two non-root interfaces. PIM-DM uses two mechanisms to solve these problems, which will be discussed next.%RV 17/6/2019

\medskip
\noindent \textbf{Assert protocol} On shared links, all non-root interfaces attached to the link will initially transmit data packets, since they are all in FORWARDING state. To avoid that this duplication keeps happening, PIM-DM elects one non-root interface to become the single link forwarder, called the \textit{Assert Winner} (AW) interface. The process that performs this election is the Assert protocol. Specifically, once a non-root interface receives a data packet, it multicasts on the shared link an Assert message containing the RPC of the sending router. The non-root interface that advertises the lowest RPC becomes the AW and the remaining non-root interfaces become \textit{Assert Losers} (ALs); in case of a tie, the interface with the highest IP address wins. The ALs are placed in PRUNED state, and the AW may be placed in FORWARDING state, depending on the interest of downstream devices. All interfaces store the IP address of the AW and its RPC. The Assert protocol is also used in point-to-point links.%RV 31/1/2020

Assert messages with an infinite cost are transmitted occasionally, to provoke the reelection of the AW. This happens, for example, when the AW interface changes from non-root to root.%RV 31/1/2020

The operation of the Assert protocol is actually controlled by a state machine with three states: NoInfo, AW, and AL. The state machine is slightly different in root and non-root interfaces. The NoInfo state is the initial state, prior to having received any data packet. For non-root interfaces, when a data packet is received in this state, the interface must transmit an Assert message advertising its RPC. The AW election is based on the information contained in the Assert messages. If the interface is non-root, it elects the AW based on its own RPC and the RPCs of its neighbors. The interface transitions to AW state if it elects itself as AW and to AL state otherwise. A root interface elects the AW based only on the RPCs of its neighbors, since it cannot be AW; in this case, the interface can only transition to AL state. Non-root interfaces in AW and NoInfo states are placed in FORWARDING state depending on the downstream interest of routers and hosts; an interface in AL state is placed in PRUNED state irrespective of downstream interest. A data packet received in AW state also triggers the transmission of Assert messages.%RV 31/1/2020

\begin{figure}[t!]
	\centering
	\includegraphics[scale=0.8]{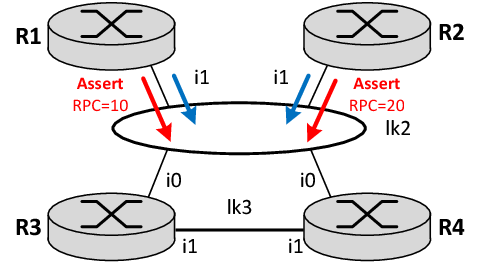}
	\caption{Assert protocol in PIM-DM.} 
	\label{fig:pimdm_assert}
\end{figure}

\medskip
\noindent \textbf{Example} In Figure \ref{fig:pimdm_assert} we illustrate what happens in link lk2 of Figure \ref{fig:pimdm_initialflood}. The non-root interfaces i1-R1 and i1-R2 are initially in NoInfo state. When they receive the data packet transmitted by the other, they reply with an Assert message where they advertise their RPC: i1-R1 advertises an RPC of 10 and i1-R2 an RPC of 20. Thus, i1-R1 is elected AW and stays FORWARDING, and i1-R2 becomes AL and is placed in PRUNED state. In this way, the next data packet transmitted by the source will only be transmitted once at shared link lk2, by interface i1-R1.%RV 26/1/2020

%RV 22/10/2018

\medskip
\noindent \textbf{Pruning of two non-root interfaces attached through a point-to-point link} When a point-to-point link connects two non-root interfaces, no data packets need to be transmitted on the link, since there is no root interface on the other side. However, data packets are actually transmitted initially, since the interfaces are initially placed in the FORWARDING state. To solve this problem, when a non-root interface receives a data packet, it replies with a Prune message which, when received on the other side, places the receiving non-root interface in PRUNED state. In this way, both non-root interfaces become PRUNED.%RV 31/1/2020

\begin{figure}[t!]
	\centering
	\includegraphics[scale=0.45]{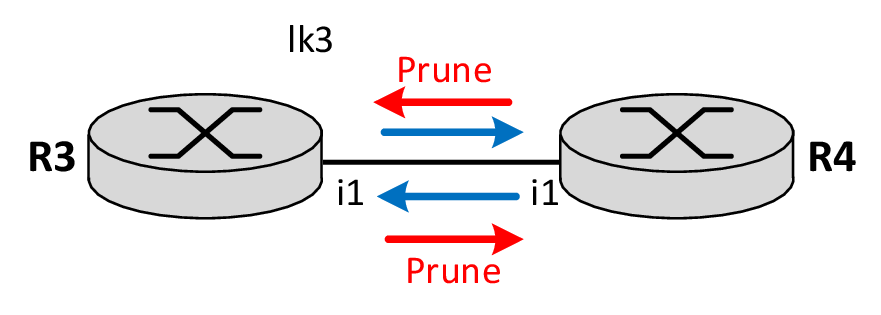}
	\caption{Pruning of two non-root interfaces attached through a point-to-point link in PIM-DM.} 
	\label{fig:pimdm_p2pprune}
\end{figure}

\medskip
\noindent \textbf{Example} In Figure \ref{fig:pimdm_p2pprune} we illustrate what happens in link lk3 of Figure \ref{fig:pimdm_initialflood}. When one interface receives a data packet, it replies with a Prune message which, when received on the other side, places the receiving interface in PRUNED state. Since both interfaces initially send data packets, both of them become PRUNED.%RV 26/1/2020

%RV 22/10/2018

\begin{figure}[b!]
	\centering
	\includegraphics[scale=0.45]{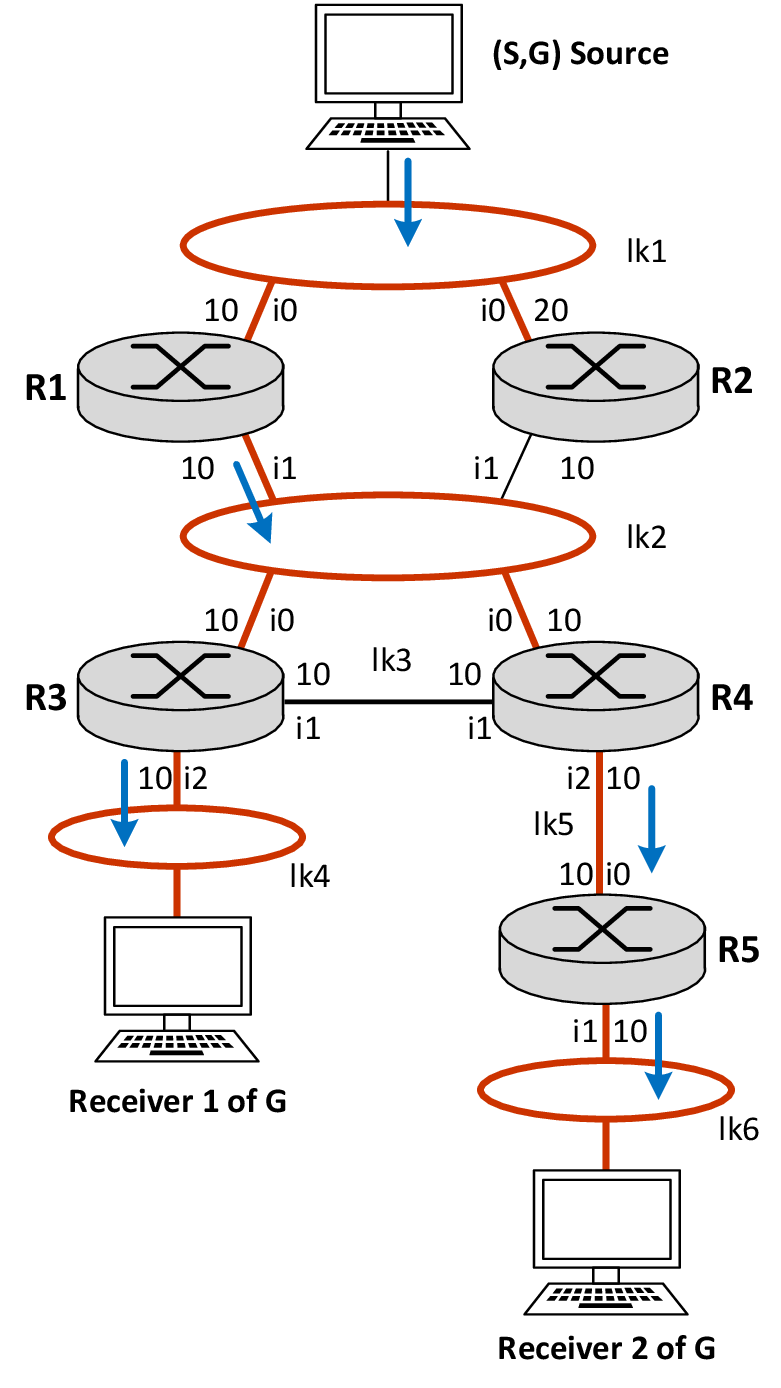}
	\caption{Broadcast tree obtained through flood-and-prune in PIM-DM (represented in dark red).} 
	\label{fig:pimdm_broadcasttree}
\end{figure}

\medskip
\noindent \textbf{Flood-and-prune} The RPF technique, enhanced with the Assert protocol and the pruning of two non-root interfaces attached through a point-to-point link, is called \textit{flood-and-prune}, and allows building of a broadcast tree pruned of redundant interfaces. Interestingly, the process is performed through a mixture of data packets and control messages. Figure \ref{fig:pimdm_broadcasttree} shows the broadcast tree of the network of Figure \ref{fig:pimdm_initialflood}, illustrating that data packets arrive at all routers and links without being transmitted more than once on each link.%RV 31/1/2020

%RV 22/10/2018

\medskip
\noindent \textbf{From the broadcast to the multicast tree} The broadcast tree is then pruned taking into account the tree interest of routers. Specifically, a router is not interested in receiving multicast data from an (S,G) tree if it has no multicast receivers or routers connected to its non-root interfaces (downstream routers) interested in receiving multicast data from this tree. The interest of multicast receivers is signaled to the routers they are directly attached to using the IGMP protocol.%RV 31/1/2020

\medskip
\noindent \textbf{Prune and Graft messages} When a router becomes not interested in receiving multicast data, it sends a Prune message through its root interface. This message is then transmitted upwards the tree, through the root interfaces of upstream routers, until it finds an interested router (at which point it is discarded). Likewise, when a router becomes interested again it sends a Graft message through its root interface, and this message is transmitted upwards the tree until it finds an interested router (at which point it is discarded). The transmission of Graft messages is ACK protected on each link using Graft-ACK messages, and both messages are unicasted on shared links. On point-to-point links, a non-root interface is placed in PRUNED state when it receives a Prune message, and is placed in FORWARDING state when it receives a Graft message. On shared links, the behavior is more complex, and will be discussed later; ALs are kept in PRUNED state irrespective of the reception of Prune or Graft messages.%RV 26/1/2020

\begin{figure}[t!]
	\centering
	\includegraphics[scale=0.45]{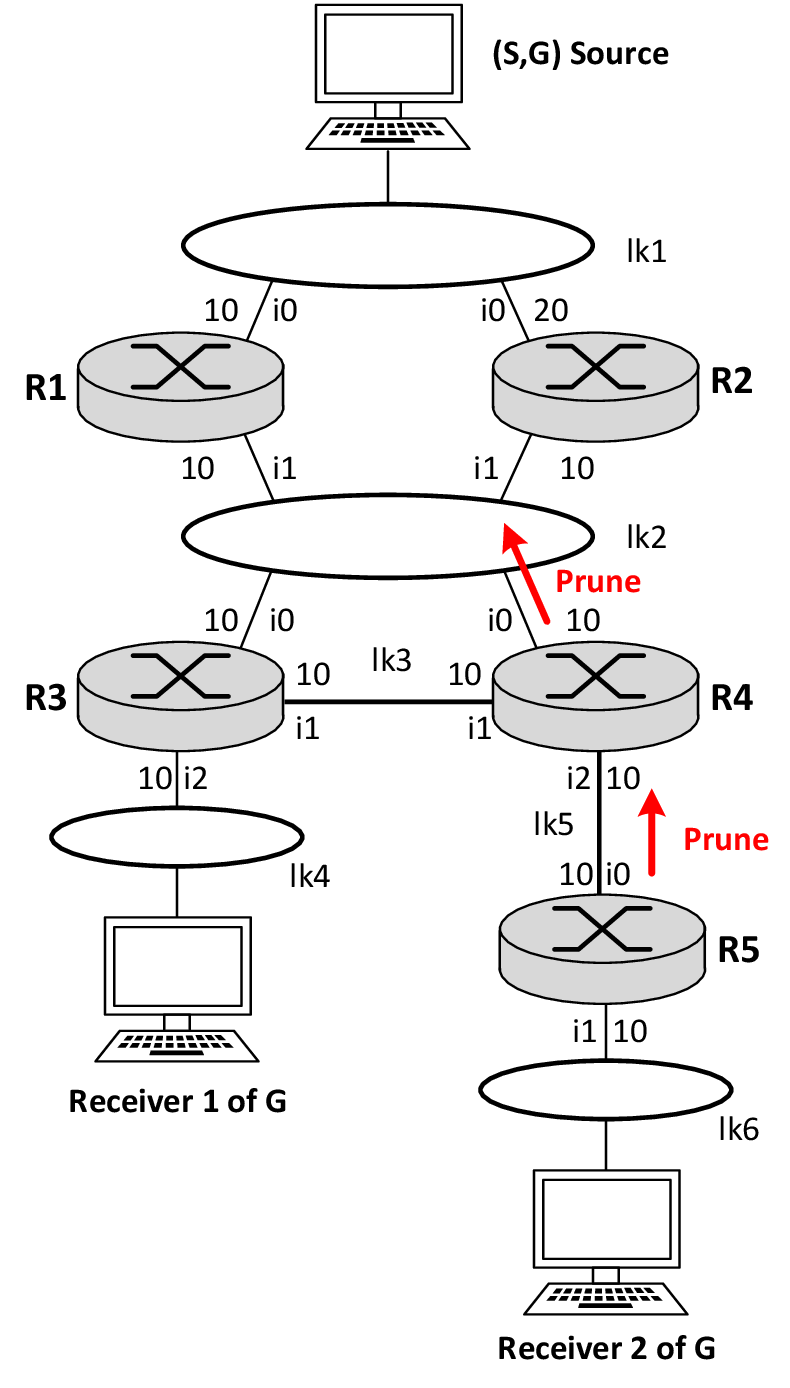}
	\caption{Pruning the broadcast tree due to lack of group interest, in PIM-DM.} 
	\label{fig:pimdm_pruneinterest}
\end{figure}

\medskip
\noindent \textbf{Example} Suppose that, in the network of Figure \ref{fig:pimdm_initialflood}, Receiver 2 ceases being interested in receiving multicast data of tree (S,G). What happens in this case is shown in Figure \ref{fig:pimdm_pruneinterest}. Receiver 2 signals R5 of this fact through the IGMP protocol. Then, R5 places its non-root interface in PRUNED state, becomes not interested since it has no downstream device (receiver or router) interested in receiving multicast data, and sends a Prune message through its root interface. When R4 receives this message, it puts the receiving interface (i2-R4) in PRUNED state and, similarly to R5, becomes not interested and sends a Prune message through its root interface, i.e. to the shared link lk2. The AW interface of the shared link, i.e. interface i1-R1, will not be pruned in this case, since Receiver 1 maintains its interest in receiving multicast data. This behavior will be addressed next. If later, Receiver 2 becomes again interested in receiving multicast data, it signals R5 of this intent using the IGMP, R5 and R4 send Graft messages through their root interfaces, and the non-root interfaces that were previously PRUNED become again FORWARDING.%RV 26/1/2020

%RV 22/10/2018

\medskip
\noindent \textbf{Prune Override} As mentioned above, if a router attached to a shared link through a root interface becomes not interested, it sends a Prune message through this interface. However, there may be other routers attached to the link that are still interested. To handle this possibility, when the AW receives a Prune message, instead of pruning immediately the interface, it waits some time to allow that interested routers express their interest. Interested routers do so by sending a Join message to the link. When the AW receives a Join, it terminates the waiting period and remains in FORWARDING state. If no Join message is received during the waiting period, the AW is placed in the PRUNED state. This mechanism is called \textit{Prune Override}. The timer that controls this waiting time is called \textit{Prune Pending Timer} (PPT) and has a default value of 3 seconds. Note that the routers that may be interested in receiving multicast traffic at a shared link are the ones connected to the link through root interfaces; non-root interfaces have no interest in receiving multicast traffic, since they can only act as traffic forwarders.%RV 26/1/2020

Note that the AW doesn't need to receive Join messages from all interested routers to determine that it needs to stay FORWARDING; it only needs to receive one. Thus, the Prune Override includes a mechanism for suppressing Join messages, to avoid excessive transmissions. When a root interface receives a Prune message and its router is still interested, instead of sending immediately the Join message, the interface delays the transmission for a random time between 0 and 2.5 seconds. The timer that controls this waiting period is called \textit{Override Timer} (OT). If a Join is received during the waiting period, the interface suppresses the transmission of its Join; otherwise, if the timer expires and no Join was received, the interface transmits its own Join.%RV 26/1/2020

Besides the above two features, when the AW is placed in the PRUNED state it sends a Prune message, to advertise its change of state and give a second opportunity to the root interfaces attached to the link for expressing their interest via Join messages. As will be explained later, this minimizes the impact of the loss of Join messages.%RV 31/1/2020

\begin{figure}[t!]
	\centering
	\includegraphics[scale=0.75]{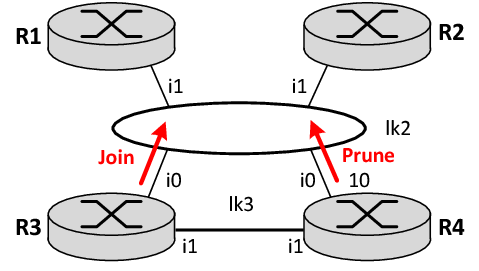}
	\caption{Prune Override in PIM-DM.} 
	\label{fig:pimdm_pruneor}
\end{figure}

\medskip
\noindent \textbf{Example} Figure \ref{fig:pimdm_pruneor} illustrates the Prune Override mechanism, using again the network of Figure \ref{fig:pimdm_initialflood}. Router R4 sends a Prune message, due to Receiver 2 becoming not interested. When the link AW (i.e. interface i1-R1) receives this message, it starts the Prune Pending Timer. However, when R3 receives the Prune message, it sends a Join message, after waiting a random time controlled by the Override Timer, since it is still interested in receiving multicast data, due to Receiver 1 being interested. The AW cancels the Prune Pending Timer when it receives the Join message and stays in the FORWARDING state.%RV 26/1/2020

%RV 22/10/2018

\medskip
\noindent \textbf{Initial expression of interest on shared links} To check whether, initially on a shared link, there are downstream routers interested in receiving multicast data, ALs must send a Prune message on the link upon being placed in this state. As discussed above, this message triggers a waiting period at the AW, forcing the AW to become PRUNED if no root interface expresses interest through a Join message, or keeping it in FORWARDING state otherwise. Moreover, a non-root interface that has no neighboring routers (as detected by the Hello protocol) stays in the NoInfo state but is placed in PRUNED state if there are no downstream multicast receivers interested.%RV 26/1/2020

\medskip
\noindent \textbf{Building the multicast tree} One can then conclude that the multicast tree is built in two steps: (i) a broadcast tree is formed through the flood-and-prune mechanism, which starts at the originator routers and proceeds downwards; (ii) the non-interested routers express their lack of interest upwards, in the direction of the originators, to prune the broadcast tree.  Data packets are transmitted downwards the tree, received at the root interface of interested routers and transmitted through their FORWARDING non-root interfaces. Thus, the multicast tree is formed by the root interfaces of interested routers and their FORWARDING non-root interfaces.%RV 26/1/2020

\medskip
\noindent \textbf{Periodic reconstruction of the tree} The mechanisms described above are not prepared to deal with network changes that might dictate the modification of the tree (e.g. router failures, changes in interface costs, or the source being switched off). PIM-DM addresses this problem by forcing the periodic reconstruction of the tree. There are two features forcing this reconstruction.%RV 26/1/2020

First, the AW and AL states of non-root interfaces are made temporary. After a period of time, controlled by the \textit{Assert Timer} (AT), the interface must return to the NoInfo state, with a reelection being triggered by the arrival of data messages. The Assert Timer has a default value of 3 minutes.%RV 26/1/2020

Second, the lack of downstream interest at the non-root interfaces must be renewed. Specifically, after a period of time of becoming not downstream interested, the interfaces consider that there is again downstream interest. This period is controlled by the \textit{Prune Timer} (PT) which has a default value of 3 minutes.%RV 26/1/2020

Thus, every 3 minutes, the non-root interfaces are placed in the FORWARDING state, since they return to the NoInfo state and become downstream interested, and the process of building the multicast trees starts all over again triggered by the transmission of multicast packets.%RV 26/1/2020

\medskip
\noindent \textbf{Sending Prune messages in reaction to data packets} A root interface that is not interested in receiving multicast data must keep replying to data packets with Prune messages. This reinstates the lack of interest in case Prune messages are lost or the broadcast tree is rebuilt. However, to avoid excessive transmission of Prune messages, its rate is limited. The transmission is regulated by the \textit{Prune Limit Timer}, which has a default value of 210 seconds.%RV 26/1/2020

\begin{figure}[t!]
	\centering
	\includegraphics[scale=0.45]{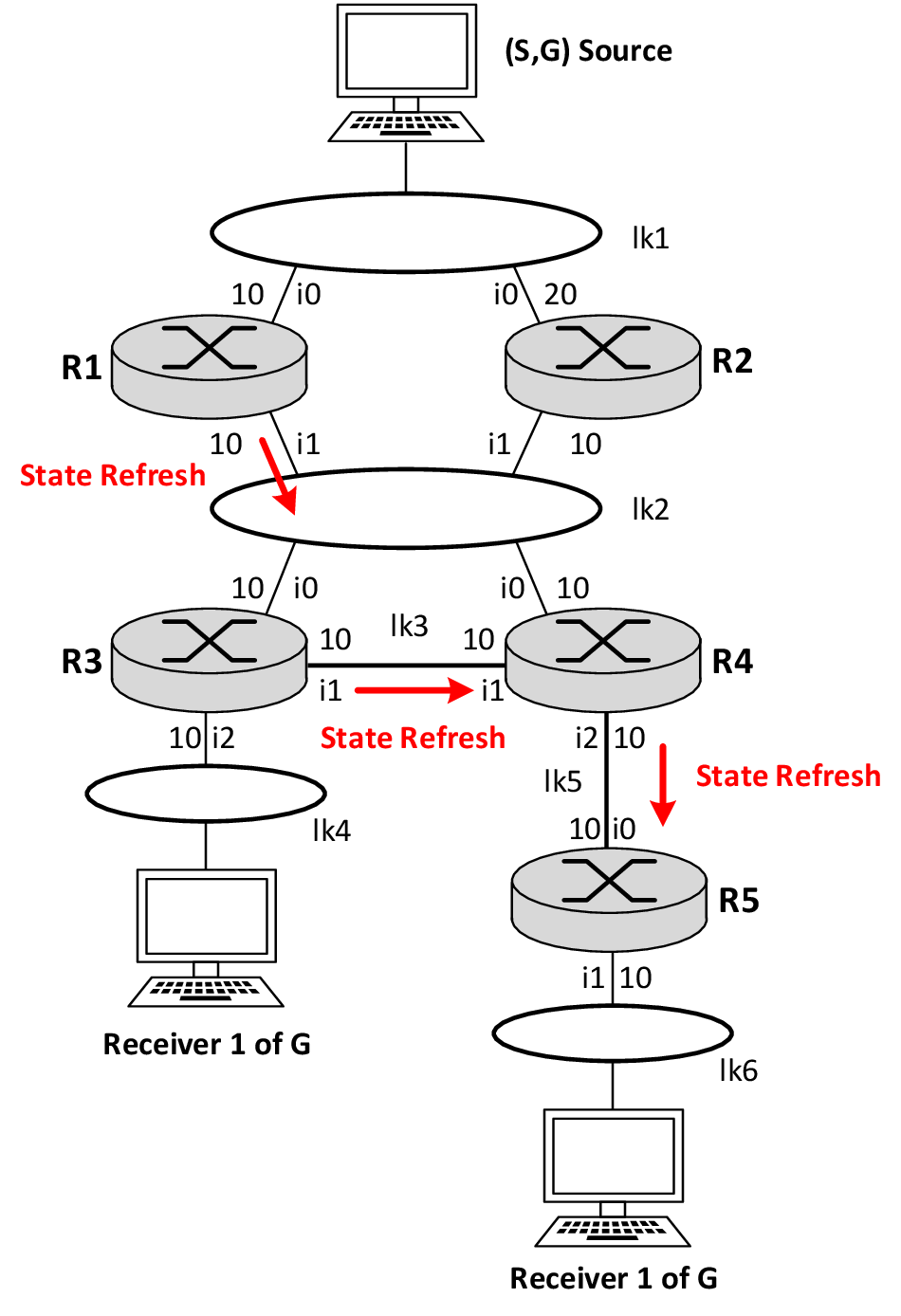}
	\caption{Flooding of State Refresh messages in PIM-DM.} 
	\label{fig:pimdm_staterefresh}
\end{figure}

\medskip
\noindent \textbf{State Refresh mechanism} To avoid the periodic reconstruction of the multicast tree, the State Refresh mechanism was added to PIM-DM. Its use is not mandatory and its support is signaled through a field in Hello messages. According to this mechanism, the originator routers periodically initiate the transmission of State Refresh messages, as long as the source remains active. These messages, when received at the root interface of a router, are transmitted by all its non-root interfaces that are AWs (in both shared and point-to-point interfaces), irrespective of their forwarding state. In this way, State Refresh messages are transmitted to all routers and links, even the ones not interested in receiving multicast traffic. The periodicity of the State Refresh transmissions is controlled by the \textit{State Refresh Timer} (SRT), which has a default value of 1 minute. Figure \ref{fig:pimdm_staterefresh} illustrates the flooding of State Refresh messages in the network of Figure \ref{fig:pimdm_initialflood}.%RV 31/1/2020

%RV: The RFC does not make a distinction between point-to-point and shared link interfaces. Thus, an AW is elected in both cases.

The State Refresh mechanism uses the \textit{Source Active Timer} (SAT) at originator routers to determine if a source remains active. The SAT has a default value of 210 seconds and is reset whenever a new data packet is received. When the SAT expires the router ceases considering itself originator and ceases transmitting State Refresh messages for the corresponding source.%RV 26/1/2020

When a State Refresh message arrives at a router, the router resets the Prune Timer and the Assert Timer of its non-root interfaces, preventing these interfaces from becoming FORWARDING while the source remains active.%RV 26/1/2020

Moreover, State Refresh messages include the RPC of the sending router, which enables the reconfiguration of the tree when the shortest paths to the source's subnet change. When a State Refresh message is received with an RPC higher than the RPC of the receiving router (indicating that the AW increased its RPC), the receiving router sends an Assert message to trigger the AW reelection.%RV 26/1/2020

If State Refresh messages cease from being received, the Prune Timer and the Assert Timer are not reset and will expire, causing the PRUNED interfaces to become FORWARDING.%RV 26/1/2020

State Refresh messages include a flag, called \textit{Prune Indicator}, indicating whether the sending interface is PRUNED or FORWARDING. This allows new routers that attach to a link (shared or point-to-point) where multicast traffic is currently not being transmitted, i.e. where the AW is PRUNED, to express their interest and start receiving multicast data. Specifically, if a root interface receives a State Refresh indicating that the sending interface, i.e. the AW of its upstream router, is PRUNED and the receiving router is interested, the interface sends a Join. Moreover, if the root interface receives a State Refresh indicating that the AW is FORWARDING and the receiving router is not interested, the interface sends a Prune, under the constraint of the Prune Limit Timer.%RV 26/1/2020

%RV: This is an optimization since a Join could always be sent in response to State Refresh.

%PO: Relativamente aos testes GNS3 dá-me a sensação que a implementação Cisco não está de acordo com o RFC.
%(i) Todas as interfaces non-root transmitem State Refresh, quando apenas o AW é que deveria transmitir.
%(ii) Dá a sensação que o State Refresh apenas refresca o estado de Assert de interfaces root, porque mesmo com todas as interfaces non-root a encaminhar mensagens de State Refresh, periodicamente vejo mensagens de Assert a reeleger o AW.
%(iii) Periodicamente a mensagem de State Refresh do AW indica que está Pruned (possivelmente para verificar se ainda continua a haver interesse em dados). Se ainda há routers interessados estes fazem Join (prune override). De acordo com o RFC a mensagem deveria dizer que está Pruned apenas quando a máquina de estados Downstream Interest encontra-se no estado Pruned (a interface nunca deveria estar neste estado com routers interessados).
%(iv) Após morte da única interface root interessada, o router envia mensagens de State Refresh mas incluindo informação que a interface está Pruned (e continua a enviar dados durante um periodo de tempo - possivelmente o tempo de override). Depois ao não ouvir mensagem de Join fica pruned e deixa de encaminhar dados.

%RV 23/10/2018

\section{PIM-DM issues}\label{sec:pimdmissues}

PIM-DM has a number of issues that limit its performance (i.e. convergence time) and impair its use in high-speed networks. These issues are caused by the lack of reliability and the lack of preservation of transmission order of control messages, as well as some other intrinsic protocol features.%RV 26/1/2020

\medskip
\noindent \textbf{Lack of reliability in the transmission of Join and Prune messages} The transmission of Join and Prune messages is not protected and this may cause problems during a Prune Override. Suppose that a Prune message is transmitted on a shared link, and that the Join messages sent by the root interfaces that remain interested in receiving multicast data are lost. This would be the case of Figure \ref{fig:pimdm_pruneor}, if the Join message sent by R3 is lost. In this case, the AW will become PRUNED upon expiration of the Prune Pending Timer, and the downstream root interfaces attached to the link will cease receiving multicast data. Moreover, all routers that are downstream of the affected ones on the multicast tree will also cease receiving multicast data. This problem will only be solved when the tree is reconstructed, which can take as long as 3 minutes, or when a State Refresh message arrives, which can take as long as 1 minute; in the meantime, several zones of the multicast tree may not receive multicast traffic.%RV 26/1/2020

When the tree is reconstructed, data packets are again transmitted at the affected link and the root interfaces will have a new opportunity to express their interest. In the case of State Refresh messages, these messages include a flag indicating whether the AW that sent the message is FORWARDING or PRUNED. A root interface that is interested, but receives a State Refresh message indicating that the sending AW is PRUNED, transmits a Join message on the link which, hopefully, places the AW in the FORWARDING state.%RV 26/1/2020

To mitigate this problem, the AW is forced to send a Prune message when it changes to PRUNED state. This message is a confirmation that the interface became PRUNED and gives a second chance for interested root interfaces to express their interest, by sending new Join messages. However, this does not completely solve the problem, since both Prune and Join messages can be lost without notice.%RV 26/1/2020

\medskip
\noindent \textbf{Delay introduced by the Prune Override mechanism} On shared links, the AW is only pruned after the expiration of the Prune Pending Timer, which has a default value of 3 seconds. This introduces a delay in the pruning process, which gets worse when the change of interest causes a chain of AW interfaces upwards the tree to become PRUNED. The total delay in pruning the interfaces is 3 seconds times the number of interfaces that need to be pruned and, in the meantime, multicast traffic keeps being transmitted on links that are no longer interested in it.%RV 26/1/2020

\begin{figure}[t!]
	\centering
	\includegraphics[scale=0.45]{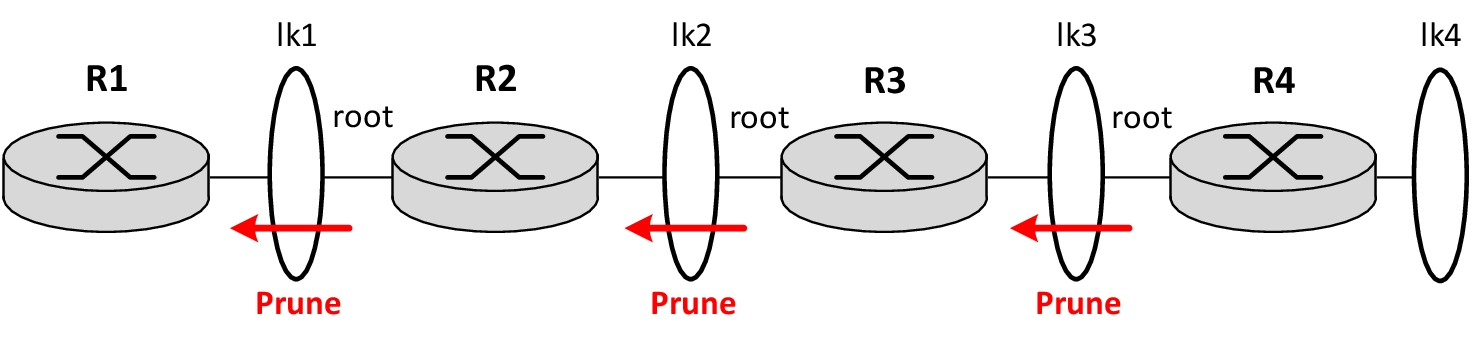}
	\caption{Delay introduced by the Prune Override in PIM-DM.} 
	\label{fig:pimdm_podelay}
\end{figure}

Consider the network of Figure \ref{fig:pimdm_podelay}, showing a chain of four routers connected via shared links. Suppose that R4 has lost interest, e.g. because the receivers connected to its non-root interface, i.e. to link lk4, ceased to have interest, and suppose also that upwards the tree there is no interest from links lk1, lk2, and lk3. In this case, the non-root interface of R1 will only become PRUNED 9 seconds after R4 has lost interest.%RV 26/1/2020

%We may need to include a second router in each link to avoid confusion with the idea of next paragraph

PIM-DM includes an optimization in case a router is only attached to one neighbor on a shared link, as detected by the Hello protocol. In this case, when the router receives a Prune message on a non-root interface, it immediately prunes the interface, without setting the Prune Pending Timer, and thus avoiding the Prune Override delay.%RV 26/1/2020

%RV 23/10/2018

\medskip
\noindent \textbf{New routers are unaware of previously formed trees} When a new router is switched on and attaches to the network, it may not receive immediately information about the currently active trees. This will happen when the router attaches a link that is currently not receiving multicast data, due to lack of interest of the neighboring routers. The router will have to wait until either the tree is reconstructed, another downstream router sends a Graft message, or a State Refresh message is received.%RV 26/1/2020

\begin{figure}[t!]
	\centering
	\includegraphics[scale=0.4]{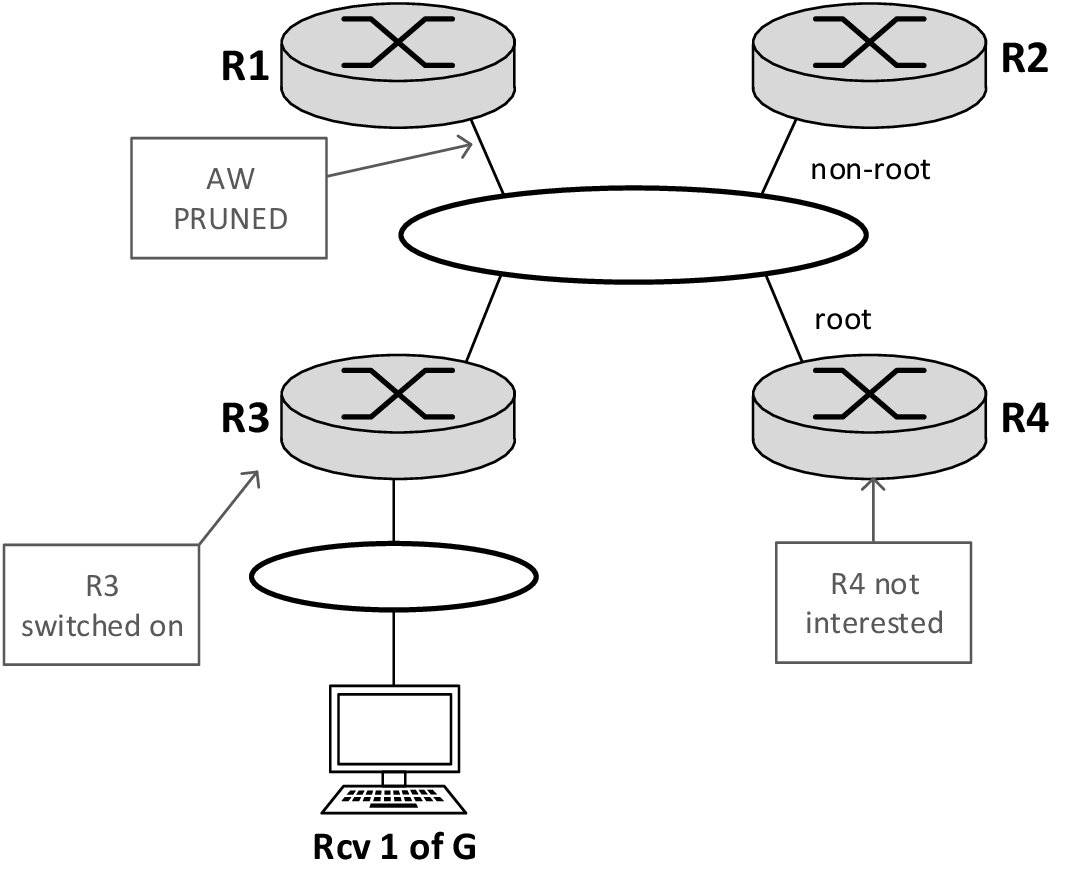}
	\caption{New routers unaware of previously formed trees in PIM-DM.} 
	\label{fig:pimdm_isolated}
\end{figure}

Consider the network of Figure \ref{fig:pimdm_initialflood} where, initially, on link lk2, R4 is not interested and R3 is switched off. In this case, the AW is PRUNED and no multicast data is transmitted on the link. This is represented in Figure \ref{fig:pimdm_isolated}. In this scenario, if R3 is switched on and Receiver 1 is interested in the (S,G) tree, R3 has no means to know if the tree is active. As mentioned above, it will have to wait until the tree is reconstructed, R4 sends a Graft, or a State Refresh message is received. If a State Refresh is received, R3 becomes aware of the tree and, since a flag included in the message indicates that the AW is PRUNED, it sends a Join to the link which places the AW in FORWARDING state.%RV 26/1/2020

%RV 23/10/2018

To mitigate this problem, PIM-DM allows the replay of State Refresh messages to downstream neighbors upon the establishment of neighborhood relationships. Specifically, when two routers become neighbors through the Hello protocol, if a router considers the neighbor as downstream regarding some tree and has a stored State Refresh message of that tree, it may replay the message to the neighbor. This feature is optional. Moreover, it does not completely solve the problem, since the transmission of State Refresh messages is not protected.%RV 26/1/2020

\medskip
\noindent \textbf{Slow reaction to RPC changes} In PIM-DM, the tree is not reconfigured when the RPC changes, since the state machine of the Assert protocol does not react to this event (routers do not send Assert messages when their RPC changes). Thus, when the RPC changes the tree remains unchanged and may no longer provide an optimal forwarding of multicast data (i.e. along reverse shortest paths). The problem is only solved when the tree is rebuilt or when a State Refresh message is received. State Refresh messages are transmitted by the AW and include its RPC. If an AL receives a State Refresh with an RPC higher than its own (indicating that it should now become the AW), then it moves to the NoInfo state and becomes FORWARDING. Thus, subsequent data packets received by the AL are transmitted on the link, and this will trigger the reelection of the AW. Recall that non-root interfaces in AW or NoInfo states react to the reception of data packets by sending an Assert message.%RV 26/1/2020

Consider again the network of Figure \ref{fig:pimdm_initialflood} and suppose that interface i0-R1 fails. In this case, i0-R1 becomes non-root and i1-R1 becomes root. Since i1-R1 was the previous AW of link lk2, it sends an Assert message with a metric of infinity, triggering the election of the new AW, which can only be i1-R2. If this Assert message is lost, no State Refresh message is transmitted on this link since other routers still believe that i1-R1 is the AW. This problem is only solved when the tree is rebuilt, triggering the election of the new AW.%RV 26/1/2020

Suppose now that the cost of i0-R1 changes to 25, instead of 40. In this case, there is no interface role change, but R1 offers now a worse RPC than R2 to the source. However, there will be no immediate reaction, and multicast packets will keep being forwarded on the shared link through i1-R1. This problem is only solved when the tree is rebuilt or a new State Refresh message is transmitted by R1. This message includes the new RPC of R1, which will trigger the AW reelection.%RV 26/1/2020

\medskip
\noindent \textbf{AW unaware of interested neighbors} The AW does not store information on the downstream routers that are interested in receiving multicast data; it only knows whether some router is interested. Thus, if the last interested router fails, the AW keeps believing that there are downstream routers interested, and keeps transmitting multicast data on the link. The way this problem is solved depends on the type of interfaces that remain connected to the link.%RV 26/1/2020

If only non-root interfaces remain connected and State Refresh is not used, the problem is only solved when the tree is reconfigured. In this case, the ALs will send a Prune to the AW, which will place the AW in PRUNED state (since no Join is transmitted on the link). The PIM-DM specification does not address the actions that should be taken in case State Refresh is used. However, at least in some implementations, when the sending interface is FORWARDING, the State Refresh messages it sends, periodically, have the Prune Indicator flag set (every three State Refresh transmissions) indicating that the AW is PRUNED, with the purpose of forcing interested interfaces (if any) to restate their interest.%RV 26/1/2020

If non-interested root interfaces remain connected to the link, the problem can be solved by the reception of data packets or State Refresh messages. In the first case, the interface replies with a Prune message; in the second case, it also replies with a Prune message since the State Refresh message indicates that the AW is FORWARDING. In both cases, the transmission of Prune messages is rate-limited, which could delay the pruning of the AW.%RV 26/1/2020

For example, suppose that in the network of Figure \ref{fig:pimdm_initialflood}, R3 is initially interested but R4 is not, and R3 fails. Since the AW only knows that there is some downstream router interested in receiving traffic, when R3 fails it stays in FORWARDING state and continues transmitting multicast data on the link. Eventually, R4 reacts with a Prune message and places the AW in PRUNED state, but this may take some time.%RV 26/1/2020

\medskip
\noindent \textbf{Lack of message ordering guarantees} In PIM-DM there is no guarantee that the transmission order of control messages is preserved. This is problematic in case the order of Prune and Join/Graft messages is reversed. If a downstream router transmits a Prune and then a Join/Graft, but the Prune is the last one to arrive, the AW becomes PRUNED while there are still interested downstream routers. In the opposite case, the AW becomes FORWARDING while there are no downstream interested routers.%RV 26/1/2020

%% file: cc_hpimdm_v20.tex
\section{HPIM-DM specification}\label{sec:hpimdm}

HPIM-DM (Hard-state Protocol Independent Multicast - Dense Mode) is a hard-state multicast routing protocol designed for dense networks that overcomes the problems of PIM-DM and is fully optimized to provide fast convergence while minimizing the amount of stored state and completely avoiding replay attacks.%RV 31/1/2020

HPIM-DM builds and maintains a broadcast tree (spanning all network routers), for each active (S,G) pair. As in PIM-DM, the broadcast tree is a reverse shortest path tree supported on a unicast routing protocol. To build and maintain the broadcast tree, routers store information about all their upstream neighbors, i.e. neighbors from which multicast data can be received and, based on this information, control whether they are connected to the tree or not. Moreover, as in PIM-DM, a non-root interface is elected at each link to become the link forwarder (i.e. the AW). The state information kept by routers to build and maintain the broadcast tree is then of three types: upstream (if a neighbor is upstream), tree (if a router is connected to the tree), and assert (if a non-root interface is AW). In order to keep this information updated, routers exchange control messages generically called upstream messages.%RV 31/1/2020

The broadcast tree can be pruned from not interested routers and router interfaces; the pruned version of the broadcast tree is called the multicast tree. Ultimately, the multicast tree is determined by the set of non-root interfaces (over all network routers) selected to transmit multicast data, i.e. the interfaces that are in forwarding state. The forwarding state of a non-root interface depends on its assert state (only AW interfaces can forward multicast data) and on the interest of the downstream devices reachable through the interface (multicast receivers or other routers). Thus, each non-root interface keeps information on the interest of its neighbors, and use this information to determine if the interface has any interested device; this is called the downstream interest state of the interface. The interest of a router in receiving multicast data is determined by the assert state and downstream interest state of its non-root interfaces. In order to keep this information updated, routers exchange control messages generically called interest messages.%RV 31/1/2020  

Unlike PIM-DM, where the broadcast tree is formed through the transmission of multicast data packets using the RPF technique, in HPIM-DM the construction and maintenance of the broadcast and multicast trees is done exclusively using control messages. All control messages are transmitted reliably and with ordering guarantees, i.e. they are always received and processed according to the order they were transmitted. Control messages need not be transmitted periodically (as in soft-state protocols), but rather in reaction to events susceptible of changing the configuration of the trees. The sequencing and reliability of message transmissions is an essential ingredient to guarantee the correctness and fast convergence of HPIM-DM.%RV 31/1/2020

To ensure that routers joining the network quickly start receiving multicast data, HPIM-DM includes a synchronization process where a joining router obtains information from each neighbor on the active trees for which the neighbor can act as an upstream router; becoming synchronized is also a requirement for establishing neighborhood relationships. This is implemented through the exchange of control messages known as synchronization messages, and synchronization state that controls the various phases of the synchronization process.%RV 2/2/2020

As in PIM-DM, routers maintain neighborhood relationships using the Hello protocol and Hello messages. However, unlike PIM-DM, the Hello protocol is not used to establish a relationship (i.e. to declare a router as neighbor), since this requires that the two neighbors synchronize.%RV 2/2/2020

\begin{figure}[t!]
	\centerline{\includegraphics[scale=0.8]{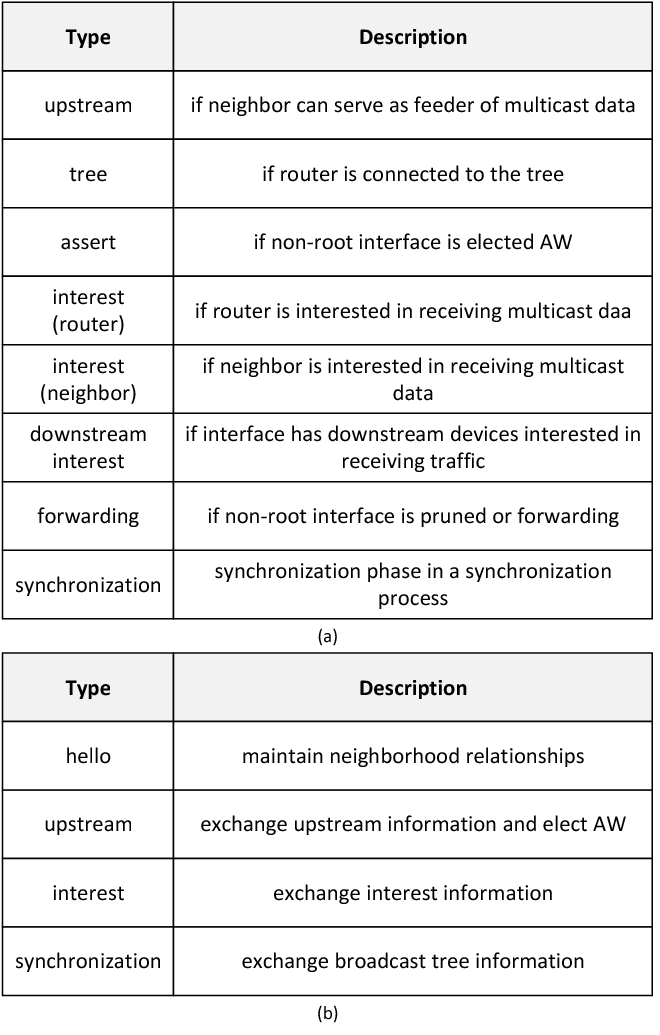}}
	\caption{Type of (a) state information and (b) control messages of HPIM-DM.}
	\label{fig:hpimdmstatesmessages}
\end{figure}

The type of state information and control messages used by HPIM-DM is summarized in Figure \ref{fig:hpimdmstatesmessages}.%RV 31/1/2020

%RV 30/1/2020

The following sections explain in detail the operation of the protocol. Section \ref{sec:hello} addresses the Hello protocol, used to maintain neighborhood relationships. Section \ref{sec:treemaintenance} discusses how the broadcast tree is built and maintained, and section \ref{sec:interestmaintenance} explains how this tree is pruned according to the interest of multicast receivers, leading to the multicast tree. Section \ref{sec:dataflooding} explains how data is forwarded initially, while the multicast tree is being formed. Section \ref{sec:messagesequencing} discusses the sequencing of control messages and section \ref{sec:initsync} explains the synchronization process. Section \ref{sec:messagereliability} discusses how the reliability of control message transmissions is provided. Section \ref{sec:securityissues} addresses the protocol mechanisms that avoid security attacks. Section \ref{sec:messageformat} presents the format of the control messages.  Finally, section \ref{sec:storedstate} summarizes the state information that needs to be stored at routers.%RV 25/6/2019

The detailed specification of HPIM-DM, including its state machines, is presented in \cite{hpim_state_machines}.%RV 30/1/2020

\subsection{Hello protocol}\label{sec:hello}%RV 12/2/2019

A mechanism is needed to maintain router neighborhood relationships, whereby routers discover which are their active neighbors that want to participate in the multicast routing process. This mechanism is used in many other routing protocols, e.g. OSPF and EIGRP, and is known as the Hello protocol.%RV 30/1/2020

There are three phases in a neighborhood relationship: detection of a neighbor, establishment of relationship, and maintenance of relationship. In HPIM-DM, the neighbor detection can be performed by any control message, and not just by Hello messages. This contrasts with other protocols, and aims at speeding up the protocol convergence when a new neighbor is found. Moreover, the relationship establishment (i.e. when a neighbor is considered active) requires that the two neighbors synchronize with each other their broadcast tree information; the synchronization process is discussed in section \ref{sec:initsync}. Finally, the relationship maintenance (i.e. knowing if the neighbor is still alive) is carried out by the Hello protocol via the periodic transmission of Hello messages.%RV 30/1/2020

The Hello protocol works as follows:
\begin{itemize}
	\item A router keeps periodically sending Hello messages on each of its links to all its neighbors. On shared links, the transmission of Hello messages is multicasted to all neighbors (using multicast addresses reserved for HPIM-DM), such that only a single message needs to be transmitted.
	\item After establishing a relationship with a neighbor (which requires synchronization), a router keeps monitoring the Hello messages transmitted by the neighbor. A neighbor is declared dead when a predefined time period (typically four times the periodicity of Hello transmissions) has elapsed without having received any Hello from it.
\end{itemize}%RV 25/6/2019

Thus, the Hello protocol requires one timer to regulate the transmission of Hello messages, called \textit{Hello Timer} (HT), and one timer per neighbor to monitor the Hello messages received from the neighbor, called \textit{Neighbor Liveness Timer} (NLT). We kept the nomenclature of PIM-DM.%RV 25/6/2019

In other routing protocols, such as OSPF and IS-IS, there is a clear separation between the establishment of neighborhood relationships and the synchronization of routing information; in OSPF and IS-IS, routers synchronize their link state databases. In these protocols, the establishment of neighborhood relationships is performed exclusively by the Hello protocol. The Hello messages transmitted by a router carry the addresses of all its neighbors on the link, and a neighbor is considered active when bidirectional communication with it has been established. Specifically, a router considers a neighbor active when it sees its address listed in the messages transmitted by the neighbor. This solution is prone to replay attacks: an attacker may record the initial Hello message transmitted by a router, where no neighbor is declared, and then use this message later to destroy neighborhood relationships. HPIM-DM avoids this problem by declaring a neighbor active only when a successful synchronization with it is completed. Thus, contrarily to OSPF and IS-IS, the exchange of Hello messages is not sufficient to declare a neighbor active.%RV 25/6/2019

\begin{figure}[t!]
	\centerline{\includegraphics[scale=0.45]{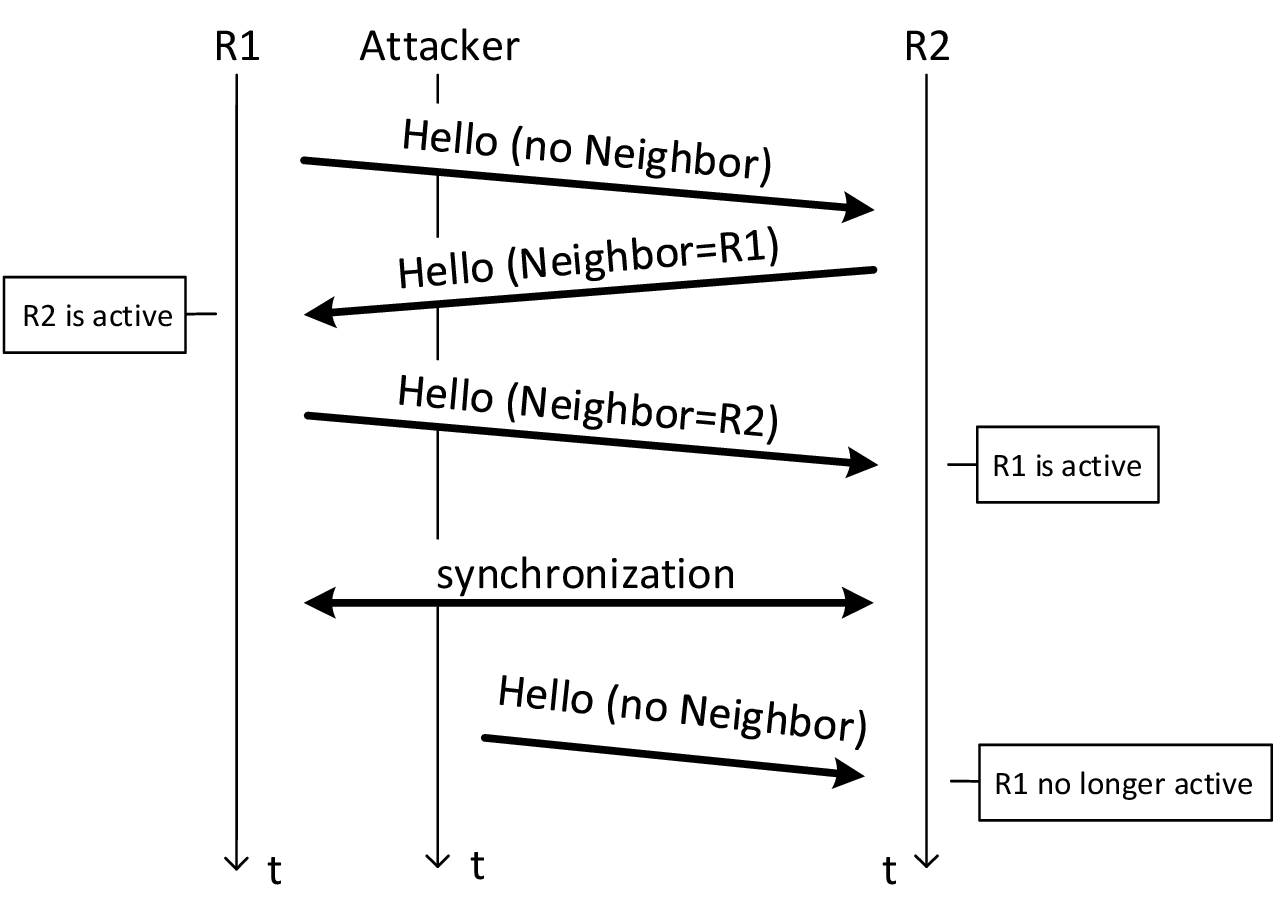}}
	\caption{Hello replay attack.}
	\label{fig:helloattack}
\end{figure}

Figure \ref{fig:helloattack} illustrates the replay attacks. The attacker stores the first Hello message sent by R1, which has no neighbors listed. Router R2 considers R1 active when it receives the second Hello from R1 with its own address. If later the attacker replays the first Hello sent by R1, R2 ceases from considering R1 active.%RV 25/6/2019

%RV 21/1/2019

\subsection{Broadcast tree maintenance}\label{sec:treemaintenance}

The broadcast tree allows multicast data to be transmitted from an originator (a router directly attached to the multicast source) to all other network routers. After being formed, this tree may be pruned from routers or router interfaces not interest in receiving multicast data, an issue that will be discussed in section \ref{sec:interestmaintenance}. In HPIM-DM, a router always keeps information on the broadcast tree (i.e. who its parent in the tree is), irrespective of its interest in receiving multicast data, and as long as the corresponding source remains active. This is performed without the need for the periodic circulation of control messages or data. In PIM-DM, the broadcast tree is built through the circulation of multicast data, and the tree is maintained at a router as long as data or State Refresh messages keep arriving. Thus, PIM-DM is a soft-state protocol whereas HPIM-DM is a hard-state protocol.%RV 31/1/2020 

Similarly to PIM-DM, the formation and maintenance of the broadcast tree relies on an unicast routing protocol. The unicast routing protocol allows determining, at a router, the root interface (the one that provides the shortest path to the source and should receive multicast data) and the non-root interfaces (the ones that may transmit multicast data) for a given tree.%RV 31/1/2020

\subsubsection{Basic aspects}

\begin{figure}[t!]
	\centerline{\includegraphics[scale=1]{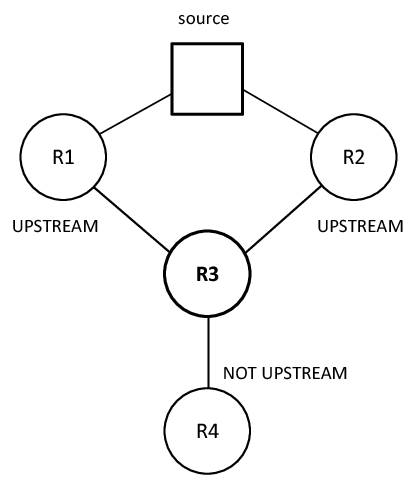}}
	\caption{The concept of upstream neighbor.}
	\label{fig:upstreamnei}
\end{figure}

\medskip
\noindent \textbf{The concept of upstream neighbor} An upstream neighbor is a neighbor connected to the source from which multicast data can be received; it is a next-hop router with a unicast route to the source. In HPIM-DM, a router stores information on \textit{all} its upstream neighbors. This allows fast reconfiguration of the broadcast tree, in case the upstream neighbor currently being used to receive multicast data loses connection with the tree. Figure \ref{fig:upstreamnei} gives an illustration. Consider the case of router R3. Its upstream neighbors are R1 and R2, since they can both feed R3 with multicast data; R4 is not an upstream neighbor (it would be if there was a direct link connecting it to the source, R1 or R2). Suppose that R1 is currently being used to receive multicast data. If R1 fails, R3 switches immediately to R2, since it has information on all its upstream routers.%RV 2/2/2020

%RV 31/1/2020

\medskip
\noindent \textbf{Control messages} Two control messages are used in the broadcast tree maintenance procedure: IamUpstream and IamNoLongerUpstream. We will refer to these messages as the \textit{upstream} messages. The upstream messages are transmitted downwards the tree, i.e. through non-root interfaces, except when an interface changes from non-root to root. In this case, the interface that just became a root must transmit an IamNoLongerUpstream message. The upstream messages are transmitted between neighbors, i.e. with link-local scope, using as IP destination address the IP multicast address reserved for HPIM-DM. There are different upstream messages for each (S,G) tree; each message includes the S and G addresses that identify the tree, i.e. the IP address of the source (the S address) and the IP multicast address identifying the multicast group (the G address). The transmission of upstream messages is ACK protected, to ensure they are delivered to all intended receivers. The IamUpstream message indicates to a receiving router that the sending router can act as a feeder of multicast data (i.e. is an upstream neighbor), and the IamNoLongerUpstream message indicates the opposite. The IamUpstream message includes the RPC of the sending router, to elect the AW at non-root interfaces and to eliminate routing loops.%RV 31/1/2020

\medskip
\noindent \textbf{Upstream state of neighbors} For each tree, each router stores information on whether a neighbor is UPSTREAM or NOT UPSTREAM, i.e. if it can act as a feeder of multicast data or not. A neighbor is UPSTREAM at a router if the last upstream message received from that neighbor was an IamUpstream message, and is NOT UPSTREAM if the last message received from that neighbor was an IamNoLongerUpstream message. Initially, before receiving an upstream message the neighbor is considered NOT UPSTREAM. Storing information on all upstream neighbors, and not just on the parent router, is a distinguishing feature of HPIM-DM that allows fast convergence when the tree changes. The parent router is selected among the UPSTREAM routers.%RV 31/1/2020

\medskip
\noindent \textbf{Parent router} The parent of a router in a broadcast tree is the UPSTREAM neighbor reachable through the root interface that provides the lowest RPC. Moreover, an UPSTREAM neighbor is only considered parent if it provides an RPC lower than the router itself (i.e. if it is closer to the source). This condition is used in other routing protocols to prevent the existence of routing loops; it is called \textit{feasibility condition} \cite{RFC6126,dsdv,dual,eigrp}. In our case, it is used to ensure that routers do not maintain trees indefinitely in the presence of routing loops. This will be illustrated in section \ref{sec:examplestree}.%RV 28/1/2020

A router will only transmit an IamUpstream message downwards the tree if it has a parent on the tree. In this way, when a router receives an IamUpstream message on a root interface, it can be certain that all routers between itself and the source form a tree structure that can forward multicast data.%RV 28/1/2020

\medskip
\noindent \textbf{Installing and removing a tree} When a source becomes active, the originators initiate the construction of the broadcast tree by transmitting IamUpstream messages through all its non-root interfaces. These messages are forwarded downwards the tree and reach all network routers. Specifically, when a router receives the first IamUpstream message from its parent on the tree, it transmits IamUpstream messages through all its non-root interfaces. Note that receiving an IamUpstream message from a non-parent router (e.g. when the message is received at a non-root interface) does not trigger the transmission of IamUpstream messages downwards; it just sets the state of the sending router as an UPSTREAM neighbor.%RV 29/1/2020

Likewise, when a source becomes inactive, the originators transmit IamNoLongerUpstream messages through all its non-root interfaces, and these messages are forwarded downwards the tree, reaching all network routers and removing the tree state. The contents of the IamUpstream messages is modified as they travel downwards the tree: the RPC information must be updated to include the RPC of the sending router.%RV 28/1/2020

A router resorts to the unicast routing protocol and the RPC information to determine if an IamUpstream message was transmitted by its parent on the tree. First, the router determines if the message was received at its root interface, based on the IP address of the source carried in the message and on its unicast routing table. Second, based on the RPC information (the one contained in the message and the one stored at the router for each UPSTREAM neighbor) it determines if the neighbor that transmitted the message provides the lowest RPC and observes the feasibility condition.%RV 28/1/2020

\begin{figure}[t!]
	\centerline{\includegraphics[scale=1]{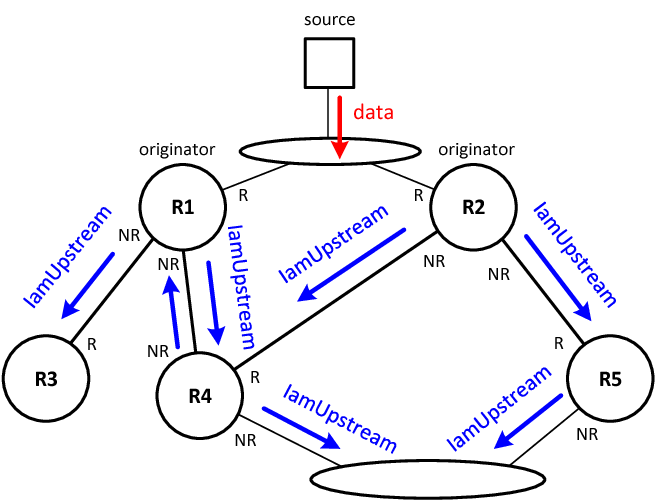}}
	\caption{Installing a tree (NR = non-root; R = root).}
	\label{fig:buildtree}
\end{figure}

The process of installing a tree is illustrated in Figure \ref{fig:buildtree}. When the source starts sending data, the originator routers transmit IamUpstream messages through their non-root interfaces and these messages are propagated downwards the tree. A non-originator router will transmit IamUpstream messages downwards the tree, when it receives an IamUpstream message from its parent on the tree, i.e. when the message is received on its root interface and the feasibility condition is observed. For example, at R4, the reception of the IamUpstream message coming from R2, triggers the transmission of the IamUpstream messages to R1 and to R5 (via the shared link), since the message was received at R4's root interface (meaning that R2 is R4's parent); contrarily, the IamUpstream message received from R1, does not trigger any IamUpstream transmissions, it just sets R1 as R4's UPSTREAM neighbor.%RV 2/2/2020

%RV 2/2/2020

\begin{figure}[t!]
	\centerline{\includegraphics[scale=1]{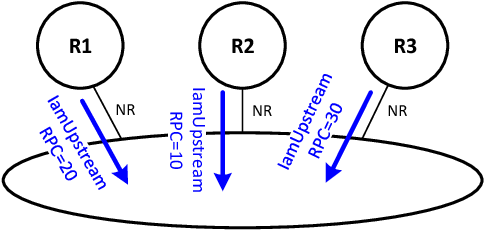}}
	\caption{Proving information to select the AW interface (NR = non-root).}
	\label{fig:assert}
\end{figure}

\medskip
\noindent \textbf{Assert protocol} The AW is elected in the same way in point-to-point and shared links. The election is a simple process, since IamUpstream messages carry the RPC of the sending router, and routers keep information on who their UPSTREAM neighbors are and what are their RPCs. When a source becomes active and a broadcast tree is built, all non-root interfaces attached to shared links transmit IamUpstream messages (after the corresponding routers have received IamUpstream messages from their tree parents). Based on this information, each interface (root or non-root) elects the AW as the non-root interface that advertised the lowest RPC. If only one non-root interface transmits an IamUpstream message, that interface will become AW. If two or more interfaces offer the same lowest RPC, the one with highest IP address becomes AW. This process is illustrated in Figure \ref{fig:assert}. In this case, R2 will be elected AW since it has the lowest RPC.%RV 28/1/2020

%RV 28/1/2020

\subsubsection{Router tree states}\label{sec:treestates}

A \textit{router} is considered to be in one of three states regarding a broadcast tree: ACTIVE, UNSURE, or INACTIVE. These states will be called the \textit{tree states}. In the ACTIVE state, the source is considered to be active and in INACTIVE state it is considered to be inactive. The UNSURE state is a transient state, where the router keeps information about the tree but is unsure on whether the source is active or not; this state was introduced for protocol correctness and to speedup convergence. The definitions of the tree states are slightly different in originator and non-originator routers. The definition is the following for non-originator routers: 
\begin{itemize}
	\item ACTIVE - A router is in ACTIVE state if it has a parent on the tree.
	\item UNSURE - A router is in UNSURE state if it has no parent on the tree, but has at least one UPSTREAM neighbor on some interface.
	\item INACTIVE - A router is in INACTIVE state if it has no UPSTREAM neighbor on any interface.
\end{itemize}
Note that a neighbor that is UPSTREAM regarding a router is necessarily in ACTIVE state, since only in this state could it have sent an IamUpstream message to the router.%RV 28/1/2020

\begin{figure}[t!]
	\centerline{\includegraphics[scale=0.85]{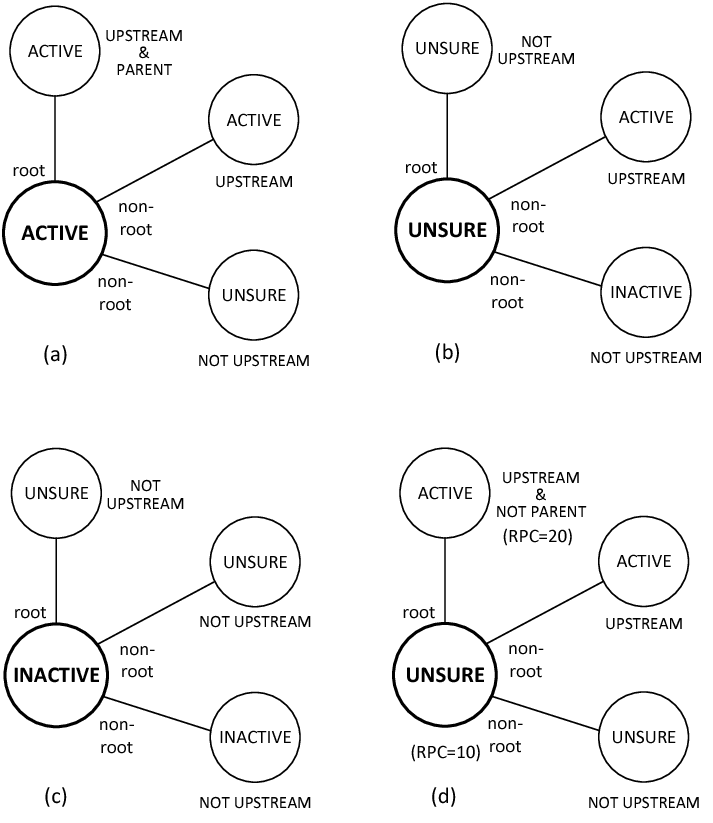}}
	\caption{Relationship between tree states and upstream states.}
	\label{fig:treestates}
\end{figure}

Figure \ref{fig:treestates} illustrates the router trees states and its relationship with the upstream state of neighbors. In the figure there is only one neighbor per interface; however, if the interface is to a shared link, there can be more than one. In Figure \ref{fig:treestates}.a the router represented in bold is ACTIVE since it has a parent on the tree, i.e. it has one UPSTREAM neighbor connected to its root interface that respects the feasibility condition. The router in Figure \ref{fig:treestates}.b is UNSURE since it has no UPSTREAM neighbor connected to its root interface, but has one connected to a non-root interface. In Figure \ref{fig:treestates}.c the router is INACTIVE since it has no UPSTREAM neighbor connected to any of its interfaces. Finally, the scenario of Figure \ref{fig:treestates}.d is similar to that of Figure \ref{fig:treestates}.a, except that the UPSTREAM neighbor connected to root interface does not respect the feasibility condition (it offers an RPC higher than the router itself) and, therefore, can not be parent; in this case, the router becomes UNSURE, since it has an UPSTREAM neighbor connected to a non-root interface.%RV 2/2/2020

%RV 28/1/2020

For originator routers, a router is ACTIVE if the source is considered active, i.e. if it is transmitting multicast data. A source is considered inactive if it ceases transmitting data for a period of time. The router is UNSURE if the source is considered inactive but the router has at least one UPSTREAM neighbor, and is INACTIVE if the source is considered inactive and the router has no UPSTREAM neighbor on any interface.%RV 2/2/2020

\medskip
\noindent \textbf{Importance of the UNSURE state} The UNSURE state is required both for correctness and to speedup convergence, and is a distinguishing feature of HPIM-DM. In fact, since the protocol is hard-state, a router can only remove information regarding a tree when it is completely sure that the tree is no longer active. Thus, a router must keep maintaining state about a tree as long as it has UPSTREAM neighbors for that tree on any interface. If a router has an UPSTREAM neighbor, that neighbor previously informed the router of this state through an IamUpstream message. Since the protocol is hard-state, the neighbor will not send again this information. Thus, the router cannot delete the upstream information even if it loses its parent on the tree. Otherwise, it will not be able to reconnect to the tree, e.g. when its root interface changes and the UPSTREAM neighbor becomes its parent. This is precisely the purpose of the UNSURE state.%RV 28/1/2020

While in UNSURE state, a router can either lose all its UPSTREAM neighbors, in which case it will become INACTIVE, or reconnect to a parent (e.g. due to a change in its root interface or the appearance of a new UPSTREAM neighbor at its root interface), in which case it will become ACTIVE. In the latter case, the election of the AW is immediate since the router already has information on the UPSTREAM neighbors connected to its interfaces.%RV 28/1/2020

\subsubsection{AW definition}\label{sec:awdefinition}

As mentioned above, the AW is the non-root \textit{interface} that provides the lowest RPC at a link. Root interfaces have no assert state (AW or AL). However, any interface, either root or non-root, must be aware of who the AW is at its link. A root interface determines that the link AW is the interface of the UPSTREAM neighbor that offers the lowest RPC to the source. A non-root interface may itself be the link AW; the interface determines that the AW is interface that, among itself and the interfaces of the UPSTREAM neighbors connect to the link, offers the lowest RPC to the source.%RV 28/1/2020

There is a close relationship between the notions of parent and AW. If a router has a parent, the parent is necessarily the AW of its root interface (it can not be an AL). If the AW is not a parent, e.g. because the feasibility condition is violated, then the ALs are also not parents.%RV 28/1/2020

The AW is defined not only for routers in ACTIVE state, but also for routers in UNSURE and INACTIVE states. This is motivated by the data forwarding issue, which will be discussed in section \ref{sec:dataflooding}. Specifically, a non-root interface of a given tree and router is considered AW in the following conditions:
\begin{itemize}
	\item The router is ACTIVE and its RPC is better than the RPCs offered by its UPSTREAM neighbors on the link. In case of tie, the IP address is used to break the tie, and the highest IP address wins.
	\item The router is UNSURE and no neighbor on the link is considered UPSTREAM.
	\item The router is INACTIVE.
\end{itemize}%RV 28/1/2020

If the router is ACTIVE, a non-root interface becomes AW if it provides the lowest RPC, considering all UPSTREAM neighbors connected to the interface. Recall that the RPC is included in all IamUpstream messages to enable the AW election. If a router is INACTIVE, all its non-root interfaces become AWs. If a router is UNSURE, all its non-root interfaces become AWs, except the interfaces where there is at least one UPSTREAM neighbor; at these interfaces, one of the UPSTREAM neighbors will surely become AW and forward data on the link. These options minimize the loss of data during transient periods.%RV 28/1/2020

\begin{figure}[t!]
	\centerline{\includegraphics[scale=0.9]{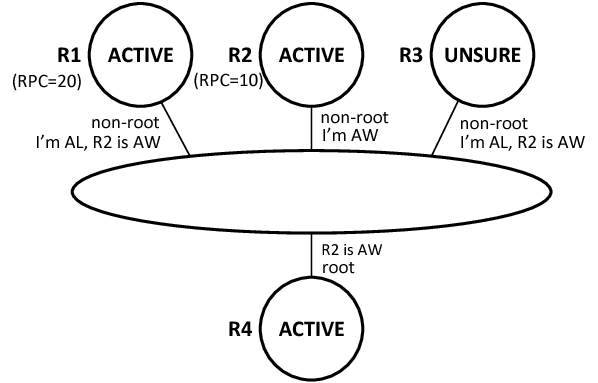}}
	\caption{Relationship between tree states and assert states.}
	\label{fig:treeassert}
\end{figure}

Figure \ref{fig:treeassert} illustrates the relationship between tree states and assert states at a link. For simplicity, we refer to routers as being AWs and ALs; however, keep in mind that, contrarily to the tree states, the assert states are states of (non-root) interfaces, and not of routers. In the figure there are three non-root interfaces and one root interface attached to the link. R2 considers itself AW since the router is ACTIVE and its UPSTREAM neighbors on the link (i.e. R1) provide higher RPCs. R1 recognizes R2 as the AW since R2 is the UPSTREAM neighbor that provides the best RPC (and this RPC is better than its own RPC); R1 places itself in AL state, since it is a non-root interface. R4 also recognizes R2 as AW; however, the interface has no assert state since it is a root interface. Finally, R3 places itself in AL state, since the router is UNSURE but has UPSTREAM neighbors on the link (both R1 and R2).%RV 28/1/2020 

%RV 28/1/2020

\subsubsection{Events and actions}\label{sec:btmaintevents}

\begin{figure}[t!]
	\centerline{\includegraphics[scale=0.45]{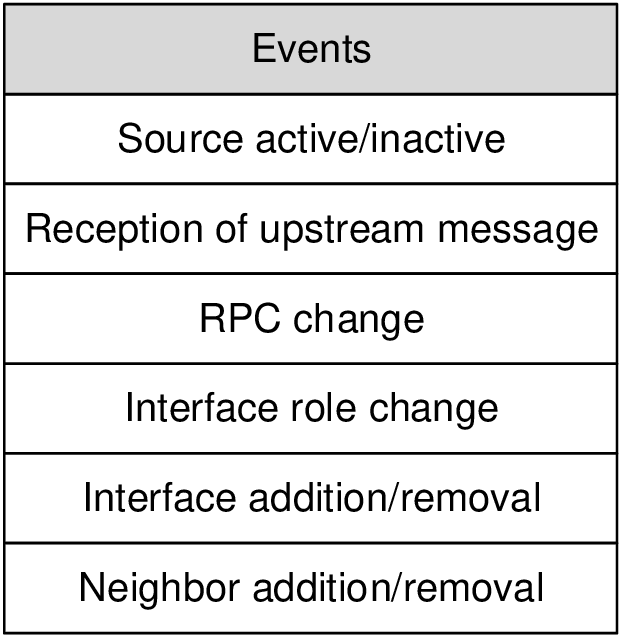}}
	\caption{Events of the broadcast tree construction and maintenance.}
	\label{fig:broadstate_broadeventsactions}
\end{figure}

The broadcast tree is constructed and maintained through event driven actions. The events susceptible of changing the broadcast tree are of six types: (i) source becoming active/inactive, (ii) reception of upstream messages, (iii) RPC change, (iv) interface role change, (v) interface addition/removal, and (vi) neighbor addition/removal. These events are listed in Figure \ref{fig:broadstate_broadeventsactions}.%RV 30/1/2020

%RV 28/1/2020

\begin{figure}[t!]
	\centerline{\includegraphics[scale=0.45]{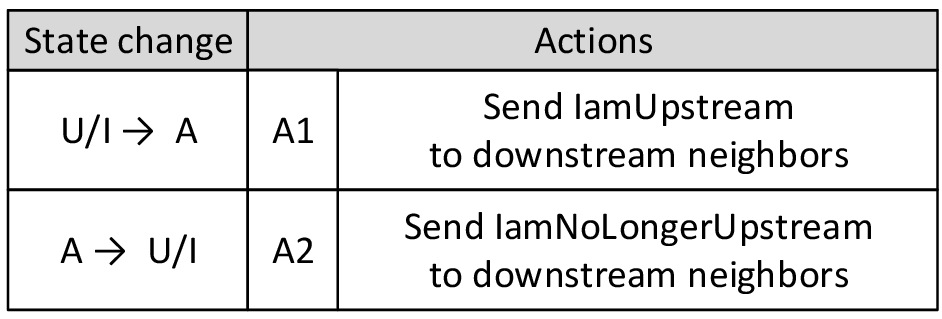}}
	\caption{Message transmissions triggered by tree state transitions.}
	\label{fig:broadstate_messagetx}
\end{figure}

\medskip
\noindent \textbf{Message transmissions triggered by state transitions} The tree state transitions at a router may trigger the transmission of IamUpstream or IamNoLongerUpstream messages downwards the tree. Specifically, when a router changes from UNSURE or INACTIVE to ACTIVE, it sends an IamUpstream message to all its neighbors on non-root interfaces, to signal downstream routers that it became connected to the tree; this will be designated by action A1. Contrarily, when a router changes from ACTIVE to UNSURE or INACTIVE it sends an IamNoLongerUpstream message to all its neighbors on non-root interfaces, to signal downstream routers that it can no longer ensure connectivity to the tree; this will be designated by action A2. Finally, no message is transmitted when a router stays in the same state or when it changes from the UNSURE to INACTIVE state or vice-versa. An exception to this behavior is for non-root interfaces directly connected to the source (which only occurs when an originator router has two or more parallel interfaces attached to the source's link); these interfaces are not allowed transmit upstream messages. This behavior is summarized in Figure \ref{fig:broadstate_messagetx}.%RV 30/1/2020

%RV 3/7/2019

\medskip
\noindent \textbf{Changing to UNSURE or to INACTIVE?} When a router has to change either to the UNSURE or the INACTIVE states, the decision on the actual state depends on the number of UPSTREAM routers that are left: if none are left, the transition is to INACTIVE; if at least one is left, the transition is to UNSURE.%RV 3/7/2019

\begin{figure}[t!]
	\centerline{\includegraphics[scale=0.85]{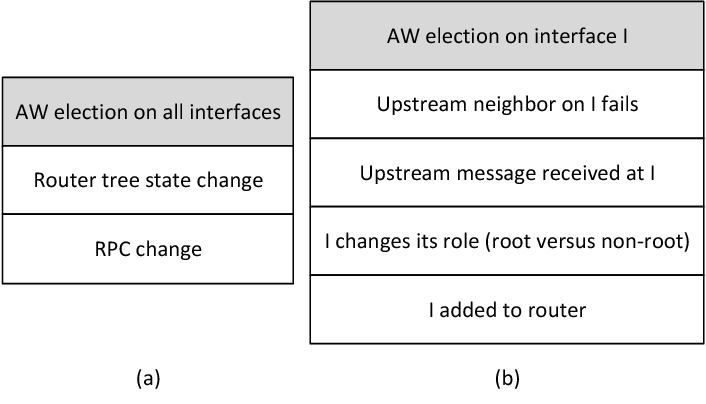}}
	\caption{Events triggering the AW election (a) on all interfaces and (b) on a specific interface.}
	\label{fig:broadstate_triggerawelection}
\end{figure}

\medskip
\noindent \textbf{Triggering the AW election} In non-root interfaces, the AW election aims at determining if the own interface is AW or AL and, in case it is AL, which upstream neighbor interface is AW; in root interfaces, it just aims at determining which upstream neighbor interface is AW (this interface will become parent if the feasibility condition is verified). Note that only a non-root interface (i.e. an interface considered upstream by its neighbors) can be AW. There are several circumstances where the election of the AW must be performed. It can be performed on a specific interface or on all router interfaces. The events that trigger the AW election are listed in Figure \ref{fig:broadstate_triggerawelection}. The AW needs to be reevaluated on all router interfaces when (i) the router tree state changes or (ii) the router RPC changes. It needs to be reevaluated on interface I if the router tree state and RPC do not change but (i) an upstream neighbor in I fails, (ii) an upstream message is received at I, (iii) I changes its role (root versus non-root), or (iv) I is added to the router.%RV 2/2/2020

%RV 2/2/2020

\begin{figure}[t!]
	\centerline{\includegraphics[scale=0.5]{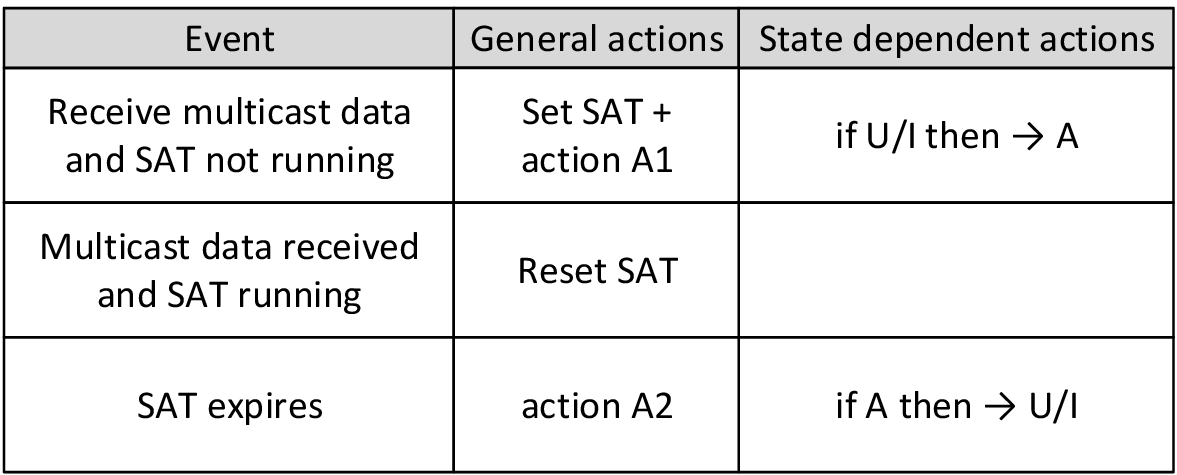}}
	\caption{Source becomes active/inactive (originator routers).}
	\label{fig:broadstate_multicastdata}
\end{figure}

\medskip
\noindent \textbf{Source becoming active/inactive (originator routers)} When the first multicast data packet arrives at the root interface of an originator, that router becomes ACTIVE, sends IamUpstream messages through all its non-root interfaces (action A1), and sets a timer to control the liveness of the source; this timer is called \textit{Source Active Timer} (SAT). The timer is reset whenever a new multicast data packet arrives. Timer expiration means that the source ceased to transmit data. In this case, the router changes to UNSURE or INACTIVE, depending on whether it still has UPSTREAM neighbors, and sends IamNoLongerUpstream messages to all its downstream neighbors (action A2). Note that, as mentioned above, upstream messages are not transmitted on non-root interfaces directly connected to the source. This behavior is summarized in Figure \ref{fig:broadstate_multicastdata}. In the figure, the states are identified by their initial letters, i.e. A for ACTIVE, U for UNSURE, and I for INACTIVE.%RV 2/2/2020

%RV 11/2/2019

\medskip
\noindent \textbf{Reception of upstream messages} The reception of an upstream message always sets the upstream state the neighbor that sent the message. When an IamUpstream message is received the neighbor is considered UPSTREAM, and when an IamNoLongerUpstream message is received the neighbor is considered NOT UPSTREAM. Moreover, receiving an upstream message triggers the election of the AW. The behavior upon the reception of upstream messages at the root interface of a non-originator router depends on whether the router has a parent following the reception of the upstream message. The router may keep having a parent, may cease having a parent, may start having a parent, or may keep not having a parent. Recall that the parent is the UPSTREAM neighbor reachable through the root interface that provides the lowest RPC, subject to the restriction that the router RPC must be higher than the parent RPC (the feasibility condition).%RV 4/7/2019

\begin{figure}[t!]
	\centerline{\includegraphics[scale=0.5]{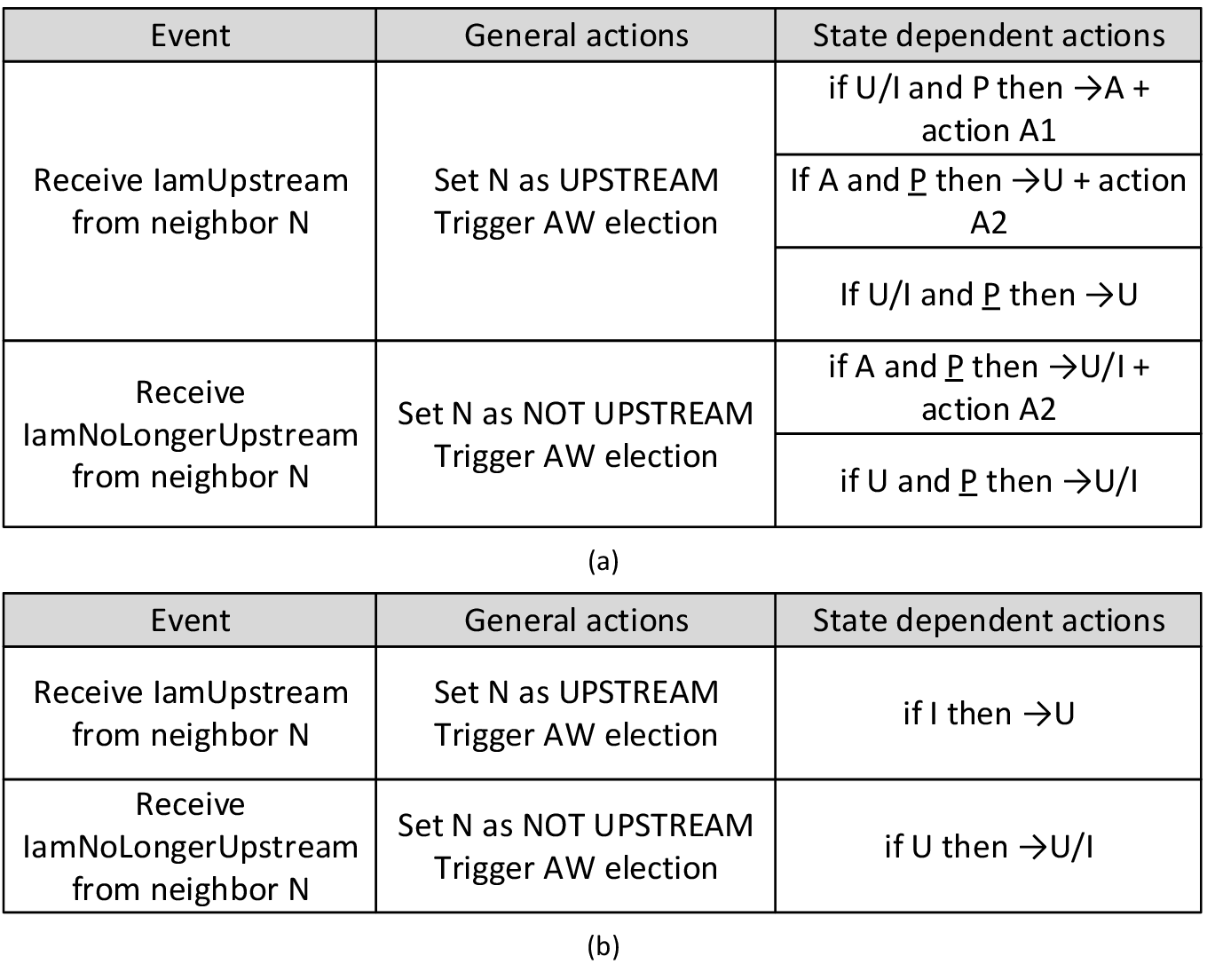}}
	\caption{Reception of upstream messages at (a) root interface of non-originator router and (b) non-root interface of originator or non-originator router.}
	\label{fig:broadstate_upmessage}
\end{figure}

Figure \ref{fig:broadstate_upmessage} summarizes the behavior upon the reception of upstream messages. In the figure, P means that the router has a parent following the reception of the upstream message, and $\underline{\text{P}}$ means the opposite.%RV 4/7/2019
\begin{itemize}	
	\item \textbf{Reception of upstream message at root interface} When an IamUpstream message is received at a root interface, if the router is ACTIVE and keeps having a parent, the router stays ACTIVE (and no upstream message is transmitted downwards); otherwise, if the router ceases having a parent, the router changes to UNSURE (and sends an IamNoLongerUpstream message downwards - action A2). The latter case can happen when the parent changes its RPC such that it now violates the feasibility condition and there is no other UPSTREAM neighbor at the root interface. Likewise, if the router is UNSURE or INACTIVE and starts having a parent, the router changes to ACTIVE (and sends an IamUpstream message downwards - action A1). Finally, if the router is UNSURE or INACTIVE and keeps not having a parent, the router stays UNSURE or changes to UNSURE (and no upstream message is transmitted downwards).%RV 4/7/2019
	
	When an IamNoLongerUpstream message is received on a root interface, if the router is ACTIVE and ceases having a parent, the router changes to UNSURE or INACTIVE (and sends an IamNoLongerUpstream message downwards - action A2). If the router is UNSURE, it may stay UNSURE or change to INACTIVE depending on whether it still has UPSTREAM neighbors, and if it is INACTIVE it stays INACTIVE (in either case no upstream message is transmitted downwards).%RV 4/7/2019
	
	When an upstream message is received at a root interface, the Assert protocol is executed and the AW associated with the interface can change in the following cases: (i) reception of IamNoLongerUpstream message from current AW, (ii) reception of IamUpstream message from current AW with RPC higher than the RPC of its UPSTREAM neighbors, and (iii) reception of IamUpstream message from neighbor with RPC lower than the one of the current AW. Note that, in this case, the interface itself cannot become AW since it is a root interface.%RV 4/7/2019
	
	\item \textbf{Reception of upstream messages at non-root interface} If a router is INACTIVE and receives an IamUpstream message, it changes to UNSURE, since it just discovered an UPSTREAM neighbor on a non-root interface. Likewise, if the router is UNSURE and receives an IamNoLongerUpstream message from the last known UPSTREAM neighbor, it changes to INACTIVE. If the router is ACTIVE, it remains in the same state irrespective of the reception of IamUpstream or IamNoLongerUpstream messages at non-root interfaces.%RV 3/7/2019
	
	When an upstream message is received at a non-root interface, the Assert protocol is executed and AW associated with the interface can change in the following cases: (i) reception of IamNoLongerUpstream message from current AW, (ii) reception of IamUpstream message from current AW with RPC higher than the RPC of its UPSTREAM neighbors, (iii) reception of IamUpstream message from current AW with RPC higher than the router's RPC and router is ACTIVE (and, therefore, the receiving non-root interface is UPSTREAM), (iv) reception of IamUpstream message from neighbor with RPC lower than the one of the current AW.%RV 3/7/2019
\end{itemize}

%RV 11/2/2019

\begin{figure}[t!]
	\centerline{\includegraphics[scale=0.5]{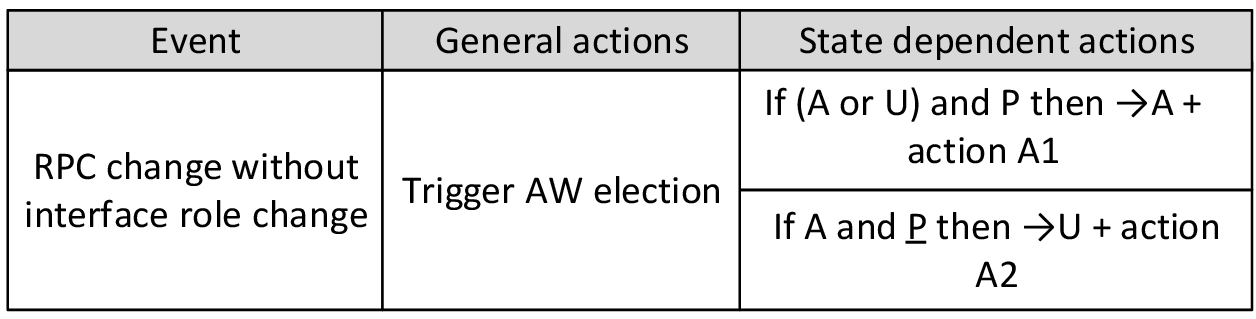}}
	\caption{Change of RPC of non-originator router without interface role change.}
	\label{fig:broadstate_rpc}
\end{figure}

\medskip
\noindent \textbf{Change of RPC of non-originator router without interface role change} When a router is ACTIVE and its RPC changes, if the router remains ACTIVE it sends an IamUpstream message to all its downstream neighbors to inform them of its new RPC value. Moreover, the AW must be reelected at the non-root interfaces of the router. In addition, if the router is ACTIVE and loses its parent then it becomes UNSURE, and if the router is UNSURE and gains a parent then it becomes ACTIVE. This behavior is summarized in Figure \ref{fig:broadstate_rpc}.%RV 4/7/2019

%RV 7/2/2019

\begin{figure}[t!]
	\centerline{\includegraphics[scale=0.5]{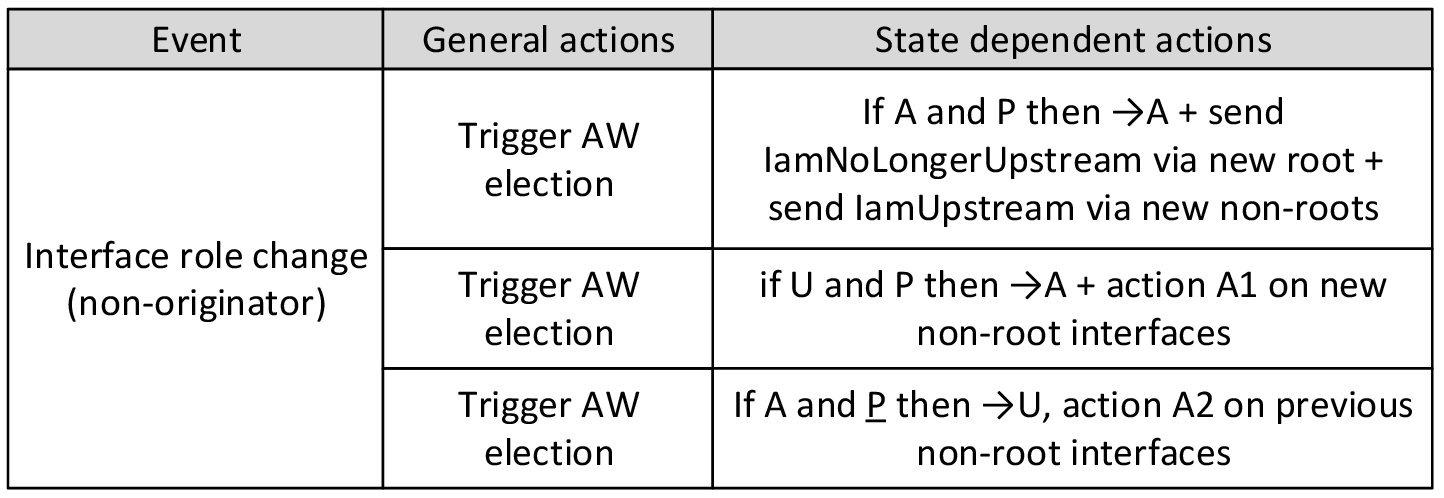}}
	\caption{Change of interface role.}
	\label{fig:broadstate_introle}
\end{figure}

\medskip
\noindent \textbf{Change of interface role at non-originator router} When a router is ACTIVE and its root interface changes to non-root, if the router keeps having a parent, it remains ACTIVE, sends an IamNoLongerUpstream to its neighbors on the new root interface and sends an IamUpstream message to its neighbors on the new non-root interfaces. These messages notify the neighbors of the router in these interfaces of its new role (as being UPSTREAM or NOT UPSTREAM). In the case of non-root interfaces that kept its role, an IamUpstream message needs only to be transmitted if there was, simultaneously, a change in the router RPC. If the router was UNSURE and a parent is found at the new root interface, it changes to ACTIVE (and sends an IamUpstream message on its new non-root interfaces - Action A1). If the router was ACTIVE and there is no parent at the new root interface, it changes to UNSURE (and sends an IamNoLongerUpstream message on all previous non-root interfaces - Action A2). Moreover, the change in interface role triggers the AW election on all interfaces. This behavior is summarized in Figure \ref{fig:broadstate_introle}.%RV 2/2/2020

%RV 7/2/2019

\begin{figure}[t!]
	\centerline{\includegraphics[scale=0.55]{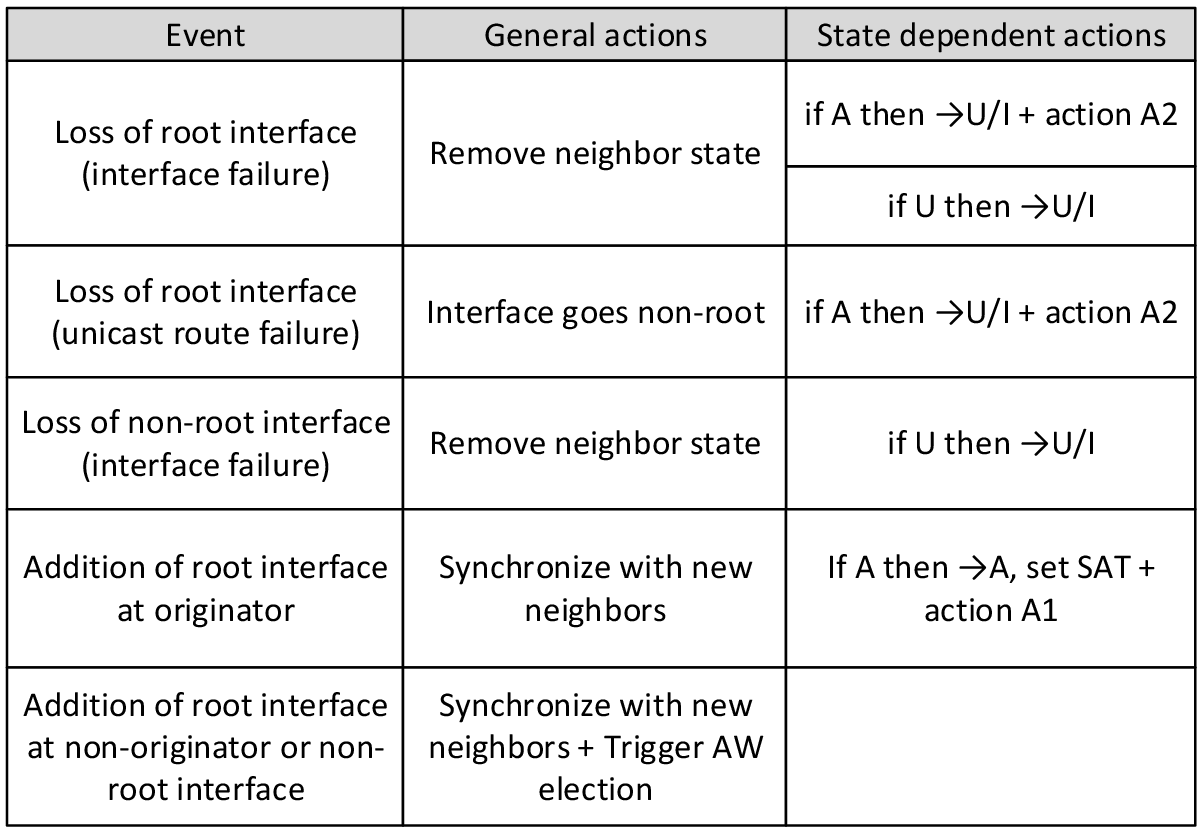}}
	\caption{Interface addition/removal.}
	\label{fig:broadstate_addremint}
\end{figure}

\medskip
\noindent \textbf{Interface addition/removal} Interfaces can be added to or removed from routers. When an interface is added, the router must establish neighborhood relationships with its new neighbors at the interface (if any), and this may trigger a change of state. Regarding removals, we consider two types: physical removals due to a failure or configuration action, and removals due to loss of unicast route (as signaled by the unicast routing table). In the case of physical removals, all state related to the removed interface must be deleted. When a root interface is removed, all other interfaces will continue to be considered non-root. The behavior regarding the addition/removal of interfaces is summarized in Figure \ref{fig:broadstate_addremint}.%RV 4/7/2019
\begin{itemize}
	\item \textbf{Removal of root interface at non-originator} If the router was ACTIVE, it changes to UNSURE or INACTIVE, depending on whether it still has UPSTREAM neighbors (and sends an IamNoLongerUpstream message to all its previous downstream neighbors - action A2). If the router was UNSURE there are two cases: (i) if the interface failed the router may remain UNSURE or change to INACTIVE, depending on whether it still has UPSTREAM neighbors; (ii) if the unicast route was lost, the router remains UNSURE.%RV 4/7/2019	
	\item \textbf{Removal of root interface at originator} In this case, the router becomes non-originator and follows the procedure described above for non-originator routers affected by an interface failure of root interface.%RV 4/7/2019
	\item \textbf{Removal of non-root interface} If the router was UNSURE, it changes to INACTIVE or remains UNSURE, depending on whether it still has UPSTREAM neighbors. If the router was ACTIVE, no action is taken.%RV 4/7/2019
	\item \textbf{Addition of root interface at (new) originator} When a router enables an interface directly connected to the source, it changes its role from non-originator to originator. If the router was ACTIVE, it remains ACTIVE, sets the Source Active Timer, and sends an IamUpstream message to all its previous downstream neighbors (action A1). If the router was UNSURE it remains in this state until it receives multicast data, and no upstream message transmission is required.%RV 4/7/2019
	\item \textbf{Addition of root interface at non-originator or of non-root interface} In these cases the router will have to synchronize with the new neighbors and trigger the AW election at the added interface. The synchronization may force a change of state. For example, if a router is INACTIVE and synchronizes with a neighbor that declares itself as UPSTREAM (for the corresponding tree), the router changes to ACTIVE if the neighbor is on the root interface and offers an RPC respecting the feasibility condition, or changes to UNSURE otherwise.%RV 4/7/2019
\end{itemize}
\noindent Note that whenever the above actions force a router tree state change or an RPC change the AW election is triggered at all router interfaces (see Figure \ref{fig:broadstate_triggerawelection}).%RV 4/7/2019

%RV 7/2/2019

\begin{figure}[t!]
	\centerline{\includegraphics[scale=0.5]{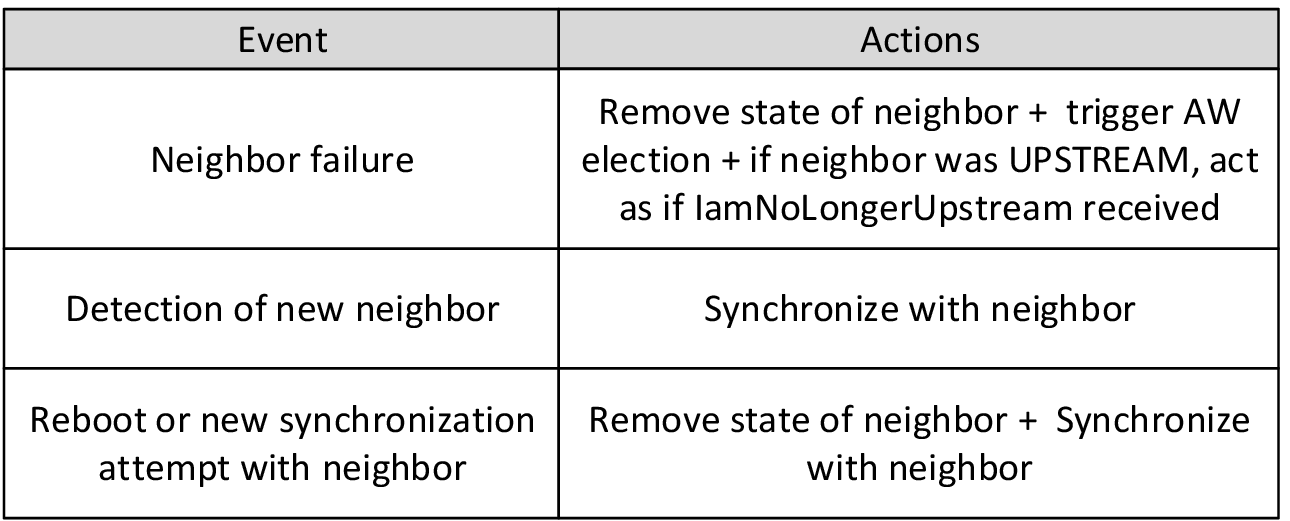}}
	\caption{Neighbor addition/removal.}
	\label{fig:broadstate_nei}
\end{figure}

\medskip
\noindent \textbf{Neighbor addition/removal} The addition/removal of a neighbor may occur in three cases: (i) neighbor failure, (ii) detection of new neighbor, and (iii) reboot or synchronization attempt of known neighbor. The behavior related to the addition/removal of neighbors is summarized in Figure \ref{fig:broadstate_nei}.%RV 5/7/2019
\begin{itemize}
	\item \textbf{Neighbor failure} When a neighbor fails (as signaled by the Hello protocol), all state information relative to this neighbor must be removed. If the failed neighbor was UPSTREAM, this may trigger a change of state, and we must follow the same procedure as if the router received an IamNoLongerUpstream message from the neighbor. In addition, the AW election must be triggered.%RV 5/7/2019
	\item \textbf{Detection of new neighbor} When a router connects to a new neighbor, the two routers must synchronize with each other. During this process, the neighbor provides information on the trees for which it can serve as UPSTREAM router. The synchronization process will be explained in section \ref{sec:initsync}.%RV 5/7/2019
	\item \textbf{Reboot or synchronization attempt from known neighbor} If a known neighbor reboots or initiates a new synchronization (e.g. because it was unable to communicate with the router for a while), the tree information stored for that neighbor may no longer be valid. In this case, all information regarding this neighbor must be removed, and a new synchronization with the neighbor must be performed. This will be addressed in section \ref{sec:initsync}.%RV 5/7/2019
\end{itemize}

%RV: Say something related to AW election in the last to cases?

%RV 7/2/2019

\subsubsection{Examples}\label{sec:examplestree}

\begin{figure}[t!]
	\centerline{\includegraphics[scale=0.75]{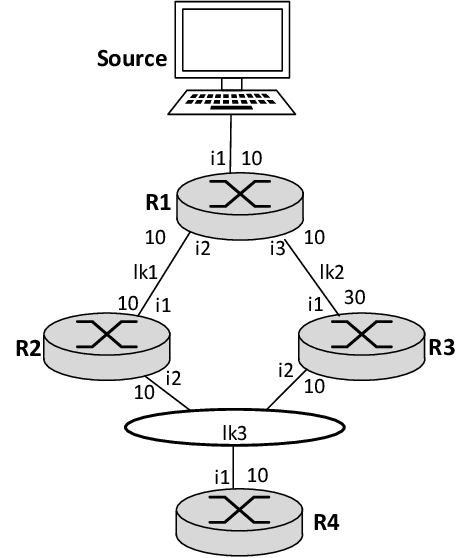}}
	\caption{Broadcast tree construction and maintenance example.}
	\label{fig:broadcast_tree_maintenance_example}
\end{figure}

Consider the network of Figure \ref{fig:broadcast_tree_maintenance_example}, comprising 4 routers and three links; lk1 and lk2 are point-to-point links, and lk3 is a shared link. R1 is connected to a multicast source and, therefore, is an originator router. The figure indicates the interface costs of the underlying unicast routing protocol. The root interfaces are i1-R1 (since it is directly attached to the source's subnet), i1-R2 with RPC=20, i2-R3 with RPC=30, and i1-R4 with RPC=30.%RV 29/1/2019

\medskip
\noindent \textbf{Initial formation of the broadcast tree} When the source becomes active, R1 also becomes ACTIVE and sends IamUpstream messages through its non-root interfaces, i.e. to routers R2 and R3. When these messages are received, both R2 and R3 set R1 as an UPSTREAM neighbor. However, since R3 receives the message through a non-root interface, it stops forwarding it. Moreover, R3 becomes UNSURE since, at this point, it has no UPSTREAM neighbor on its root interface (i2-R3). Contrarily, since R2 received the message on its root interface, it transmits an IamUpstream message through its non-root interface (i.e. on the shared link) containing its own RPC, which is 20. Moreover, R2 becomes ACTIVE since it has an UPSTREAM neighbor on its root interface (i1-R2). The message multicasted by R2 on the shared link is received by R3 and R4, and both routers set R2 as an UPSTREAM neighbor, and both become ACTIVE since R2 is reachable through their root interfaces (i2-R3 and i1-R4). Finally, R3 transmits an IamUpstream message in its non-root interface (i1-R3). This message will not trigger further message transmissions at R1 since it was received at a non-root interface. The AW at the shared link is R2, since only R2 sent an IamUpstream message on this link. By the same reason, R1 is the AW at link lk1. At link lk2, both routers transmitted IamUpstream messages, and R1 becomes the AW since it offered a lower RPC.%RV 29/1/2019

\medskip
\noindent \textbf{Change of interface role} Suppose now that the interface cost of i1-R3 is changed to 5. Then, there will be a change in interface roles: i1-R3 becomes root and i2-R3 becomes non-root. Since R3 keeps having a parent (now R1), it stays ACTIVE and sends IamUpstream through its new non-root interface (i2-R3); it also sends and IamNoLongerUpstream through its new root interface (i1-R3). When R2 and R4 receive the IamUpstream (with RPC=15), they set R3 as an UPSTREAM neighbor and elect R3 as the new AW. When R1 receives the IamNoLongerUpstream, it sets R3 as a NOT UPSTREAM neighbor, and R1 keeps being AW.%RV 3/2/2020

\medskip
\noindent \textbf{AW failure} Suppose now that the AW (currently R3) fails. In this case, R2 and R4 remove the state information relative to R3, and both select R2 as the new AW of the shared link, based only on the stored information. No control messages circulate in this case. Note that R4 has information that R2 is an UPSTREAM neighbor offering an RPC of 20; R2 considers itself AW since it has no UPSTREAM neighbor on the shared link.%RV 29/1/2020

\medskip
\noindent \textbf{Originator failure} In the scenario with the previous interface costs (cost of i1-R3 equal to 5), suppose that R1, the originator router, fails. When R2 and R3 detect the failure, they become UNSURE (since each one considers the other an UPSTREAM neighbor) and each transmits an IamNoLongerUpstream message at the shared link. When R2 receives the message from R3 it becomes INACTIVE, and when R3 receives the message from R2 it also becomes INACTIVE. R4 remains ACTIVE when it receives the first message, since it still has one UPSTREAM neighbor with an RPC lower than the RPC of R4, but changes to INACTIVE when it receives the second message.%RV 29/1/2020

\medskip
\noindent \textbf{Failure of link lk1} Consider again the scenario with the initial interface costs (cost of i1-R3 equal to 30), and suppose that link lk1 fails. When R2 detects the failure it becomes INACTIVE since it no longer has UPSTREAM neighbors and transmits an IamNoLongerUpstream message at the shared link. Suppose that R3 receives this message before the unicast routing protocol finds the new path to the source (which implies a change in the role of its interfaces). In this case, R3 becomes UNSURE, since it still has one UPSTREAM neighbor (R1), but no longer in its root interface (i2-R3). It will also send an IamNoLongerUpstream message through its non-root interface, i.e. to R1. At this point, R4 becomes INACTIVE, since it lost its sole UPSTREAM neighbor (which was R2). When finally the interface role changes at R3 (i1-R3 becomes root and i2-R3 becomes non-root), R3 changes to ACTIVE state again and sends an IamUpstream message in its new non-root interface (i2-R3). At this point, R4 returns to ACTIVE state, since a new UPSTREAM neighbor appeared at its root interface. R2 also changes to ACTIVE if in the meantime its interface i2-R2 became root; otherwise, it changes to UNSURE, moving only to ACTIVE when i2-R2 changes to root (due to the unicast routing protocol).%RV 29/1/2020

It is also possible that, in R3, the unicast routing protocol triggers the interface role change before the arrival of the IamNoLongerUpstream message from R2. In this case, R3 keeps being ACTIVE but sends an IamNoLongerUpstream message through the previous non-root interface (i.e. to R1) and an IamUpstream message through the new non-root interface (i.e. to the shared link). The behavior of R4 depends on the order by which the messages sent by R2 and R3 are received: it becomes temporarily INACTIVE if the message from R2 is received first, or keeps being ACTIVE if the message received from R3 is received first.%RV 29/1/2020

\begin{figure}[t!]
	\centerline{\includegraphics[scale=0.45]{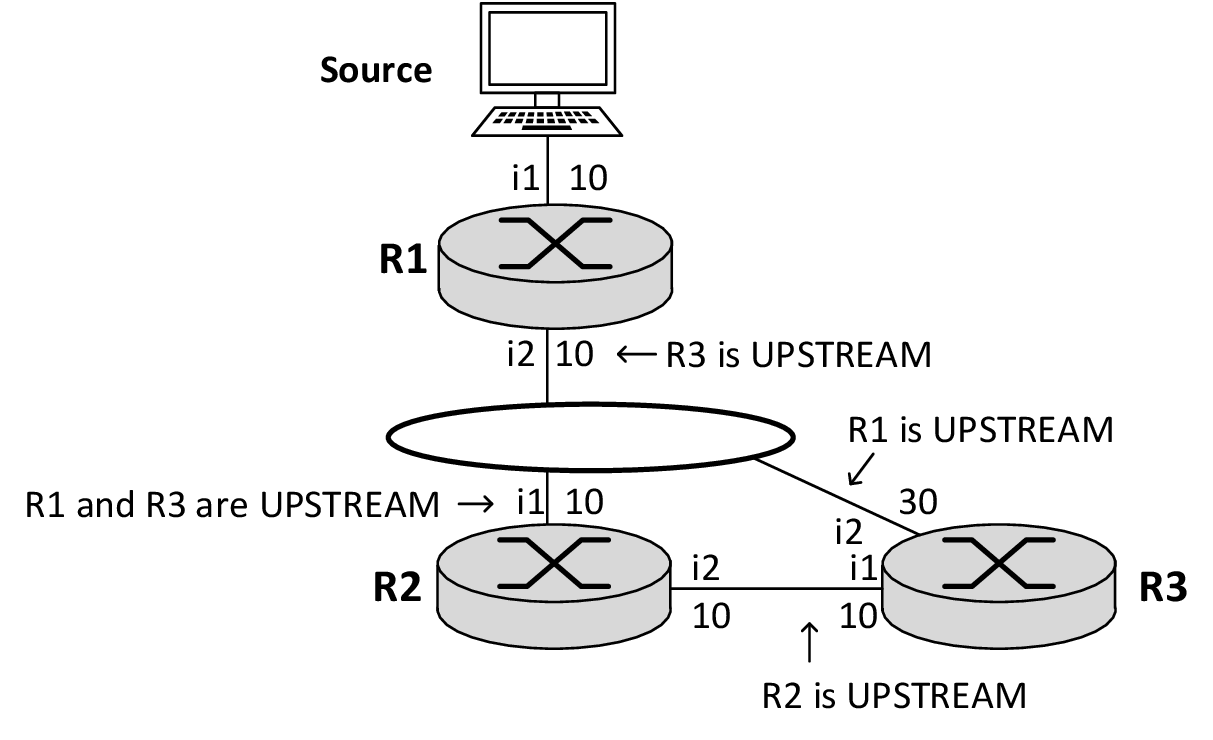}}
	\caption{Importance of feasibility condition.}
	\label{fig:broadcast_tree_maintenance_loop_example}
\end{figure}

\medskip
\noindent \textbf{Importance of feasibility condition} We illustrate now the importance of the feasibility condition. Consider the network of Figure \ref{fig:broadcast_tree_maintenance_loop_example}. When the source is active the root interfaces are i1-R1, i1-R2 and i1-R3. Moreover, due to the initial transmission of IamUpstream messages, R3 considers R1 UPSTREAM at interface i2-R3 and R2 UPSTREAM at interface i1-R3, R2 considers R1 and R3 UPSTREAM at interface i1-R2, and R1 considers R3 UPSTREAM at interface i2-R1. The RPCs advertised by R1, R2, and R3, in their IamUpstream messages are 10, 20, and 30, respectively. Now suppose that the source stops transmitting multicast data. Consider first that the feasibility condition is ignored, in which case the ACTIVE state is only defined by having UPSTREAM neighbors at the root interface. In this case, R1 changes to UNSURE and transmits IamNoLongerUpstream at the shared link. When R2 receives this message, it remains ACTIVE since it still has an UPSTREAM neighbor at its root interface (i1-R2), which is R3. Thus, it transmits no message to R3. When R3 receives the message, it removes R1 as UPSTREAM at interface i2-R3, but it remains ACTIVE since it has an UPSTREAM neighbor at its root interface (i1-R3), which is R2. Thus, both R2 and R3 remain ACTIVE, despite the source being no longer active. The problem is that R2 believes that it is possible to reach the source via R3, and R3 believes it is possible via R2: a routing loop is formed!%RV 5/7/2019

Consider now that the feasibility condition is included. When the IamNoLongerUpstream message sent by R1 arrives at R2, R3 is still considered UPSTREAM. However, R3 offers an RPC higher than R2: R3 offers 30 and R2 offers 20. Thus, R3 cannot be considered a parent of R2, and R2 becomes UNSURE and sends IamNoLongerUpstream through i2-R2. When R3 receives this message it becomes INACTIVE and sends IamNoLongerUpstream through i2-R3. Finally, when R1 and R2 receive this message they both remove R3 as UPSTREAM neighbor and become INACTIVE. Thus, all routers understand that the source stopped transmitting multicast data.%RV 5/7/2019

%RV 5/7/2019

\subsubsection{Summary of the broadcast tree maintenance}

The broadcast tree is a logical structure that allows the delivery of multicast data from a source to \textit{all} network routers. This tree may be pruned according to the interest of routers, such that pruned routers do not receive multicast data; this issue is discussed in section \ref{sec:interestmaintenance}. However, in HPIM-DM, all routers maintain the broadcast tree information, irrespective of their interest.%RV 29/1/2020

For the broadcast tree maintenance, a router stores several types of state information regarding the tree. The tree states give indication about the source liveness, as perceived locally by a router: in ACTIVE state the source is considered active, in INACTIVE state is considered inactive, and in UNSURE state the router is not certain about the source liveness. Routers also store the upstream state of each neighbor: an UPSTREAM neighbor is a router from which multicast data can be received, and a NOT UPSTREAM neighbor is the opposite.%RV 29/1/2020

An originator router considers the tree ACTIVE if it is receiving multicast data from its source. A non-originator router considers the tree ACTIVE if it has a parent on the tree. A parent is an UPSTREAM neighbor connected to the root interface which respects the feasibility condition. The feasibility condition aims at avoiding loops and, to be observed, the parent of a router must have an RPC lower than the one of router. On a non-originator router, the tree is considered UNSURE if the router has no parent but has at least one UPSTREAM router, and INACTIVE if it has no UPSTREAM router. On an originator router, the tree is considered UNSURE or INACTIVE if the router is not receiving multicast data; it is UNSURE if the router has at least one UPSTREAM neighbor and INACTIVE otherwise. The UNSURE state is a transient state required for correctness and to speed up convergence.%RV 29/1/2020

To set the upstream state, routers transmit two types of control messages: IamUpstream and IamNoLongerUpstream. In general, these messages are transmitted through non-root interfaces to signal downstream routers whether the sender can serve as an upstream neighbor or not.%RV 29/1/2020

When a source becomes active, the originator transmits IamUpstream messages to all its downstream neighbors (i.e. neighbors reachable through non-root interfaces), and successive downstream routers repeat this behavior as soon as they find a parent on the tree (and become ACTIVE). In this way, all routers become connected to source through a chain of parent routers leading to the originator, from which multicast data can be received. When the source becomes inactive, the same process is repeated but now with IamNoLongerUpstream messages.%RV 29/1/2020

The IamUpstream messages carry the RPC of the sending routers, and the upstream state of neighbors stored at routers includes their RPCs. In this way, the election of the AW at a link is a simple process since routers maintain locally all the information required to determine who the AW is. Specifically, the AW elected by an interface at a link (shared or point-to-point) is the interface at that link, selected among the UPSTREAM neighbors and the own interface (if the selecting interface is non-root) or just among the UPSTREAM neighbors (if the selecting interface is root), that provides the lowest RPC. The AW interface is the one selected for forwarding multicast data on a link, if there are downstream devices interested.%RV 29/1/2020

To ensure that the state information is kept correct at all times, the protocol must react correctly to all events susceptible of changing this information. These events are: (i) source becoming active/inactive, (ii) reception of upstream messages, (iii) RPC change, (iv) interface role change, (v) interface addition/removal, (vi) neighbor addition/removal. In HPIM-DM, the actions triggered by these events have been defined such that the protocol converges fast to the correct states.%RV 29/1/2020

\subsubsection{Comparison with PIM-DM}

PIM-DM is a soft-state protocol where the broadcast tree either (i) needs to be rebuilt periodically or (ii) relies on the periodic transmission of control messages (i.e. State Refresh messages). Moreover, PIM-DM suffers from slow convergence, leading to loss of data or excessive data transmission when control messages are lost or disordered, when the interest of routers change, and when routers join the network.%RV 5/7/2019

HPIM-DM is a hard-state protocol where routers store information on their upstream neighbors, i.e. on the neighbors from which multicast data can be received, and this information is kept updated through the exchange of IamUpstream and IamNoLongerUpstream messages, which are transmitted reliably and with ordering guarantees. In this way, routers keep always the sense of the correct broadcast trees. Moreover, the protocol converges fast in the presence of events susceptible of changing the configuration of the trees.%RV 5/7/2019

\subsection{Interest maintenance}\label{sec:interestmaintenance}%RV 12/2/2019

The broadcast tree must be pruned from routers not interested in receiving multicast data and router interfaces not interested in sending it, to save resources. In this section, we explain the protocol behavior that allows maintaining a multicast tree free of not interested routers and router interfaces.%RV 29/1/2020

\subsubsection{Interest and forwarding states}\label{sec:intereststate}

\medskip
\noindent \textbf{Interest state of routers} Regarding the interest in receiving multicast data of an (S,G) tree, a router can be in one of two states: INTERESTED, if the router is interested in receiving multicast data, or NOT INTERESTED, if the router is not interested.%RV 31/1/2020

\medskip
\noindent \textbf{Downstream interest state of non-root interfaces} The interest of a router depends on the assert state (AW or AL) and the downstream interest of its non-root interfaces. The downstream interest indicates whether an interface has downstream devices interested in receiving multicast traffic; these devices can be routers or multicast receivers. Multicast receivers signal their interest to the routers they are directly attached to using the IGMP or MLD protocols. Downstream routers signal their interest using interest control messages (see section \ref{sec:intmessages}). Regarding the downstream interest, a non-root interface can be in DOWNSTREAM INTERESTED state, if interested, or in NOT DOWNSTREAM INTERESTED state, otherwise.%RV 31/1/2020

\medskip
\noindent \textbf{Forwarding state of non-root interfaces} The forwarding state of a non-root interface is defined in terms of its assert state and downstream interest state. Specifically, a non-root interface is FORWARDING if it is both AW and DOWNSTREAM INTERESTED, and is PRUNED otherwise.%RV 31/1/2020

An exception to this definition is for non-root interfaces directly attached to the source. This case can occur if an originator router has two or more parallel interfaces attached to the source's link. These non-root interfaces are always placed in NOT DOWNSTREAM INTERESTED and PRUNED states. Otherwise, there would be the possibility of having them transmitting data, provoking a routing loop. Moreover, these interfaces are not allowed to transmit control messages.%RV 31/1/2020

\medskip
\noindent \textbf{Definition of router interest according to interface forwarding state} Finally, the interest state of a router in receiving multicast traffic is defined in terms of the forwarding state of its non-root interfaces. Specifically, a router is INTERESTED if it has at least one non-root interface in FORWARDING state (i.e. that must transmit multicast data), and is NOT INTERESTED otherwise.%RV 31/1/2020

\begin{figure}[t!]
	\centerline{\includegraphics[scale=0.54]{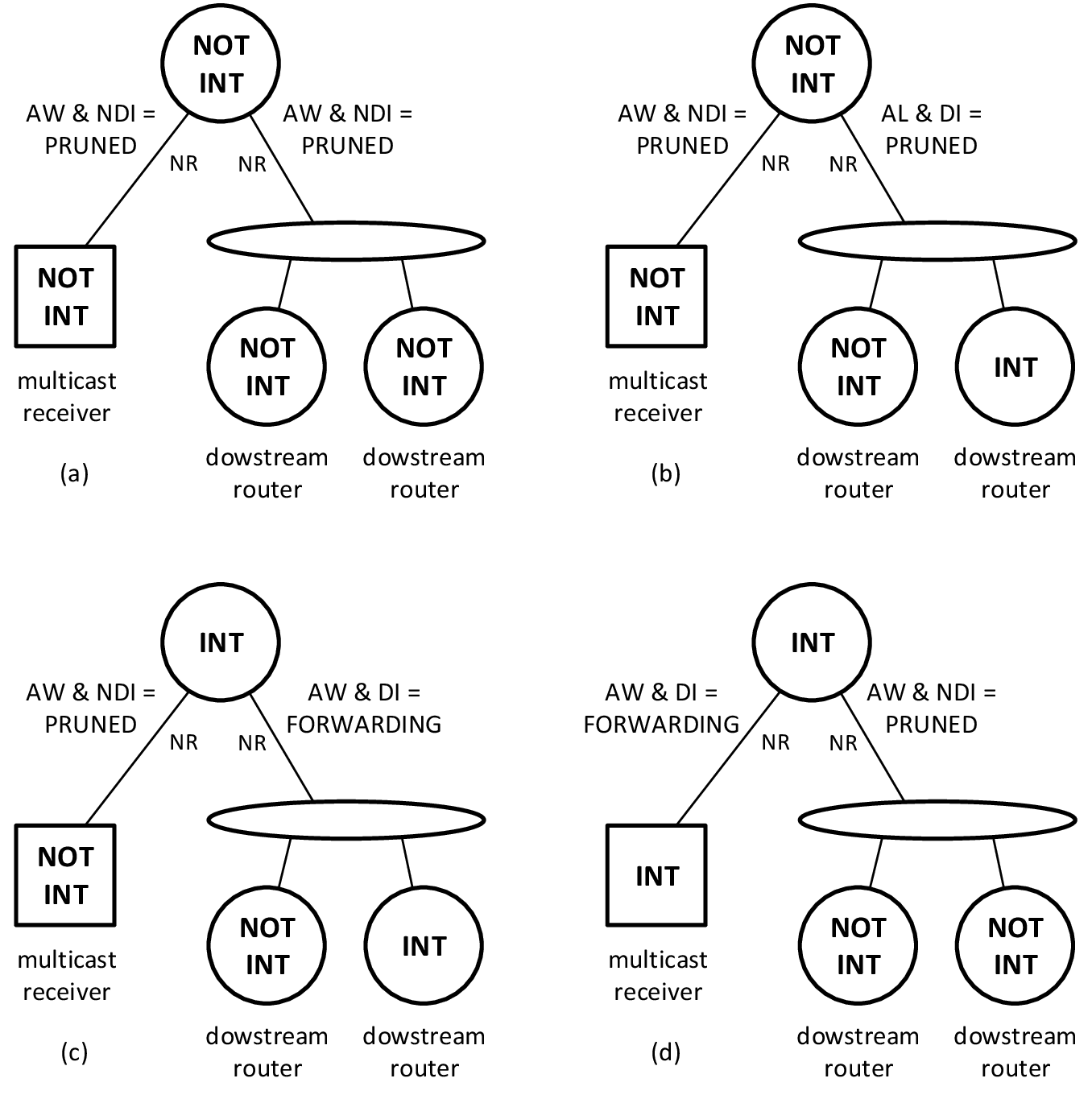}}
	\caption{Relationship between downstream interest, assert, router interest, and forwarding states.}
	\label{fig:introuter}
\end{figure}

Figure \ref{fig:introuter} gives several examples of the relationship between downstream interest, assert, router interest, and forwarding states. In the four figures, the upper router is connected to two types of devices: a multicast receiver and two downstream routers (through a shared link). In Figure \ref{fig:introuter}.a, the upper router is NOT INTERESTED (NOT INT) since all downstream devices are NOT INTERESTED and, therefore, the corresponding non-root interfaces are NOT DOWNSTREAM INTERESTED (NDI) and PRUNED. In Figure \ref{fig:introuter}.b, one of the downstream routers is INTERESTED, making the corresponding interface DOWNSTREAM INTERESTED (DI). However, that interface is now AL, which means another interface will forward multicast data on the link (the link AW). Therefore, the shared link interface stays PRUNED and the router stays NOT INTERESTED. The case of Figure \ref{fig:introuter}.c is similar to that of Figure \ref{fig:introuter}.b, except that the shared link interface is now AW. In this case, the shared link interface becomes FORWARDING and the router becomes INTERESTED. Finally, in Figure \ref{fig:introuter}.d, the upper router is INTERESTED since the interface with the multicast receiver is DOWNSTREAM INTERESTED and FORWARDING (because it is also AW), despite the shared link interface being NOT DOWNSTREAM INTERESTED.%RV 31/1/2020

%RV 29/1/2020

\medskip
\noindent \textbf{Interest according to tree state} We consider that the router interest state is only meaningful in the ACTIVE state. Thus, a router does not have to store interest information in UNSURE or INACTIVE states.%RV 31/1/2020

\begin{figure}[t!]
	\centerline{\includegraphics[scale=0.8]{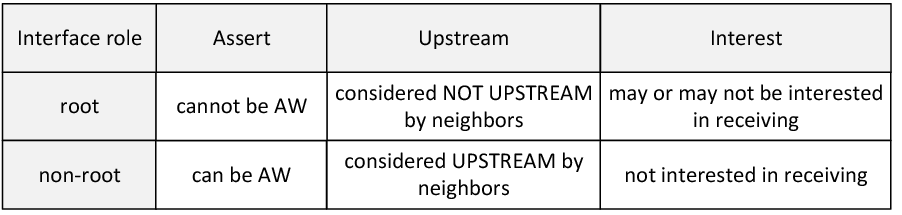}}
	\caption{Relationship between interface role, upstream, assert and interest states.}
	\label{fig:introleaui}
\end{figure}

\medskip
\noindent \textbf{Relationship between interface role, upstream, assert and interest states} Non-root interfaces are considered UPSTREAM by their neighbors on the corresponding link (under stable conditions, if they already sent and IamUpstream message). They transmit data on the link according to the interest of their downstream devices (routers or multicast receivers) and their assert state, and are never interested in receiving multicast data. Root interfaces are considered NOT UPSTREAM by their neighbors on the corresponding link. Their role is to receive multicast data, but may or may not be interested in receiving it. These relationships are summarized in Figure \ref{fig:introleaui}.%RV 2/2/2020

%RV 2/2/2020

\medskip
\noindent \textbf{What interfaces store interest information?} A non-root interface is required to store the interest state of each of its neighbors, in order to determine its downstream interest. However, this is not required at root interfaces, since these interfaces are not allowed to transmit multicast data and, therefore, do not need to know if their neighbors are interested or not. Moreover, regarding non-root interfaces, despite all interfaces having to store interest information, only the AW is required to have it completely updated. This is because only the AW is candidate for forwarding multicast data; ALs are pruned irrespective of downstream interest.%RV 19/7/2019

%PO: Um router AL não necessita de manter o estado de interesse de todos os vizinhos em todas as situações... Quando um router passa de AW->AL este pode esperar algum tempo e depois apagar o interesse de todos os vizinhos e depois enviar IamUpstream (isto se permanecer ACTIVO). Isto faz com que se algum router considerar este vizinho como AW e se este já transmitiu o interesse o facto de receber IamUpstream do AW faz com que retransmita o interesse. Se um router receber IamUpstream de um router que não é o AW este não faz nada em termos de interesse.   <- isto corresponde a um melhoramento que não foi implementado nem referido ao longo do documento nem tese, mas permite que ALs apaguem estado de interesse armazenado acerca de vizinhos

\subsubsection{Interest messages}\label{sec:intmessages}

\medskip
\noindent \textbf{Message types} The interest information is signaled through Interest and NoInterest messages. These messages are transmitted (i) through root interfaces to signal upstream neighbors about the interest of a router in receiving multicast data, and (ii) through non-root interfaces to signal other interfaces attached to the same link about their lack of interest in receiving multicast data (in some cases). Recall that non-root interfaces are not interested in receiving multicast data (and, therefore, only transmit NoInterest messages). Interest messages may be transmitted by a router in ACTIVE or UNSURE state.%RV 31/1/2020

\medskip
\noindent \textbf{Double meaning of IamUpstream messages} IamUpstream messages are only transmitted by non-root interfaces and, as noted above, these interfaces are not interested in receiving multicast data. Thus, when a router receives an IamUpstream message from a neighbor, this indicates that the neighbor is not interested in receiving multicast data (i.e. the equivalent to receiving a NoInterest message). In HPIM-DM, we allow the interest state of a neighbor to be set by the reception of IamUpstream messages.%RV 31/1/2020

\medskip
\noindent \textbf{Building a tree according to interest} When a tree is activated, the IamUpstream messages start flowing downstream and routers start discovering upstream routers. Within each link, when the first IamUpstream message is transmitted, the remaining interfaces attached to the link react by sending its interest information. Interest information must also be sent when the interest of a router changes, and upon other events described later.%RV 31/1/2020

\begin{figure}[t!]
	\centerline{\includegraphics[scale=0.9]{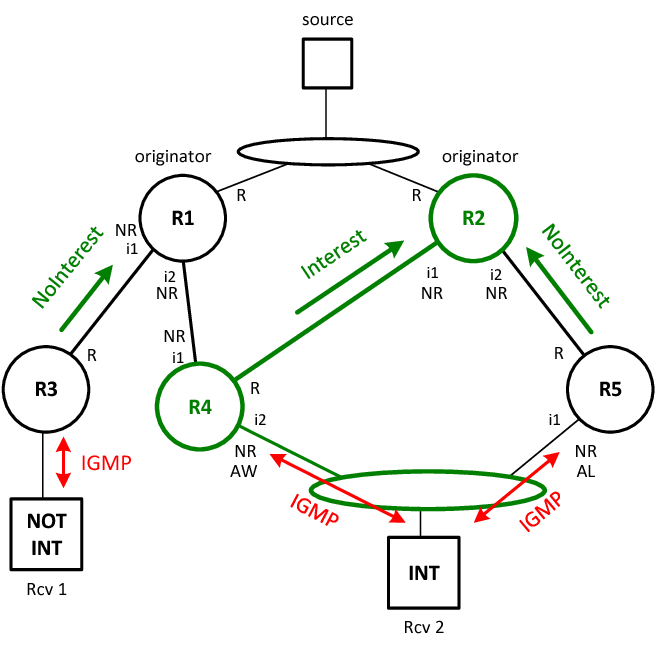}}
	\caption{Pruning the broadcast tree according to the interest of multicast receivers.}
	\label{fig:prunetree}
\end{figure}

Figure \ref{fig:prunetree} illustrates the process of pruning the broadcast tree according to the interest of multicast receivers. We reuse the network of Figure \ref{fig:buildtree} for this example. There are two receivers: Rcv1, which is not interested in receiving multicast data, and Rcv2, which is interested. The receivers express their interest to the routers they are directly attached to using the IGMP protocol, i.e. Rcv 1 signals R3 and Rcv2 signals R4 and R5. The transmission of interest messages is triggered by the reception of IamUpstream messages. When R3 receives the IamUpstream message from R1, R3 has no downstream device interested in receiving multicast data. Thus, it becomes NOT INTERESTED and sends a NoInterest message through its root interface, which places interface i1-R1 in PRUNED state. Routers R4 and R5 know through the IGMP protocol that receiver Rcv2 is interested in receiving multicast data. However, following the AW election, which occurs when IamUpstream messages are transmitted on the shared link, interface i1-R5 becomes AL and, therefore, is placed in PRUNED state. Since R5 has no other non-root interface, it becomes NOT INTERESTED, and sends a NoInterest message through its root interface, which places interface i2-R2 in PRUNED state. Following the AW election at the shared link, interface i2-R4 becomes AW and, since it has an interested downstream receiver (Rcv2) it becomes DOWNSTREAM INTERESTED and FORWARDING. Because of this, R5 becomes INTERESTED and sends an Interest message through its root interface. This messages places interface i1-R2 in FORWARDING state and router R2 in INTERESTED state. No interest messages need to be exchanged in the link between R1 and R4 since, as noted above, the interest information can, in this case, be inferred from the IamUpstream messages received during the construction of the broadcast tree. The two interfaces become NOT DOWNSTREAM INTERESTED and therefore PRUNED, which places router R1 in NOT INTERESTED state. To summarize, routers R1, R3 and R5, and interfaces i2-R2 and i1-R4 are pruned from the broadcast tree, leading to a multicast tree that distributes multicast data from the source to receiver Rcv2. Data is received by R2, transmitted on interface i1-R2 to R4, and then transmitted on interface i2-R4, to reach Rcv2.%RV 30/1/2020

%RV 2/2/2020

\medskip
\noindent \textbf{Which neighbors should receive interest information} Only the AW interface is required to have the most recent information regarding the interest of the neighbors attached to the corresponding link, since only this interface is candidate to become the link forwarder. Thus, when neighbors on a link have some indication that their interest information might not be correct at the AW, they are required to send their interest information to the AW.%RV 31/1/2020

\medskip
\noindent \textbf{Why interest messages are unicasted to the AW} Since only the AW is required to have the most recent interest information, the interest information is unicasted to the AW, and not multicasted (as in the case of upstream messages). Since interest messages are transmitted reliably through ACK-protection, unicasting saves in the number of transmitted ACKs.%RV 31/1/2020

\begin{figure}[t!]
	\centerline{\includegraphics[scale=1]{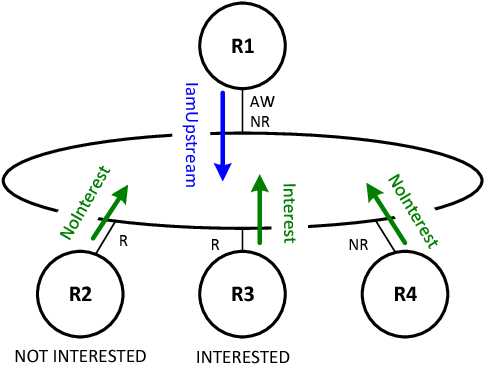}}
	\caption{Transmission of interest information to the AW.}
	\label{fig:intmessages}
\end{figure}

Figure \ref{fig:intmessages} illustrates the transmission of interest information to the AW. R1 is the AW at the shared link and transmits the first IamUpstream message on the link. The downstream routers reply indicating their interest information, which is unicasted to the AW. R2 and R3 are connected to the link through their root (R) interfaces. R2 is NOT INTERESTED and sends a NoInterest message. R3 is INTERESTED and sends an Interest message. R4 is connected to the link through a non-root (NR) interface and, therefore, the interface is not interested in receiving multicast data; thus, it sends a NoInterest message on the link.%RV 30/1/2020

%RV 30/1/2020

\medskip
\noindent \textbf{Why interest information needs to be stored at all non-root interfaces} The information regarding the interest of each neighbor in receiving multicast data must be stored at all non-root interfaces, despite the fact that only the AW is required to have the most recent information. This is needed since the AW election may not be simultaneous at all routers and, during a transient period where the AW is changing, different routers may have different views on who the AW is. Thus, it may happen that the new AW receives interest information while still believing it is AL; the router must store the interest information, since it will only be sent once by its neighbors. This issue is illustrated through an example in section \ref{sec:interestexamples}.%RV 29/1/2020

%RV: Could the AW, upon realizing it is the new AW, ask for expression of interest from their neighbors? Yes, but there is no advantadge in doing so.

\medskip
\noindent \textbf{Can a router control the interest state of neighbors?} When a router sends an Interest or a NoInterest message to a neighbor, there is no guarantee that the corresponding information is stored at the neighbor, since a neighbor stores interest information according to its tree state. For example, when a router sends interest information to an UNSURE or INACTIVE neighbor the information is not stored. This contrasts with IamUpstream and IamNoLongerUpstream messages: when a router sends these messages to a neighbor, the neighbor always stores the corresponding upstream state, independently of the tree state. Thus, while routers have full control over the upstream information stored at its neighbors, the same is not true for interest information. However, a router can detect when its interest is incorrectly stored at the AW, in which case it must send it the correct interest information.%RV 29/1/2020

\subsubsection{Events triggering the transmission of interest information}%RV 25/7/2019

\begin{figure}[t!]
	\centerline{\includegraphics[scale=0.9]{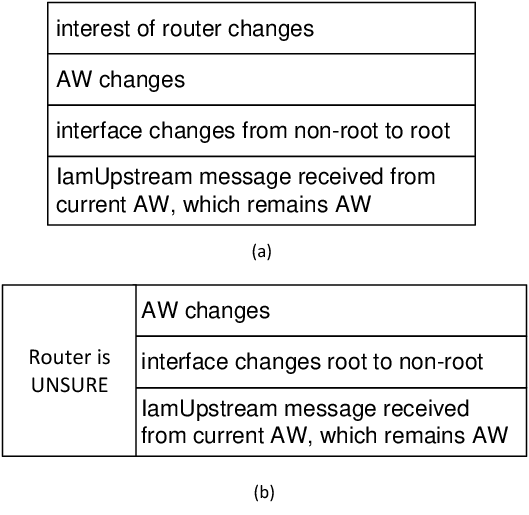}}
	\caption{Events that trigger the transmission of interest messages through (a) root interfaces and (b) non-root interfaces.}
	\label{fig:interest_events}
\end{figure}

The interest information can be sent by both root or non-root interfaces. The events that trigger the transmission of interest messages are listed in Figure \ref{fig:interest_events} and will be described next.%RV 29/1/2020

%RV 27/9/2018

\medskip
\noindent \textbf{Root interfaces} A root interface needs to send interest information on a link whenever (i) the interest of the router changes, (ii) the AW changes, (iii) the root interface changes, and (iv) an IamUpstream message is received from the current AW, which remains AW. The reason for the last event has to do with the possibility that the router to which the AW interface belongs changes its tree state (e.g. from ACTIVE to UNSURE) and loses its interest information without other routers noticing. This case is illustrated through an example in section \ref{sec:interestexamples}. The initial expression of interest (following the construction of the broadcast tree) corresponds to the AW change event: initially there is no AW and the arrival of IamUpstream messages triggers its election.%RV 29/1/2020

\medskip
\noindent \textbf{Non-root interfaces} Non-root interfaces are not interested in receiving multicast data, since they are only forwarders of multicast data. This lack of interest has to be notified to the AW in some situations (using NoInterest messages), if the interface itself is not the AW. There are two cases regarding the events that trigger the transmission of NoInterest messages, which depend on the how the upstream state of the interface is viewed by its neighbors, in particular the AW; this is directly related with the router tree state. Note that, if the interface is root, its neighbors should consider the interface as NOT UPSTREAM and, if the interface is non-root, its neighbors should consider the interface as UPSTREAM if the router is ACTIVE or NOT UPSTREAM if the router is UNSURE or INACTIVE. The two cases are then:
\begin{itemize}	
	\item \textbf{Router is UNSURE:} In this case, the interface has never transmitted an IamUpstream message or the last transmitted upstream message was IamNoLongerUpstream, and all neighbors consider the interface as NOT UPSTREAM. The AW needs to receive information on the interest state of this interface, since there is no way to distinguish it from a root interface. For this reason, a NoInterest message needs to be transmitted when: (i) an IamUpstream message is received from the AW, which remains AW, (ii) the AW changes, or (iii) the interface changes from root to non-root (and the tree state remains or becomes not ACTIVE).
	\item \textbf{Router is ACTIVE:} In this case, the last transmitted upstream message by this interface was an IamUpstream message, and all neighbors consider the interface as UPSTREAM. In this case, the interface can trust that its neighbors have the correct interest information, due to the double meaning of the IamUpstream message. Thus, there is no need to react to the above events. For example, in Figure \ref{fig:intmessages} the non-root interface wouldn't need to send the NoInterest message if it had previously sent an IamUpstream message on the link.
\end{itemize}%RV 3/2/2020

\subsubsection{Configuring the initial downstream interest of interfaces}\label{sec:configdownint}

We allow that the network manager configures initially (at startup) the downstream interest of each non-root interface (as DOWNSTREAM INTERESTED or NOT DOWNSTREAM INTERESTED), which determines if routers are initially INTERESTED or NOT INTERESTED in receiving multicast traffic. This configuration can be superseded by the IGMP or MLD protocols (indicating if the interface has interested downstream hosts) or by the reception of interest messages (indicating if the interface has interested downstream routers). The motivation for introducing this flexibility on the initial configuration of the downstream interest of non-root interfaces has to do with the forwarding of multicast data, which will be discussed in section \ref{sec:dataflooding}.%RV 29/1/2020

\subsubsection{Dealing with the possibility of losing data in transient state}\label{sec:losingdata}

When the AW changes, neighbors may need to send interest information to the AW, and this may introduce a delay in determining the correct forwarding state of the new AW. Thus, there may be temporary loss of data, if the new AW is initially NOT DOWNSTREAM INTERESTED (e.g. because it has outdated interest information) and takes some time to understand that there are downstream routers interested in receiving multicast data. This problem can be minimized by introducing some hysteresis in the FORWARDING state of the previous AW. Specifically, when a non-root interface changes from AW to AL state, it will be allowed to transmit data for an additional period of time.%RV 29/1/2020

%RV: We have still not implemented this feature.

%RV: Refer this solution above, when introducing the Assert protocol 

\medskip
\noindent \textbf{Possibility of multicasting interest messages} There are other alternatives to minimize this problem. One of them is to multicast interest messages (instead of unicasting). This increases the probability that, upon being elected, the AW has the correct downstream state. However, as noted above, this requires more message transmissions since interest messages are ACK-protected. We have selected the unicasting alternative since changes in the interest of routers occur more frequently than broadcast tree reconfigurations (e.g. changes of AW). Changes of interest are triggered by the interest of multicast receivers, and this is a common and legitimate event in multicast networks; broadcast tree reconfigurations are triggered by changes in the network itself (e.g. router failures, router additions, or changes in unicast costs), which are less frequent and less desirable.%RV 29/1/2020

%The multicast alternative described above per se does not ensure that the downstream interest is correct at all times. For example, a router that changes from ACTIVE to UNSURE and then again to ACTIVE would lose the interest state of its NOT UPSTREAM neighbors. Ensuring that the interest information stored at neighbors remains correct at all times requires multicasting interest messages and introducing additional complexity in the protocol. This solution is described in Appendix \ref{alternative_interest_solution}.%RV 12/2/2019

\subsubsection{Examples}\label{sec:interestexamples}%RV 25/7/2019

\begin{figure}[t!]
	\centerline{\includegraphics[scale=0.4]{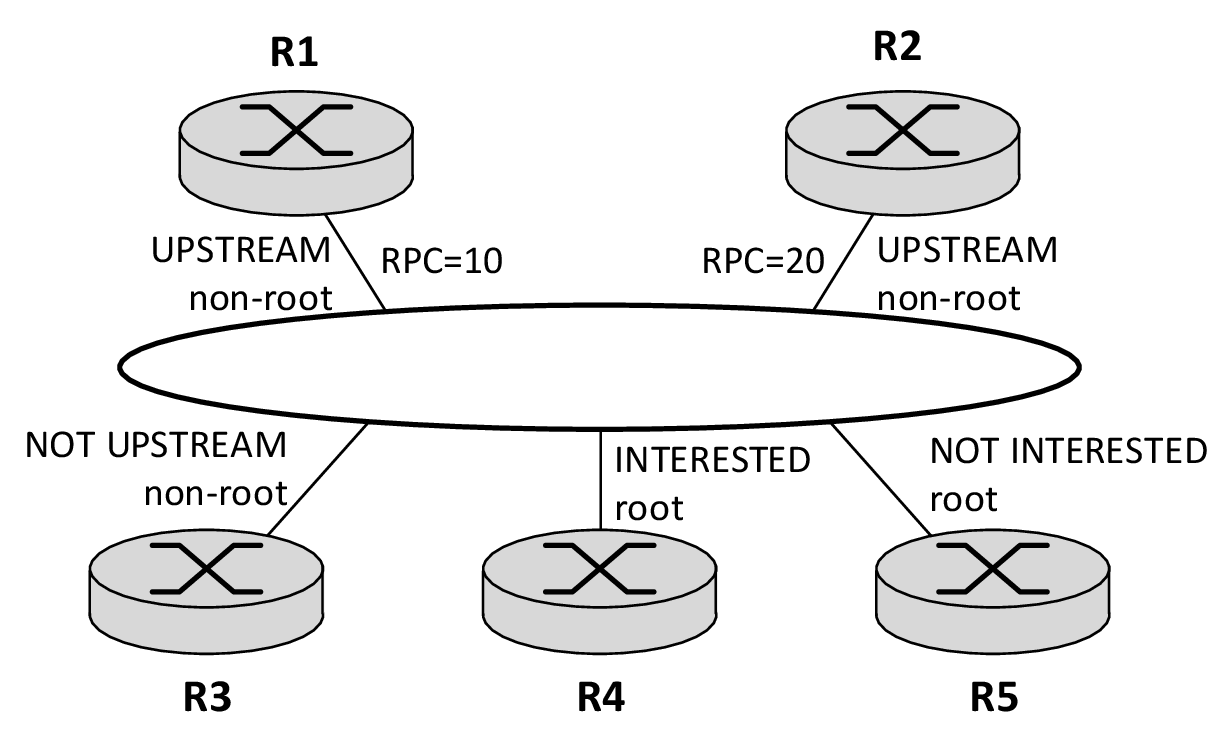}}
	\caption{Interest example (initial configuration).}
	\label{fig:interest_example}
\end{figure}

Consider the shared link represented in Figure \ref{fig:interest_example}, which is part of a larger network. Consider that R1, R2, and R3 attach to the link through non-root interfaces, R4 and R5 through root interfaces, and there are no multicast receivers attached to the link. Moreover, R1 and R2 are UPSTREAM neighbors, R3, R4 and R5 are NOT UPSTREAM, R4 is INTERESTED and R5 is NOT INTERESTED. R3 is NOT UPSTREAM since it is still not connected to the tree, and did not send an IamUpstream message on the link. Furthermore, consider that R1 provides an RPC of 10 and R2 an RPC of 20.%RV 2/2/2020

%RV 27/9/2018

\medskip
\noindent \textbf{Initial expression of interest} When the source is switched on both R1 and R2 send IamUpstream messages on the link. Suppose that R2 is the first one to send a message. In this case R2 is considered AW, and all other routers have to express their interest to it. R5 and R3 unicast a NoInterest message to R2, and R4 unicasts an Interest message to R2. R3 unicasts a NoInterest message because it is a non-root interface and it is sure that the AW considers it as NOT UPSTREAM. The behavior of R1 depends on its tree state when it receives the IamUpstream message of R2: if it is INACTIVE or UNSURE, it unicasts a NoInterest message; if it is ACTIVE, it does nothing since it already sent its own IamUpstream message on the link.%RV 29/1/2020

In any case, R1 will have to transmit an IamUpstream message on the link, which includes its RPC. When other routers receive this message, they learn that R1 is the new AW. In this case, R3, R4, and R5, restate their interest information by retransmitting the same interest messages, but now to R1. R2 sends nothing, since it is sure that its upstream state is correctly stored at R1. Since R2 already sent (reliably) an IamUpstream message, R1 knows that R2 is UPSTREAM and, because of the double meaning of this message, is NOT INTERESTED.%RV 29/1/2020

\medskip
\noindent \textbf{R4 becomes not interested} Suppose now that R4 becomes NOT INTERESTED, leaving no one interested to receive multicast data at the link. In this case, R4 unicasts a NoInterest message to R1 (the AW). Then, the non-root interface of R1 becomes NOT DOWNSTREAM INTERESTED and, therefore, PRUNED. If R1 has no other non-root interface in FORWARDING state, it changes to being NOT INTERESTED and sends a NoInterest message through its root interface, to signal its parent on the tree that it no longer needs to receive multicast data.%RV 29/1/2020

\medskip
\noindent \textbf{R1 interface becomes root} In relation to the initial scenario, suppose now that the interface of R1 becomes root. In this case, R1 multicasts an IamNoLongerUpstream message on the link, which forces the AW to change from R1 to R2. Thus, all other neighbors, including R1, have to express their interest to R2: R3 and R5 unicast a NoInterest message to R2, R4 unicasts an Interest message, and the message unicasted by R1 can be either Interest or NoInterest depending on the interest state of the router.%RV 29/1/2020

\begin{figure}[t!]
	\centerline{\includegraphics[scale=0.7]{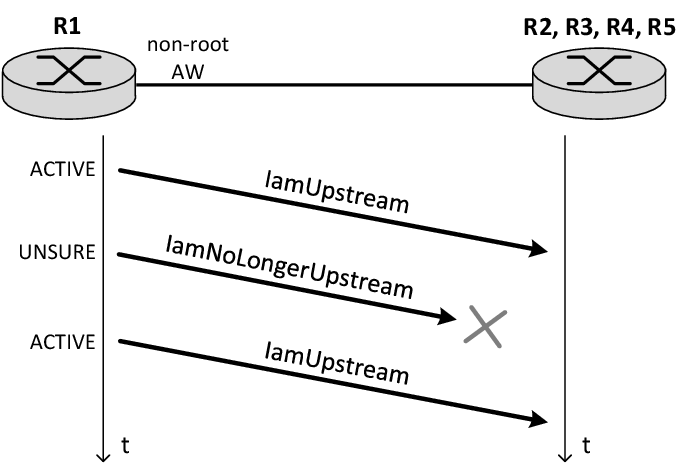}}
	\caption{AW loses interest state without other routers noticing.}
	\label{fig:interest_lastevent}
\end{figure}

\medskip
\noindent \textbf{AW loses interest state without other routers noticing} Again in relation to the initial scenario, suppose that R1 becomes UNSURE (e.g. because it lost its parent on the tree), sends an IamNoLongerUpstream message which is lost, and then changes back to ACTIVE and sends an IamUpstream message. This is illustrated in Figure \ref{fig:interest_lastevent}. In this case, R1 remains AW but all neighbors, except R2, have to retransmit their interest. This is required since R1 may remove its downstream interest information while in UNSURE state. Thus, R3, R4, and R5 have to unicast their interest information to R1 in response to the IamUpstream message. R2 doesn't have since it is sure that its upstream state is correctly stored at R1: R1 knows that R2 is UPSTREAM and, therefore, NOT INTERESTED.%RV 2/2/2020

%PO: Outra razão para ser necessário transmitir o interesse no caso do AW transmitir mensagem de upstream: Apesar das mensagens upstream serem reliably protected, por optimização apenas é necessário garantir a entrega da última mensagem transmitida. Por essa razão no exemplo não é necessário garantir que a mensagem IamNoLongerUpstream seja entregue dado que uma mensagem mais recente foi enviada (IamUpstream) e essa sim terá que ser entregue de forma fiável. Dado isto pode acontecer que nenhum router se tenha apercebido que o vizinho enviou IamNoLongerUpstream (e que tenha perdido o interesse ao transitar para UNSURE), logo o facto de se receber IamUpstream requer retransmissão de interesse visto que mensagens de upstream anteriores podem ter sido transmitidas (mas não recebidas).

%RV 1/10/2018

\begin{figure}[b!]
	\centerline{\includegraphics[scale=0.6]{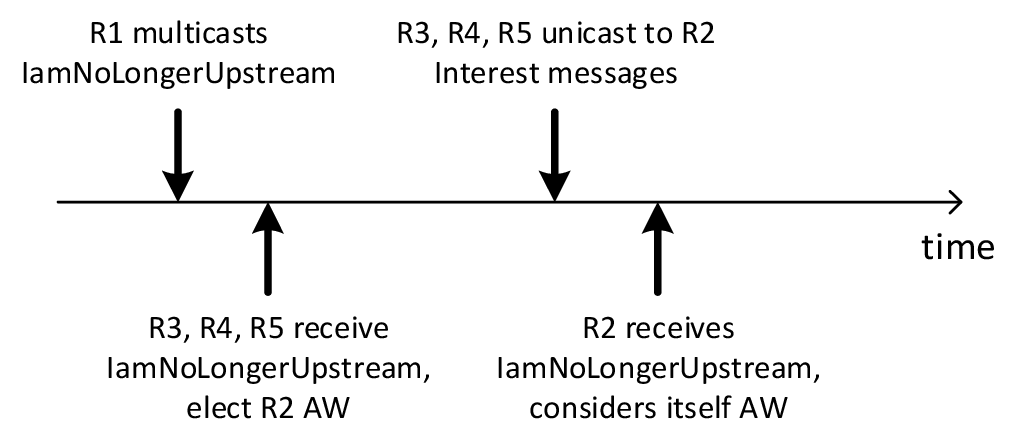}}
	\caption{Sequence of events when new AW receives interest messages before realizing it is the AW.}
	\label{fig:interest_seqawlast}
\end{figure}

\medskip
\noindent \textbf{Why interest information needs to be stored at all non-root interfaces} Consider again the initial scenario, and the sequence of events depicted in Figure \ref{fig:interest_seqawlast}. Suppose that R1 loses its parent on the tree; therefore, it changes from ACTIVE to UNSURE and multicasts an IamNoLongerUpstream message. Moreover, suppose that this message is received first by R3, R4, and R5, and only later by R2. Upon receiving the IamNoLongerUpstream message, the first routers elect R2 as the new AW and unicast immediately to R2 their interest messages. Suppose that when R2 receives these interest messages, it still did not receive the IamNoLongerUpstream message sent by R1 and, therefore, it still considers itself AL. Despite that, R2 will have to store the interest information, otherwise it will not have it when it finally receives the IamNoLongerUpstream message and realizes it is the new AW.%RV 29/1/2020

%RV 1/10/2018

\begin{figure}[t!]
	\centerline{\includegraphics[scale=0.75]{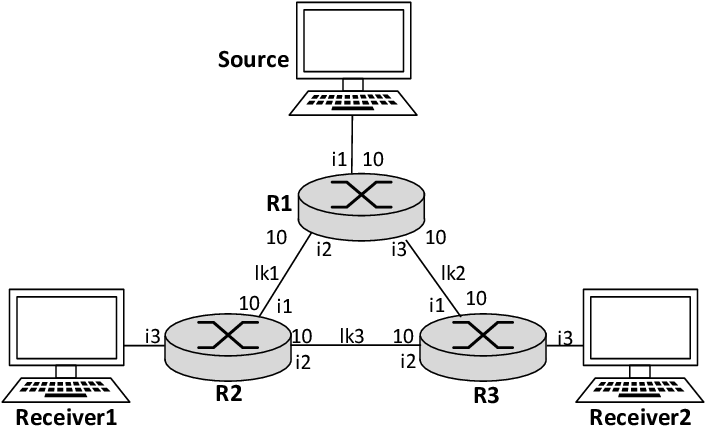}}
	\caption{Role of interest in the forwarding state of point-to-point interfaces.}
	\label{fig:interest_example_p2p}
\end{figure}

\medskip
\noindent \textbf{Role of interest in the forwarding state of point-to-point interfaces} Consider now the network of Figure \ref{fig:interest_example_p2p}, with three routers connected by point-to-point links. R1 is the originator and is the parent of both R2 and R3. In the point-to-point link connecting R2 to R3 (lk3), each router considers the other as UPSTREAM and, therefore, as NOT INTERESTED. For this reason, interfaces i2-R2 and i2-R3 are NOT DOWNSTREAM INTERESTED, and are placed in PRUNED state. The interfaces i3-R2 and i3-R3 are placed in FORWARDING state since they are connected to interested receivers. This causes R2 and R3 to transmit Interest messages through their root interfaces to the respective AW (i.e. to i2-R1 and to i3-R1). After this exchange os messages, R1 forwards traffic directly to R2 and to R3.%RV 29/1/2020

%RV 14/10/2018

In case Receiver1 becomes NOT INTERESTED in receiving multicast data, i3-R2 is placed in the PRUNED state. This causes R2 to transmit a NoInterest message through its root interface to the link AW, i.e. to i2-R1. This interface becomes NOT DOWNSTREAM INTERESTED and is placed in the PRUNED state.%RV 29/1/2020

Consider again the initial scenario (both receivers interested), but suppose that the cost of interface i1-R2 is set to 30. This causes R3 to become the parent of R2. At link lk1, each router considers its neighbor as UPSTREAM and NOT INTERESTED, and both interfaces are placed in the PRUNED state. At link lk3, i2-R3 is considered the AW, and i2-R2 is a root interface. Router R2 transmits an Interest message to i2-R3 and this interface is placed in the FORWARDING state. Router R3 also transmits an Interest message through its root interface to the corresponding link AW, which is i3-R1, and this interface is placed in the FORWARDING state. The forwarding of traffic from R1 to Receiver1 is now via R3 and R2.%RV 29/1/2020

%RV: We may enumerate the various cases regarding point-to-point links above

\subsubsection{Summary of interest maintenance}

Routers may not be interested in receiving multicast traffic (through their root interfaces) and non-root interfaces may not be interested in transmitting it, in which case they must be pruned from the multicast distribution tree. Ultimately, the interest of a router depends on the interest of its downstream devices, which can be routers or multicast receivers. Multicast receivers signal their interest to the routers directly attached to them through the IGMP (IPv4) or MLD (IPv6) protocols. For the purpose of interest maintenance, the following state information is defined: the interest of a router, INTERESTED or NOT INTERESTED, which is equivalent to the interest of its root interface in receiving multicast data; and the downstream interest of an interface in transmitting multicast data, DOWNSTREAM INTERESTED or NOT DOWNSTREAM INTERESTED. The forwarding state of an interface depends on its downstream interest and assert states. Specifically, an interface is FORWARDING if it is AW and DOWNSTREAM INTERESTED, and is PRUNED otherwise. The initial downstream interest of an interface (when a new tree is being created) can be configured by the network manager.%RV 2/2/2020

The interest of the router interfaces in receiving multicast data is signaled to its neighbors through Interest or NoInterest messages. Non-root interfaces are not interested in receiving multicast traffic and, therefore, they only transmit NoInterest messages. Root interfaces transmit Interest or NoInterest messages depending on the router interest. Within a link, the interest messages are unicasted to the AW, since only the AW needs to have the interest information updated. This is because AL interfaces are PRUNED irrespective of the interest of its downstream devices. AW interfaces need this information to determine if they become FORWARDING or PRUNED.%RV 2/2/2020

A neighboring interface that is UPSTREAM is necessarily non-root and, therefore, is not interest in receiving multicast traffic. Thus, when a router receives an IamUpstream message from a neighbor it knows that the neighbor is not interested. Because of this double meaning, when a router transmits an IamUpstream message, it doesn't need to transmit a NoInterest message.%RV 29/1/2020

Root interfaces must transmit interest messages when (i) the interest of its router changes, (ii) the AW changes, (iii) the interface just changed to root, and (iv) an IamUpstream message is received from the AW, which remains AW. A non-root interface is only required to transmit interest messages (in this case, only NoInterest messages) when they are NOT UPSTREAM, in case of events (ii), (iii), and (iv). When the interface is UPSTREAM, their neighbors can immediately understand from this state that it is not interested.%RV 29/1/2020

%PO: Indicar o siginficado de interface ser UPSTREAM ou NOT UPSTREAM em termos de estado da árvore: a interface é UPSTREAM quando a árvore é ACTIVA; interface é NOT UPSTREAM em caso contrário.

\subsubsection{Comparison with PIM-DM}

In PIM-DM, the messages that convey interest information are the Join, Prune and Graft messages. None of these messages have ordering guarantees, and only the latter is transmitted reliably. This means that interest information may not arrive or may arrive out of order at the intended destinations. In HPIM-DM, the interest messages are transmitted reliably and with ordering guarantees.%RV 29/1/2020

In PIM-DM, routers only transmit interest messages in reaction to interest changes, to AW changes, or when there is a periodic reconfiguration of the tree. A router does not know, prior to a reconfiguration, whether the AW holds the freshest interest of the link. In HPIM-DM, a router knows exactly when to transmit an interest message, even when the AW loses previously stored information.%RV 20/8/2019

In PIM-DIM, at shared links there must exist a delay between the reception of a Prune message and the actual pruning of the AW, to give opportunity for interested root interfaces to override the Prune. In HPIM-DIM, the AW knows the interest of all its neighbors and, therefore, no delay is incurred when a change of interest occurs.%RV 20/8/2019

\subsection{Data forwarding}\label{sec:dataflooding}%RV 21/8/2019

Data packets are forwarded down the multicast tree, being received by root interfaces and retransmitted by non-root interfaces in the FORWARDING state. In a well formed multicast tree, data packets are transmitted only once in each link. However, during transient periods packets may be lost or transmitted more than once at a link. Two cases require attention: data forwarding (i) during the formation of the tree and (ii) during a tree reconfiguration.%RV 31/1/2020

\medskip
\noindent \textbf{Tree reconfiguration} As discussed in section \ref{sec:losingdata}, during a tree reconfiguration there may be temporary loss of data at a link if there is an AW change and the new AW takes some time to understand that there are downstream routers interested in receiving multicast data. This problem is minimized by allowing an interface that changes from AW to AL to transmit data for an additional period of time.%RV 31/1/2020

\medskip
\noindent \textbf{Tree formation} The construction of the multicast tree is made using only control messages, and it can take non-negligible time. To avoid losing the initial data packets, the forwarding of data must be allowed even if the tree is still not formed. To enable this possibility, we take two measures.%RV 31/1/2020

First, the assert state is defined for the INACTIVE and UNSURE states. Specifically, as discussed in section \ref{sec:awdefinition}, if a router is INACTIVE all its non-root interfaces are placed in the AW state, and if it is UNSURE the non-root interfaces with no UPSTREAM neighbors are placed in the AW state. In the UNSURE state, if one or more neighbors are UPSTREAM they are also ACTIVE and one of them will surely be elected AW.%RV 31/1/2020

Second, as discussed in section \ref {sec:configdownint}, the downstream interest of non-root interfaces can be configured initially (at start-up) by the network manager. If the network manager configures all non-root interfaces as DOWNSTREAM INTERESTED, and since they are also in AW state, they become FORWARDING and data is flooded throughout the whole network. The network manager may also try to prevent the initial data flooding by configuring all non-root interfaces as NOT DOWNSTREAM INTERESTED. However, recall that this configuration is immediately superseded at an interface if there are downstream receivers attached to it, as signaled by the IGMP or MLD protocols.%RV 31/1/2020

\subsubsection{Summary of data forwarding}

Data packets may be lost in transient periods, during the formation and reconfiguration of a multicast tree. To avoid this type of loss, (i) an interface changing from AW to AL is allowed to transmit data for an additional period of time, (ii) the interfaces of UNSURE and INACTIVE routers are placed in the AW state (except the interfaces of UNSURE routers with UPSTREAM neighbors), and (iii) the initial downstream interest of routers is allowed to be configurable.%RV 31/1/2020

\subsubsection{Comparison with PIM-DM}

In PIM-DM, data is what triggers the construction and maintenance of multicast trees. Thus, in PIM-DM, the initial forwarding behavior cannot be configured, whereas in HPIM-DM it can.%%RV 31/1/2020

\subsection{Message sequencing}\label{sec:messagesequencing}%RV 21/8/2019

For the correct operation of the protocol, control messages need to be processed at the intended receivers by the order they were transmitted. For example, if an originator transmits an IamNoLongerUpstream message (e.g. because its source was switched off), and shortly after transmits an IamUpstream message (e.g. because its source was switched on again), the former must be processed before the latter at its neighbors, otherwise the neighbors will keep believing that the source remains inactive. However, the link between two neighboring routers can be a relatively complex layer-2 network, which cannot guarantee the preservation of the transmission order, e.g. during transient periods where layer-2 routing is changing. Thus, some mechanism must be introduced to ensure that messages are orderly processed at the receivers. This mechanism is called \textit{message sequencing} and will be described in this section.%RV 31/1/2020

\subsubsection{Sequence Numbers}

One way to solve the message sequencing problem is to use linear Sequence Numbers (SNs), i.e. to number the messages sequentially according to the transmission order (such that whenever a new message needs to be transmitted, conveying more recent information, its SN is incremented by one). The SNs start at SN=1 and end at a final value that depends on the number of bits used for SN encoding.%RV 31/1/2020

Moreover, we assume that the protocol is designed such that outdated information is no longer useful, in which case only the highest SN received from a neighbor needs to be stored.%RV 31/1/2020

Using linear SNs poses the problem of what to do when the final SN value is reached. This event is called \textit{SN overflow}, and the way to deal with it is addressed in section \ref{sec:snrestart}.%RV 31/1/2020

\medskip
\noindent \textbf{SNs per message type and per tree} The SN spaces are per message type, and not per individual message. The relevant message types for this purpose are two: upstream messages (IamUpstream and IamNoLongerUpstream) and interest messages (Interest and NoInterest). A receiver needs to know if a received IamUpstream is more recent than a received IamNoLongerUpstream, and the same is true for Interest and NoInterest messages. Thus, routers must use one SN space for both IamUpstream and IamNoLongerUpstream messages, and another SN space for both Interest and NoInterest messages. These two SN spaces can be independent but, as will be discussed in section \ref{sec:singlesnspaceperinterface}, it is possible that messages of different types share the same SN space. Sharing SN spaces allows saving in the amount of stored information at routers.%RV 31/1/2020

In principle, messages related with different multicast trees use separate SN spaces, but as it will also be discussed in section \ref{sec:singlesnspaceperinterface}, it is again possible that these messages share the same SN space.%RV 31/1/2020

\medskip
\noindent \textbf{Local message sequencing} Message sequencing is implemented locally between neighbors. Each router stores the highest SN received from each neighbor so far and, in this way, can determine if a received message contains more recent information. In HPIM-DM, there is no need to ensure ordering of message transmissions beyond a link, since the state communicated by upstream and interest messages is also of local nature. For example, routers do not \textit{directly} store information on whether a source, which may be several hops away, is active or not; routers only store information on whether a neighbor is UPSTREAM or not for the source, which indirectly indicates whether the source is active or not.%RV 31/1/2020

\begin{figure}[t!]
	\centerline{\includegraphics[scale=0.4]{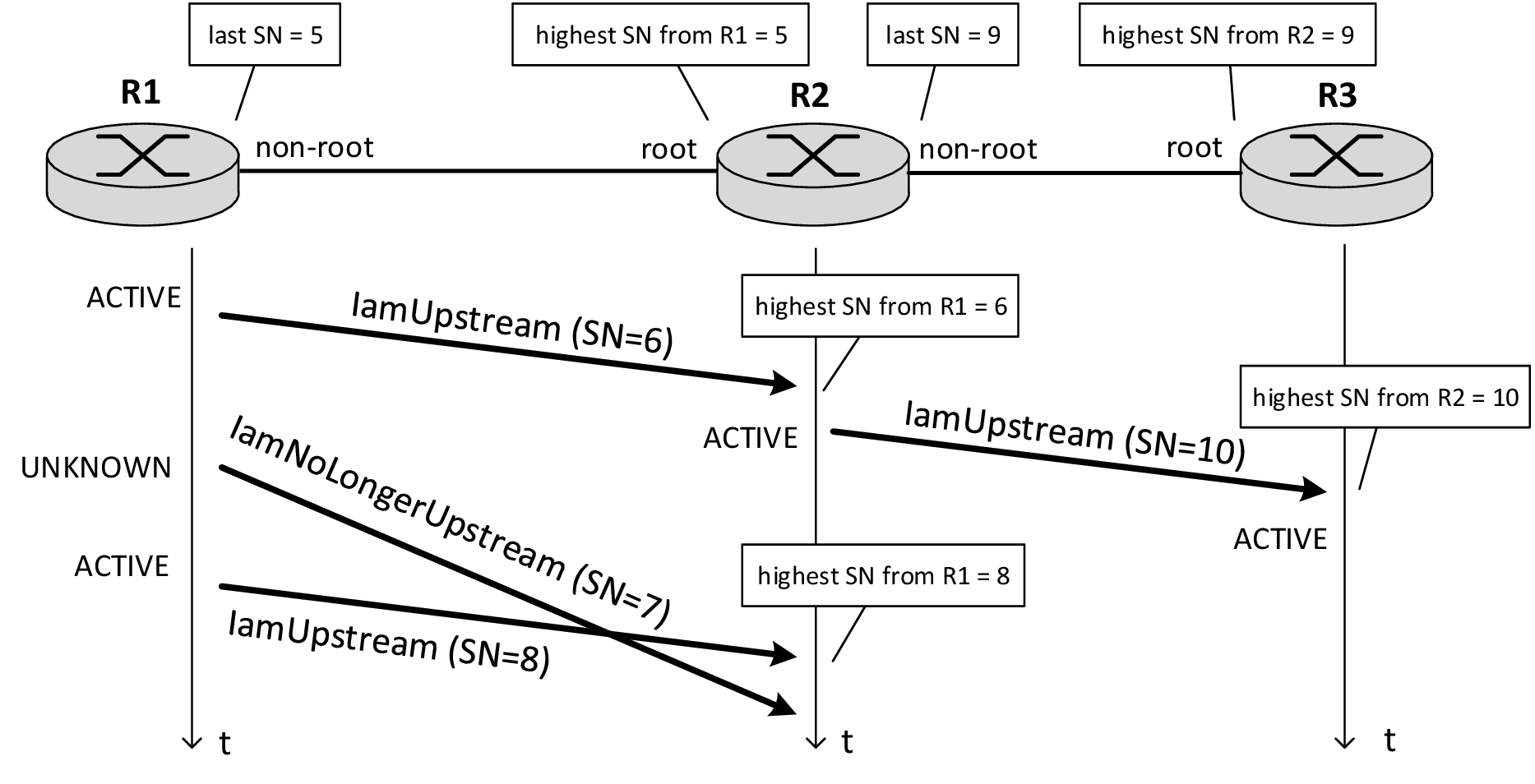}}
	\caption{Local sequence numbers.}
	\label{fig:seqnumbers}
\end{figure}

\medskip
\noindent \textbf{Example} To illustrate the use of SNs, consider the example of Figure \ref{fig:seqnumbers}. R1 is directly connected to a multicast source (not shown in the figure), and all routers are initially INACTIVE regarding the corresponding tree. In this example, the source is switched on, then switched off, and finally switched on again. Suppose that, initially, the last SN used by upstream messages was SN=5 at router R1 and SN=9 at router R2. When the source is switched on, R1 becomes ACTIVE and sends an IamUpstream message to R2 with SN=6. Upon receiving this message, R2 sets the highest SN received from R1 to SN=6, since the stored SN is lower (SN=5). Moreover, R2 becomes ACTIVE and sends an IamUpstream message to R3 with SN=10. Upon receiving this message, R3 sets the highest SN received from R2 to SN=10.%RV 31/1/2020

When the source is switched off, R1 sends an IamNoLongerUpstream message with SN=7 and, when it is switched on again, it sends an IamUpstream message with SN=8. Suppose that the order of arrival of these two messages is reversed, i.e. the IamUpstream arrives first at R2. At this point, R2 sets the highest SN received from R1 to SN=8, since the stored SN is lower (SN=6). The received message maintains the router in the ACTIVE state and, therefore, no message is transmitted to R3. When the delayed IamNoLongerUpstream message is finally received at R2, it is ignored since its SN is lower than the highest one received so far (SN=8).%RV 31/1/2020

%RV 2/10/2018

\subsubsection{Enlarging the scope of SN spaces}\label{sec:singlesnspaceperinterface}

Having to store SNs per neighbor, per message type, and per tree, requires considerable memory space. In HPIM-DM, we devised a solution where (i) the messages transmitted by a router use a single SN space per interface, irrespective of message type and tree, and (ii) each router stores, for each tree, the highest SN received from each neighbor interface, irrespective of message type. Thus, transmitted SNs are per interface and the received SNs are stored per neighbor interface and tree. In this way, the amount of sequencing information that needs to be stored is significantly reduced.%RV 21/8/2019

\begin{figure}[t!]
	\centerline{\includegraphics[scale=0.7]{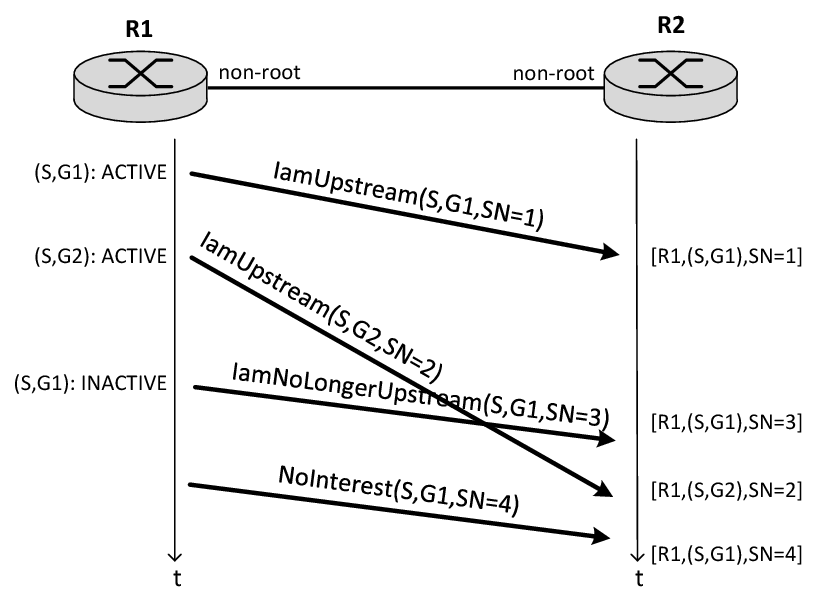}}
	\caption{Transmitted SNs per interface and received SNs stored per neighbor and tree.}
	\label{fig:store_sn_per_tree}
\end{figure}

\medskip
\noindent \textbf{Example} Figure \ref{fig:store_sn_per_tree} illustrates these optimizations, and motivates the need for storing the SNs of received messages per neighbor interface and per tree. R1 maintains information regarding two trees: (S,G1) and (S,G2). Suppose that R1 becomes ACTIVE for (S,G1), then becomes ACTIVE for (S,G2), and finally becomes INACTIVE for (S,G1). Thus, it transmits first an IamUpstream message for (S,G1) with SN=1, then an IamUpstream message for (S,G2) with SN=2, and finally transmits an IamNoLongerUpstream message for (S,G1) with SN=3 and a NoInterest message for (S,G1) with SN=4. Since the transmitted SNs are per interface, R1 increments the SN by one whenever a new message is transmitted, irrespective of message type and tree. R2 has to store the received sequencing information per neighbor and per tree. Thus, when the first message arrives it stores information about the neighbor, the tree, and the SN, i.e. [R1, (S,G1), SN=1]. As before, the stored SN is the highest SN received so far. Suppose that the second and third messages are received in reverse order. When the (S,G1) IamNoLongerUpstream message is received, R2 updates the sequencing information to [R1, (S,G1), SN=3], and when the (S,G2) IamUpstream is received, it updates the sequencing information to [R1, (S,G2), SN=2]. Since these messages belong to different trees, their sequencing information is stored independently, as if the messages were received without order reversal.%RV 21/8/2019

Suppose for a moment that the sequencing information is not stored per tree, but only per neighbor. In this case, R2 would store [R1, SN=1] when the (S,G1) IamUpstream message is received, [R1, SN=3] when the (S,G1) IamNoLongerUpstream message is received, and the (S,G2) IamUpstream message would be ignored since it has a lower SN than the stored one. Thus, the upstream information regarding (S,G2) and R1 would become incorrect at R2. This example illustrates the need for storing received SNs per neighbor and per tree.%RV 21/8/2019

%RV 2/10/2018

\medskip
\noindent \textbf{Problem introduced by merging the SN space of different message types} Merging the SN space of message types can be problematic. There are circumstances where a router needs to transmit two messages of different types, to notify its neighbors about its upstream and interest states. By having a single SN space, in case the two messages are received in reverse order, the message with the lower SN is never processed, since the message with higher SN arrives first. For example, if a router transmits an IamNoLongerUpstream message followed by a NoInterest message and the second message arrives first at the receiver, the router does not process the IamNoLongerUpstream message, and the corresponding upstream state becomes incorrect.%RV 21/8/2019

\medskip
\noindent \textbf{Increasing the semantic scope of control messages} This problem can be solved by increasing the semantic scope of some control messages. Specifically, (i) IamUpstream messages are also interpreted as NoInterest, and (ii) Interest or NoInterest messages are also interpreted as IamNoLongerUpstream.%RV 21/8/2019

The first case was already introduced in section \ref{sec:interestmaintenance}. The double meaning is possible since IamUpstream messages are only transmitted by non-root interfaces, and non-root interfaces are not interested in receiving multicast data. Moreover, due to this double meaning, the transmission of NoInterest messages is suppressed whenever an interface transmits IamUpstream messages.%RV 21/8/2019

The double meaning of the second case is made possible by the first case: since an UPSTREAM interface never transmits interest messages, receiving an interest message is an indication that the sending interface is NOT UPSTREAM. However, unlike the first case, we do not suppress the transmission of IamNoLongerUpstream messages. This is because interest messages are unicasted to the AW and, therefore, cannot convey upstream state to all neighbors. Thus, the double meaning is only used to ensure that the correct upstream state is set in case of order reversal: an IamNoLongerUpstream message is always transmitted before the Interest or NoInterest messages but, if it arrives after and gets discarded because of a lower SN, the arrival of the interest message already set the upstream state of the sending router to NOT UPSTREAM.%RV 21/8/2019

\begin{figure}[t!]
	\centerline{\includegraphics[scale=0.4]{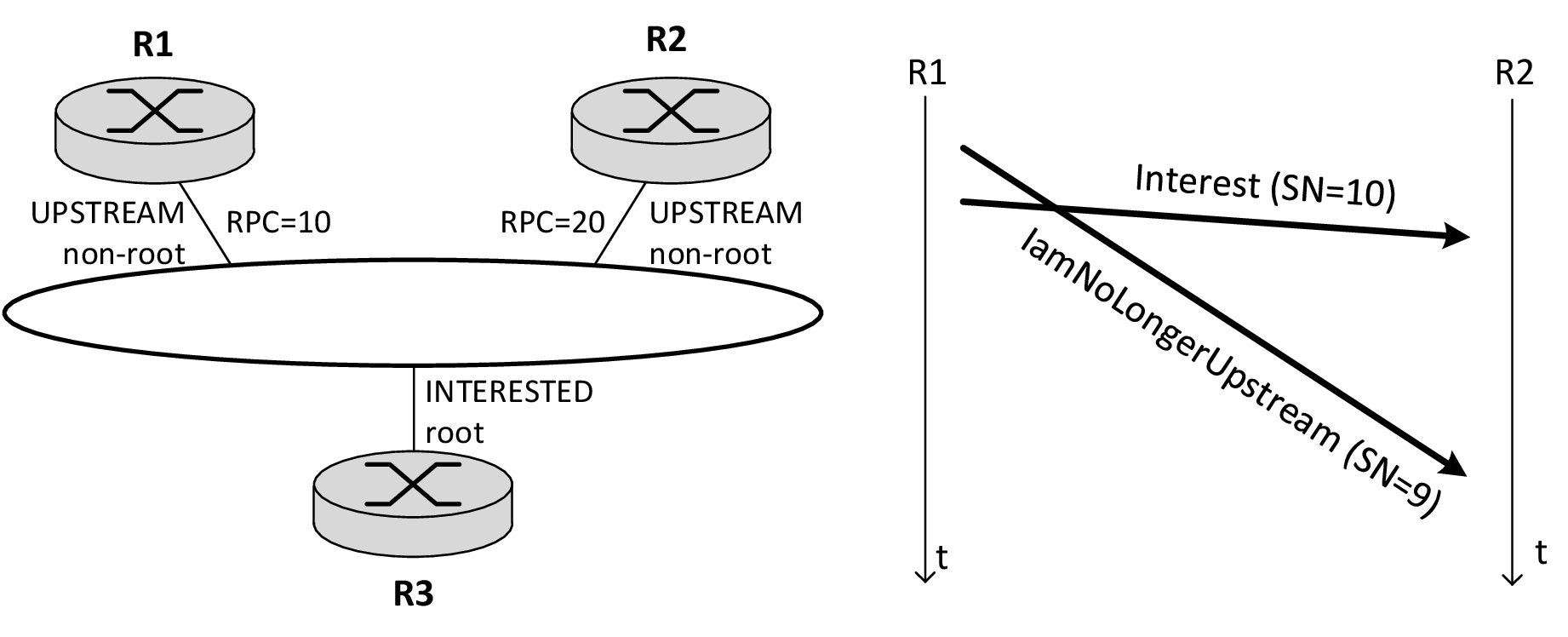}}
	\caption{Merging the semantic scope of upstream and interest messages.}
	\label{fig:messages_semantic_scope_example}
\end{figure}

\medskip
\noindent \textbf{Example} Consider the example of Figure \ref{fig:messages_semantic_scope_example}. R1 and R2 attach to the link through a non-root interface, and R3 through a root interface; R1 offers a lower RPC and, therefore, is the AW. Suppose that the interface of R1 becomes root and the router becomes interested in receiving multicast data. In this case, R1 determines that R2 is the new AW. R1 will then send an IamNoLongerUpstream message followed by an Interest message, with the Interest message having a higher SN. The IamNoLongerUpstream message is multicasted to all neighbors but the Interest message is only unicasted to the AW. Now suppose that the messages are received at R2 in reverse order. In this case, the IamNoLongerUpstream message is ignored at R2 since, when it arrives, a message with a higher SN has already been processed. However, due to the double meaning of the Interest message, the correct upstream state regarding R1 is set at R2. When R2 receives the Interest message it sets the state of R1 as being NOT UPSTREAM and INTERESTED (and realizes it is the new AW). R3 does not receive the Interest message (nor does it need, since it is not AW). When it receives the IamNoLongerUpstream message it elects R2 as the new AW and sends it an Interest message.%RV 21/8/2019

If the Interest message had not a double meaning, the upstream state of R2 regarding R1 would become incorrect: R2 would keep believing that R1 is UPSTREAM and, therefore, the AW.%RV 21/8/2019

%RV 2/10/2018 

\subsubsection{SN restart}\label{sec:snrestart}

Routers are only required to store the highest SN received so far. In this setting, a message received with an SN lower or equal than the stored one is ignored, since it is considered to include outdated information. However, there are situations where the SN needs to be restarted from the initial value, i.e. SN=1. In this case, messages with a lower SN can indeed include fresher information.%RV 21/8/2019

\medskip
\noindent \textbf{SN restart scenarios} There are two situations where the SN has to be restarted:
\begin{itemize}
	\item Reboot of router (due to configuration action or failure) - When a router reboots it loses all its information, and message sequencing has to be started from the initial SN value.
	\item SN overflow - When the SN reaches its final value, the next SN is the initial SN value.
\end{itemize}%RV 21/8/2019

\medskip
\noindent \textbf{BootTime} When there is an SN restart, messages with lower SN may be ignored despite including fresher information. One solution to this problem is to include in transmitted messages an additional sequence number that is incremented when an SN restart event occurs (reboot or SN overflow). This is similar to the Extended Sequence Numbers used by OSPFv2 \cite{RFC7474} and IS-IS \cite{RFC7602} to prevent replay attacks. This additional sequence number is called BootTime.%RV 21/8/2019

\medskip
\noindent \textbf{BootTime implementation alternatives} The BootTime sequencing must be preserved, even in the case of reboots. There are two ways of implementing it. One solution is to use the router's clock, and make the BootTime equal to the clock value when an SN restart event occurs. This solution is suggested in \cite{RFC7602}. Another solution is to rely on the router's non-volatile memory to store the last BootTime value. In this case, the BootTime is a sequence number that is incremented by one when an SN restart event occurs. Moreover, the new BootTime must replace the previous one stored in the router's non-volatile memory. This solution is suggested in \cite{RFC7602, RFC7474}. In our implementation we have used the first solution.%RV 21/8/2019

\medskip
\noindent \textbf{BootTime precedence over SN} The BootTime is transmitted in all control messages together with the SN. The first takes precedence over the former in deciding which message is fresher. Thus, a message with a higher BootTime is always considered fresher. When two messages have the same BootTime, the one with highest SN is considered fresher.%RV 21/8/2019

\medskip
\noindent \textbf{Storage of received BootTime} The receivers only have to store one BootTime per neighbor, unlike the SNs which have to be stored per neighbor and per tree.%RV 21/8/2019

\medskip
\noindent \textbf{Need for resynchronization} When a router receives a control message from a neighbor with a BootTime higher than the stored one, it must resynchronize with the neighbor, since this means that an SN restart event occurred at the neighbor. The synchronization process is described in section \ref{sec:initsync}.%RV 21/8/2019

\medskip
\noindent \textbf{BootTime per interface} The BootTime is defined per interface (and not per router), to avoid synchronizing all interfaces in case of SN overflow.%RV 21/8/2019

\begin{figure}[t!]
	\centerline{\includegraphics[scale=0.7]{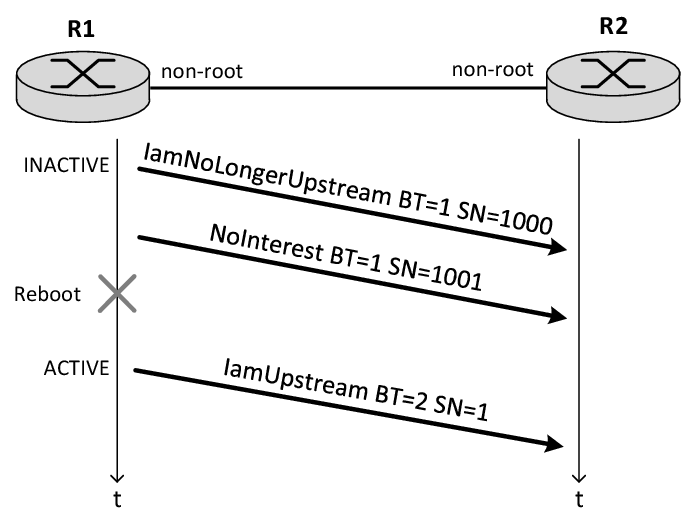}}
	\caption{Use of BootTime.}
	\label{fig:boot_time}
\end{figure}

\medskip
\noindent \textbf{Example} The example of Figure \ref{fig:boot_time} illustrates the use of the BootTime (referred as BT). In the Figure, R1 becomes INACTIVE and transmits an IamNoLongerUpstream message with SN=1000 followed by a NoInterest message with SN=1001. Both messages have BT=1 and therefore the second message is considered fresher. Based on these messages, R2 sets the state of R1 as being NOT UPSTREAM and NOT INTERESTED. Now suppose that R1 reboots and changes to ACTIVE. It will then send an IamUpstream message with the initial SN value (SN=1) but with a BootTime incremented by one (BT=2). The previous SN was lost due to the reboot, but the BootTime was not since it was stored in R1's non-volatile memory. The IamUpstream message is considered fresher than the previous one since it has a higher BootTime. Thus, router R2 sets correctly the upstream state of R1 as being UPSTREAM. Were the BootTime not used, the last message would be discarded, and R2 would keep believing, incorrectly, that R1 is NOT UPSTREAM.%RV 21/8/2019

%RV 2/10/2018

\subsubsection{Removing stored sequencing information}\label{sec:removeseq}%RV 3/10/2018

As discussed in section \ref{sec:singlesnspaceperinterface}, a router is required to maintain SN information per neighbor and per tree, which can occupy significant memory resources. Given that multicast sources (and the corresponding trees) can be switched on and off fairly frequently, an important question is for how long the SNs need to be stored at routers.%RV 21/8/2019

\begin{figure}[t!]
	\centerline{\includegraphics[scale=0.45]{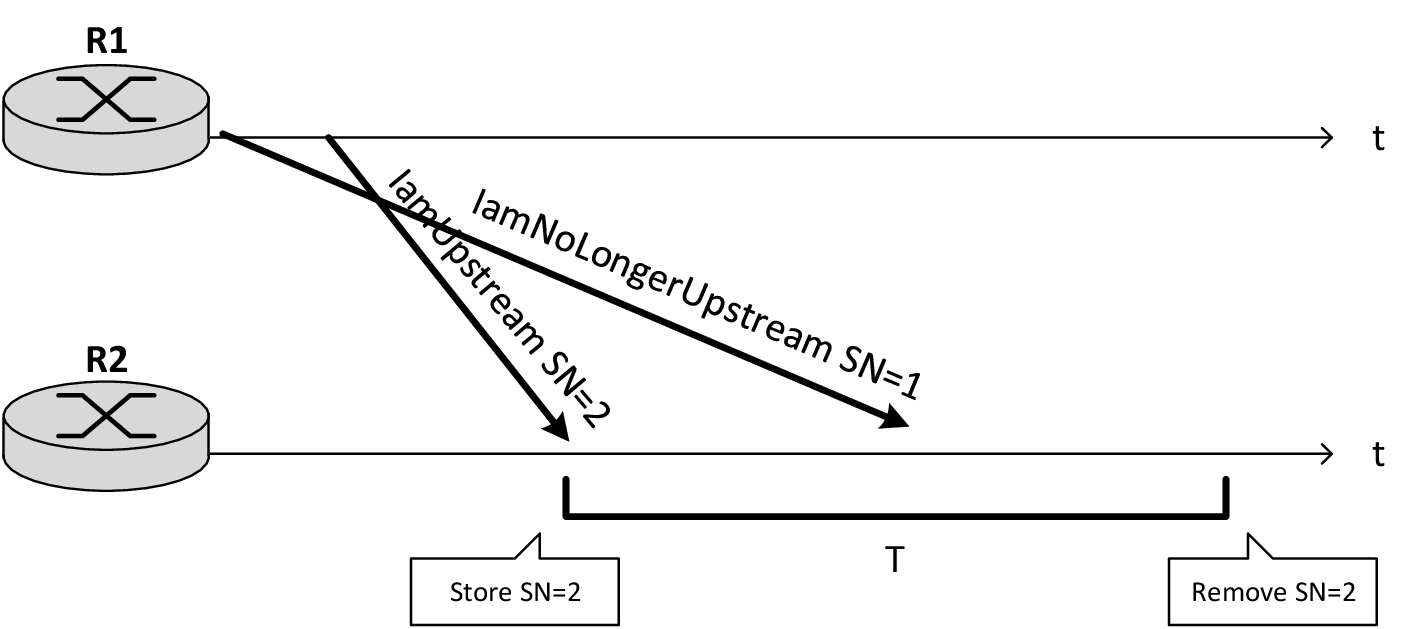}}
	\caption{Removing SNs when there is a network delay bound.}
	\label{fig:delaybound}
\end{figure}

\medskip
\noindent \textbf{Removing SNs when there is a network delay bound} SNs need to be stored as long as there is the possibility of an older message still being received. Suppose that there is an upper bound on the delay between the transmission of a message and its reception by all neighbors; we denote this bound by $T$. In this case, the SN of a message can be safely removed $T$ time units after its reception. This is illustrated in Figure \ref{fig:delaybound}. R1 transmits an IamNoLongerUpstream message with SN=1 followed by an IamUpstream message with SN=2, and they are received in reverse order. When the IamUpstream message arrives at R2, R2 stores its SN. This SN needs only be stored for $T$ time units since it is guaranteed that after this period no message with a lower SN will be received. Thus, the first message, despite arriving after the second one, arrives surely during this period, and is ignored since it has a lower SN.%RV 21/8/2019

%RV 3/10/2018

\medskip
\noindent \textbf{The possibility of replay attacks} However, the network may be subject to replay attacks. An attacker may capture a message transmitted between routers, and retransmit it later. In this case, if the message is retransmitted to a router after the removal of the SN, the router will have to accept it, and this can cause a malfunction. Unfortunately, there is no delay bound for a replay attack and thus, to avoid these attacks, SNs must be stored indefinitely.%RV 21/8/2019

\medskip
\noindent \textbf{The CheckpointSN} We devised a solution to minimize the amount of stored sequencing information. Recall that transmitters perform message sequencing per interface (irrespective of tree) but receivers are required to store SNs per transmitter and per tree. In our solution, an interface sends periodically to its neighbors information on the highest SN of the messages sent by itself and acknowledged so far by the destination interfaces, such that all SNs lower than it have also been acknowledged. This SN will be called the \textit{CheckpointSN}. Note that a message with a higher SN can be acknowledged before a message with a lower one; the CheckpointSN is such that all SNs lower than it have already been acknowledged. When a router receives the CheckpointSN, it can be sure that the transmitter will not send any message, for any tree, with a lower SN. Thus, it can remove all sequencing information of trees that have an SN lower than the CheckpointSN and, from then on, assume that the stored SN for any other tree (for which it might receive information in the future) equals the CheckpointSN.%RV 21/8/2019

\begin{figure}[t!]
	\centerline{\includegraphics[width=\linewidth]{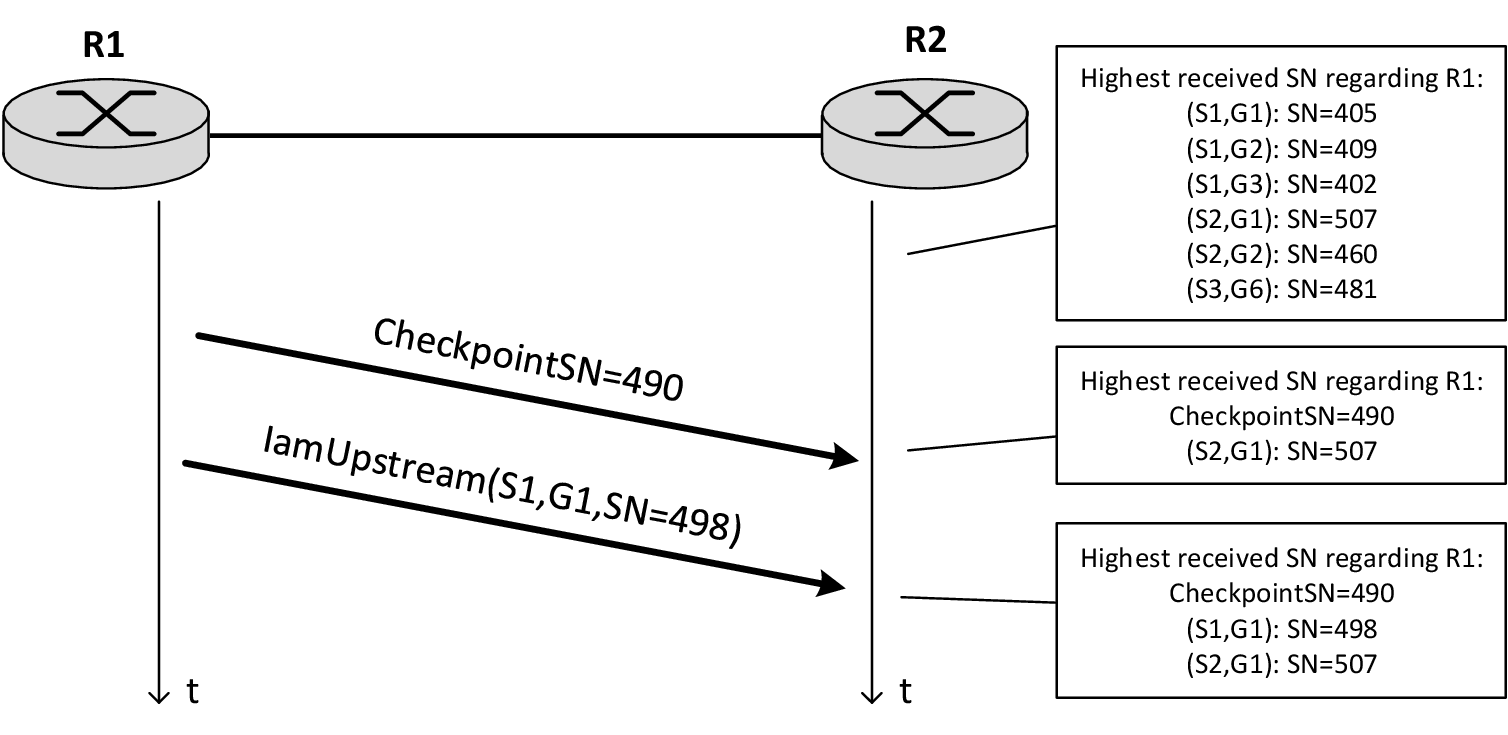}}
	\caption{Use of CheckpointSN.}
	\label{fig:checkpointsn}
\end{figure}

\medskip
\noindent \textbf{Example} Figure \ref{fig:checkpointsn} gives an example. Initially R2 stores SNs of six trees regarding R1. Assume that the CheckpointSN is 490. This means that all messages transmitted by R1 with a SN equal to or lower than 490 have already been acknowledged. When R2 receives the CheckpointSN, it can remove the sequencing information of the five trees for which the highest received SN is lower than 490. Moreover, the CheckpointSN is stored and becomes the SN of all unknown trees. If later R1 sends a message for (S1,G1) with SN=498, R2 will store again the sequencing information of this tree, since the corresponding SN is higher than the CheckpointSN.%RV 21/8/2019

%RV 3/10/2018

\medskip
\noindent \textbf{Control messages used for the CheckpointSN transmission} The CheckpointSN could be transmitted using a specific control message type. This message needs not be ACK-protected since the failure of its reception would not compromise protocol correctness. Another option is to include the CheckpointSN in Hello messages. However, since the frequency of Hello transmissions is much higher than the one required for CheckpointSN transmissions, the CheckpointSN can be included only in some Hello messages. In our implementation we have adopted the second solution.%RV 21/8/2019

\subsubsection{Summary of the sequencing mechanisms}

The operation of HPIM-DM requires that upstream and interest messages be processed at the receivers in the order they were transmitted. However, the underlying network may not preserve the transmission order, and some mechanism must be introduced to solve this problem. In HPIM-DM, the messages use sequence numbers to address this problem. Sequence numbers are such that messages with higher sequence numbers carry fresher information and must be accepted at routers for processing.%RV 21/8/2019

In HPIM-DM, we use two types of sequence numbers, one incremented on a message basis, called SN, and another incremented whenever the first needs to be restarted, called BootTime. The SN values start at SN=1 and are incremented by one. An SN restart occurs when the sending router reboots or the SN reaches its maximum value (SN overflow). The BootTime value corresponds to the router's clock when the restart event occurred and takes precedence over the SN. The message sequencing is performed locally between neighboring routers. Thus, sequence numbers have link scope.%RV 21/8/2019

In HPIM-DM, we introduced several optimizations to minimize the amount of stored sequencing information. First, the messages transmitted by a router use a single SN space per interface, irrespective of message type and tree, and each router stores the highest SN received from each neighbor and the corresponding tree, irrespective of message type. To enable this optimization, and since different message types share the same SN space, a double interpretation must be given to some messages. Specifically, IamUpstream messages are also interpreted as NoInterest, and Interest or NoInterest messages are also interpreted as IamNoLongerUpstream. Finally, to avoid that SNs are indefinitely stored at receivers we introduced the CheckpointSN. The CheckpointSN is sent periodically by a transmitting interface to its neighbors and indicates the highest SN that has been acknowledged so far, such that all SNs lower than it have also been acknowledged. In this case, receivers can store the CheckpointSN and remove sequencing information relative to trees that have an SN lower or equal than the CheckpointSN.%RV 21/8/2019

\subsubsection{Comparison with PIM-DM}

PIM-DM does not offer ordering guarantees for control messages. Thus, messages received out of order can still be processed.%RV 21/8/2019

PIM-DM is able to detect neighbor reboots through the Generation ID present in Hello messages. However, unlike HPIM-DM, it does not change previously stored state relative to the neighbor that suffered the reboot. When a reboot is detected, the router may send an immediate Hello to speed up the neighborhood relationship establishment and may replay the last State Refresh message if the neighbor is downstream.%RV 21/8/2019

\subsection{Initial synchronization of tree information}\label{sec:initsync}%12/2/2019

The initial synchronization is a procedure between two neighboring routers that have just established or reestablished connection, whereby the routers exchange upstream information, to determine for which active trees a neighbor considers itself being UPSTREAM. This is required when a router newly attaches the network and in other situations.%RV 31/1/2020

\subsubsection{Synchronization triggering events}

The synchronization is triggered by the following events: (i) new neighbor detected, (ii) known neighbor rebooted or its SN overflowed, (iii) bidirectional communication with known neighbor temporarily interrupted.%RV 31/1/2020

The first case occurs when a control message sent by an unknown router is received. Note that any control message, and not only Hello messages, can be used to detect a new neighbor. The second case occurs when the neighbor sends a control message with a BootTime higher than the one currently stored for that neighbor. Note that all control messages include the BootTime. The third case occurs if one router, say R1, considers that a neighbor, say R2, is disconnected, but the opposite is not true (e.g. because of temporary loss of Hello messages in only one direction). In this case, R1 will trigger a synchronization as soon as it resumes receiving control messages from R2.%RV 31/1/2020

\subsubsection{Sync messages}

When two neighbors synchronize, they send each other the list of trees for which they consider themselves being UPSTREAM in relation to the neighbor, and the corresponding RPCs. A router considers itself UPSTREAM regarding a tree in relation to a neighbor if (i) it is ACTIVE for that tree and (ii) it is connected to the neighbor through a non-root interface (except if this interface is directly connected to the source).%RV 31/1/2020

The list of trees is exchanged in Sync messages. When the list is large, more than one Sync message may need to be transmitted. Moreover, the transmission of Sync messages must be protected. These features are similar to the initial exchange of link state information that occurs in OSPF \cite{RFC2328}. Thus, we use a procedure similar to the so-called database description process of OSPF. This process will be detailed in section \ref{sec:syncprotocol}.%RV 31/1/2020

\subsubsection{Tree state changes}

The upstream information received during a synchronization process may change the tree state of a router. In particular, if the router was INACTIVE, it becomes UNSURE or ACTIVE if it discovers a new upstream neighbor. It becomes ACTIVE if the new neighbor is connected to its root interface and verifies the feasibility condition (i.e. it is a parent), and becomes UNSURE otherwise.%RV 31/1/2020

\subsubsection{Consistency of the synchronization process}

While a router is synchronizing with a neighbor, upstream or interest messages susceptible of changing the state of one or more trees being communicated to the neighbor may arrive at the router. Thus, measures have to be taken in order to ensure the consistency of the synchronization process.%RV 31/1/2020

\medskip
\noindent \textbf{Taking a snapshot of the tree information} When a synchronization is started with a neighbor, the router takes a snapshot of all trees for which it considers itself being UPSTREAM regarding the neighbor. It is this snapshot that will be completely transmitted to the neighbor in the synchronization process, even if the tree information changes in the meanwhile. For example, a router may initially consider itself as UPSTREAM on a tree and then become NOT UPSTREAM during the synchronization period; in this case, the initial information on the tree will still be communicated to the neighbor.%RV 31/1/2020

\medskip
\noindent \textbf{The SnapshotSN} To distinguish among different synchronization periods, we introduce a new sequence number, called \textit{SnapshotSN}. When a synchronization starts, the SN of the transmitting interface is incremented by one and this value is considered the SnapshotSN. Thus, the SnapshotSN belongs to the SN space of transmitting interfaces. The SnapshotSN along with the BootTime uniquely identify a synchronization period, and are both transmitted in Sync messages. These two SNs allow ignoring outdated Sync messages, which may have been transmitted during previous synchronization periods. As it will be discussed later, Sync messages have to include both the BootTime and SnapshotSN of the sending router and of the neighbor to which it is synchronizing.%RV 31/1/2020

Moreover, when a Sync message is received, the SnapshotSN is assigned to all trees of the transmitter interface listed in the message, as the highest SN received so far, except for trees where the receiver has already tree information with a higher SN. Thus, the SnapshotSN allows ignoring all upstream and interest messages, of all trees, received with an SN lower than SnapshotSN.%RV 31/1/2020

Once a snapshot is taken and a synchronization starts, subsequent upstream or interest messages may arrive at a router, susceptible of changing the upstream state being communicated to a neighbor. Recall that the snapshot is reliably delivered to the neighbor, as taken when the synchronization starts. The upstream state changes that occur during the synchronization period may trigger the transmission of new upstream or interest messages to the neighbor. However, these messages can be sent concurrently with the synchronization process, without correctness concerns. The messages will have an SN higher than the SnapshotSN and, therefore, will be accepted by the neighbor as having fresher information.%RV 31/1/2020

\begin{figure}[t!]
	\centerline{\includegraphics[scale=0.4]{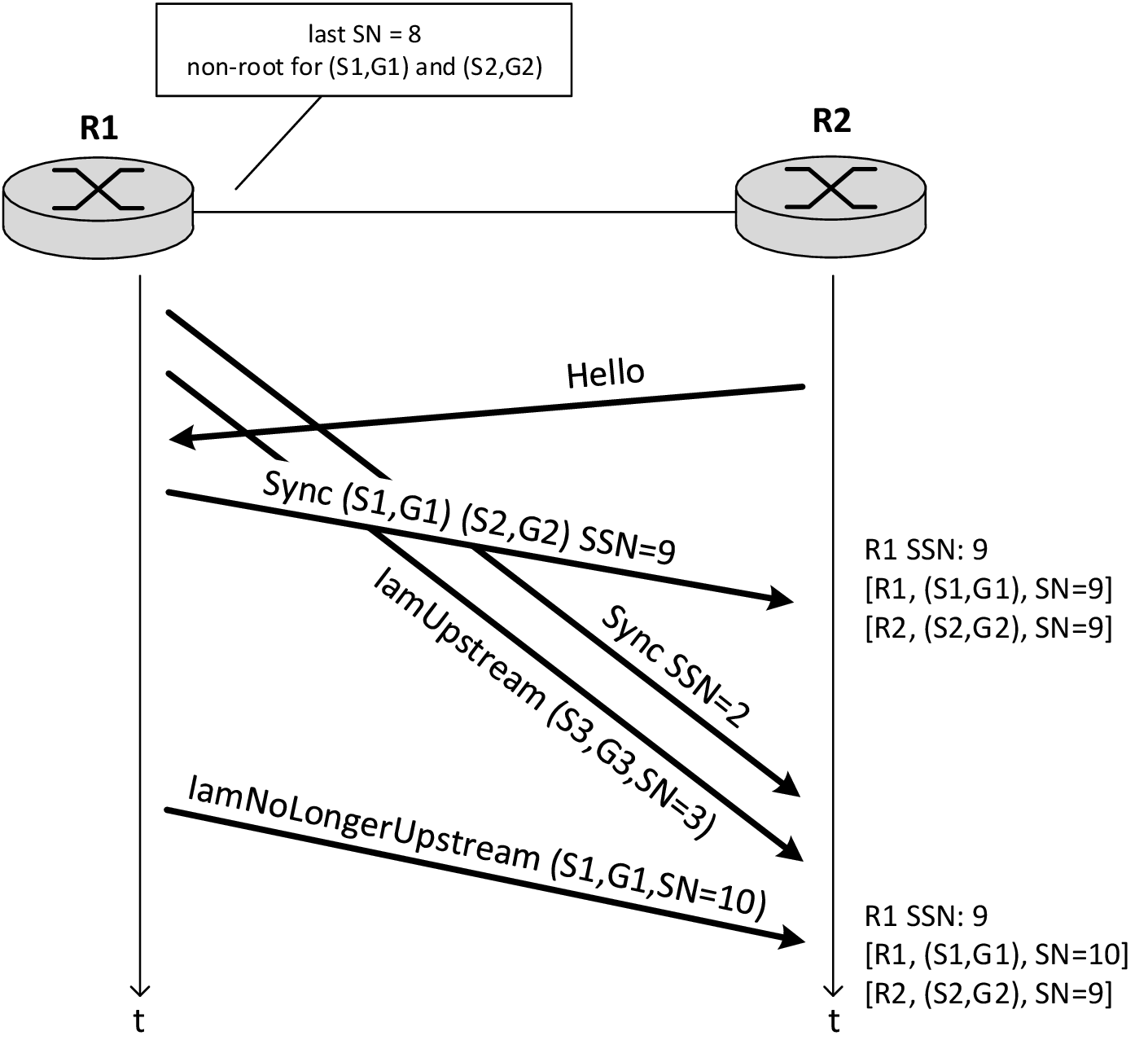}}
	\caption{The tree's snapshot and the SnapshotSN.}
	\label{fig:sync_ssn_motivation}
\end{figure}

\medskip
\noindent \textbf{Example} Figure \ref{fig:sync_ssn_motivation} illustrates the synchronization process. We denote the SnapshotSN by SSN. R1 is initially ACTIVE regarding trees (S1,G1) and (S2,G2). When the Hello message is received by R1, R1 discovers the new neighbor and initiates the synchronization process. At this point, R1 takes a snapshot of the tree information that it will send to R2. Since R1 is ACTIVE regarding trees (S1,G1) and (S2,G2), and its interface with R2 is non-root for both trees, the snapshot includes both trees. Suppose that, when the synchronization starts, the SN of R1's non-root interface is SN=8. In this case, the interface SN is incremented by one and the SSN is given the interface SN value, i.e. SSN=9. Then, R1 transmits to R2 a Sync message with the tree snapshot and SSN=9. Suppose that now R2 receives an old Sync message, transmitted in a previous synchronization period with R1, with SSN=2. This message is rejected and does not trigger a new synchronization since its SSN is lower than the stored SSN. Suppose also that R2 receives an old IamUpstream message advertising an (S3,G3) tree with SN=3. This tree is unknown at R2, but is is rejected, since the SN of the IamUpstream message is lower than the current SSN. Finally, suppose that R1 receives an (S1,G1) IamNoLongerUpstream message from its parent on the tree which makes the router to become UNSURE or INACTIVE regarding the tree, during or after the synchronization period. R1 then sends an IamNoLongerUpstream message to R2 with SN=10. This message is accepted and processed by R2, since its SN is higher than the stored SSN.%RV 31/1/2020

%RV 28/1/2019

\medskip
\noindent \textbf{Including the BootTime and SnapshotSN of the router and of the neighbor in Sync messages} The Sync message transmitted by a router must include the BootTime and SnapshotSN of the sending router, and also of the neighbor to which it is synchronizing (if known when the message is transmitted). A router receiving this message can acknowledge that the neighbor knows its BootTime and SnapshotSN. This avoids accepting outdated Sync messages when a router loses the sequencing information (BootTime and SnapshotSN) relative to previous synchronizations with a neighbor (e.g. because of temporary loss of connection). This mutual acknowledgment allows routers to ascertain that they are exchanging Sync messages belonging to the same synchronization period. We denote the BootTime and SnapshotSN of the sending router by myBT and mySSN, and the BootTime and SnapshotSN of the neighboring router by neiBT and neiSSN.%RV 22/8/2019

\begin{figure}[t!]
	\centerline{\includegraphics[scale=0.8]{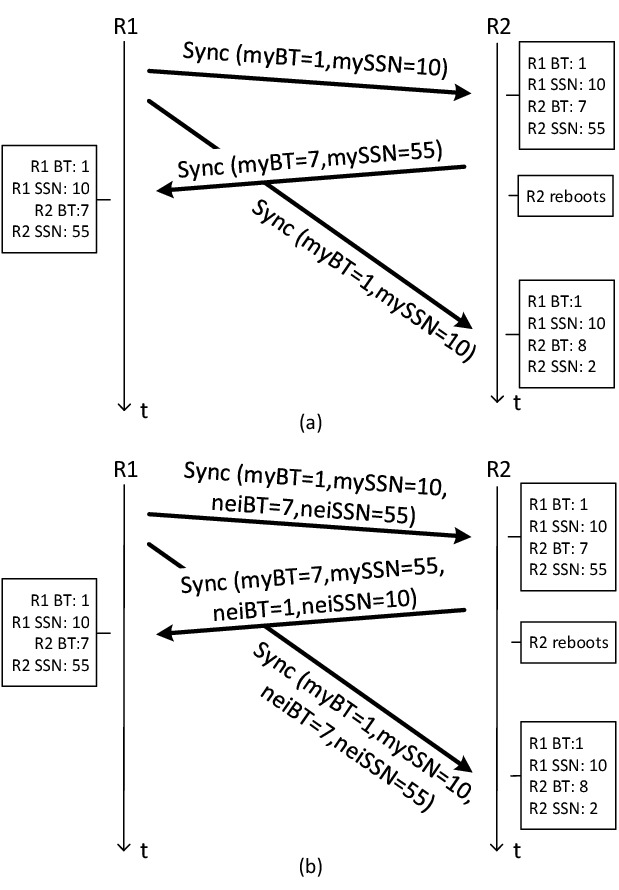}}
	\caption{Including the BootTime and SnapshotSN of the sending router and of its neighbor in Sync messages.}
	\label{fig:sync_include_sequence_numbers}
\end{figure}

Figure \ref{fig:sync_include_sequence_numbers} illustrates the consequences of not including the BootTime and SnapshotSN of the neighboring router in Sync messages. In Figure \ref{fig:sync_include_sequence_numbers}.a R1 sends initially a Sync message to R2 with myBT=1 and mySSN=10, and R2 replies with another Sync message with myBT=7 and mySSN=55. Suppose that R1 needs to retransmit its Sync message and the retransmission is delayed. In the meantime, R2 reboots and receives the retransmission from R1. At this point, R2 can only accept the message, despite being from a previous synchronization and possibly having outdated information. In Figure \ref{fig:sync_include_sequence_numbers}.b the Sync messages include the BootTime and SnapshotSN of the sending router and of its neighbor. The message retransmitted by R1 arrives with myBT=1 and mySSN=10, and with neiBT=7 and neiSSN=55. However, since R2 rebooted, its BootTime and SnapshotSN no longer coincide with the ones received in the Sync message sent by R1: the BootTime is 8 and the SnapshotSN is 2. Thus, the message is rejected.%RV 22/8/2019

%RV 8/10/2018

%RV: Include example with replay attack

\subsubsection{Relationship with interest information}

The synchronization process deals exclusively with broadcast tree information, and thus interest information is not exchanged as part of this process. This is because interest information is not useful for all routers; as discussed in section \ref{sec:interestmaintenance}, it is only relevant for the AW. Thus, the initial synchronization between routers has to be followed by the exchange of interest information. After synchronization with a neighbor, a router may discover trees for which the neighbor is the AW, and this may trigger the unicast of interest messages to the neighbor according to the procedure defined in section \ref{sec:interestmaintenance}.%RV 22/8/2019

The snapshot carried in Sync messages is equivalent to receiving (one or more) IamUpstream messages, since the neighbor is telling that it is UPSTREAM for the trees contained in the snapshot. Thus, when a Sync message is received, the router learns from the snapshot that the neighbor is not interested in receiving multicast data for the trees included in the snapshot.%RV 22/8/2019

The exchange of interest information following a synchronization between two routers, R1 and R2, has the following cases:
\begin{itemize}
	\item If R1 includes a tree in its snapshot and R2 does not, then R2 considers R1 UPSTREAM and R1 considers R2 NOT UPSTREAM for that tree. If R2 determines that R1 is the new AW for the tree, then R2 must send to R1 an interest message.
	\item If R1 and R2 include the same tree in their snapshots, each of them considers the other UPSTREAM and NOT INTERESTED, and no interest message is sent.
	\item If R1 and R2 do not include a tree in their snapshots, then they are both considered NOT UPSTREAM by their neighbors, and none of them sends an interest message because none of them is considered AW.
\end{itemize}%RV 22/8/2019

\subsubsection{Synchronization protocol}\label{sec:syncprotocol}

As referred above the protocol used for the exchange of Sync messages follows closely the database description process of OSPF. Similarly to OSPF, a router may have to transmit several Sync messages due to the size of the tree snapshot, and the message transmissions must be protected. The routers elect a Master while exchanging the first Sync messages and the Master controls the subsequent communications.%RV 3/2/2020

The Master is the router that sends the first Sync message, and the Slave is the other router. If both routers send a Sync message declaring themselves as Master, the one with highest IP address becomes Master. To control this election a Master flag is included in all Sync messages, and is set whenever the transmitting router believes being the Master.%RV 3/2/2020

The transmission of Sync messages is protected by a Stop-and-Wait protocol controlled by the Master. The Sync messages are numbered using a sequence number called SyncSN, set at SyncSN=0 when the exchange of Sync messages is initiated. The Master is responsible for incrementing the SyncSN whenever a new Sync message is transmitted. When the Slave receives a Sync message from the Master, it must reply with another Sync message containing the same SyncSN. The Master retransmits a Sync message whenever the corresponding reply from the Slave is not received within a timeout period. This ensures reliable transmission of the Sync messages sent by the Master and the Slave. The synchronization is aborted after a number of retransmission attempts without any reply.%RV 3/2/2020

The exchange of Sync messages ends when the routers have no more information to send to each other. To control this process, Sync messages include the More flag, which is set whenever the transmitting router has more information to send. Specifically, the More flag is only cleared when the reception of all transmitted snapshot fragments has been acknowledged. The synchronization ends when the Master transmits a Sync message with the More flag cleared, and the Slave replies with a Sync message with the same SyncSN and the More flag also cleared.%RV 3/2/2020

In order to guarantee the reliable transmission of the BootTime and SnapshotSN of both routers in Sync messages, the synchronization process must necessarily terminate with a SyncSN greater than 0. Thus, if the first two Sync messages exchanged by the routers (where SyncSN=0) have the More flag cleared (indicating that the routers have no information to send), a new round of Sync message transmissions is required. By having at least two rounds of Sync messages, we guarantee that each neighbor correctly acknowledges the exchanged sequence numbers.%RV 3/2/2020

\begin{figure}[t!]
	\centerline{\includegraphics[scale=0.7]{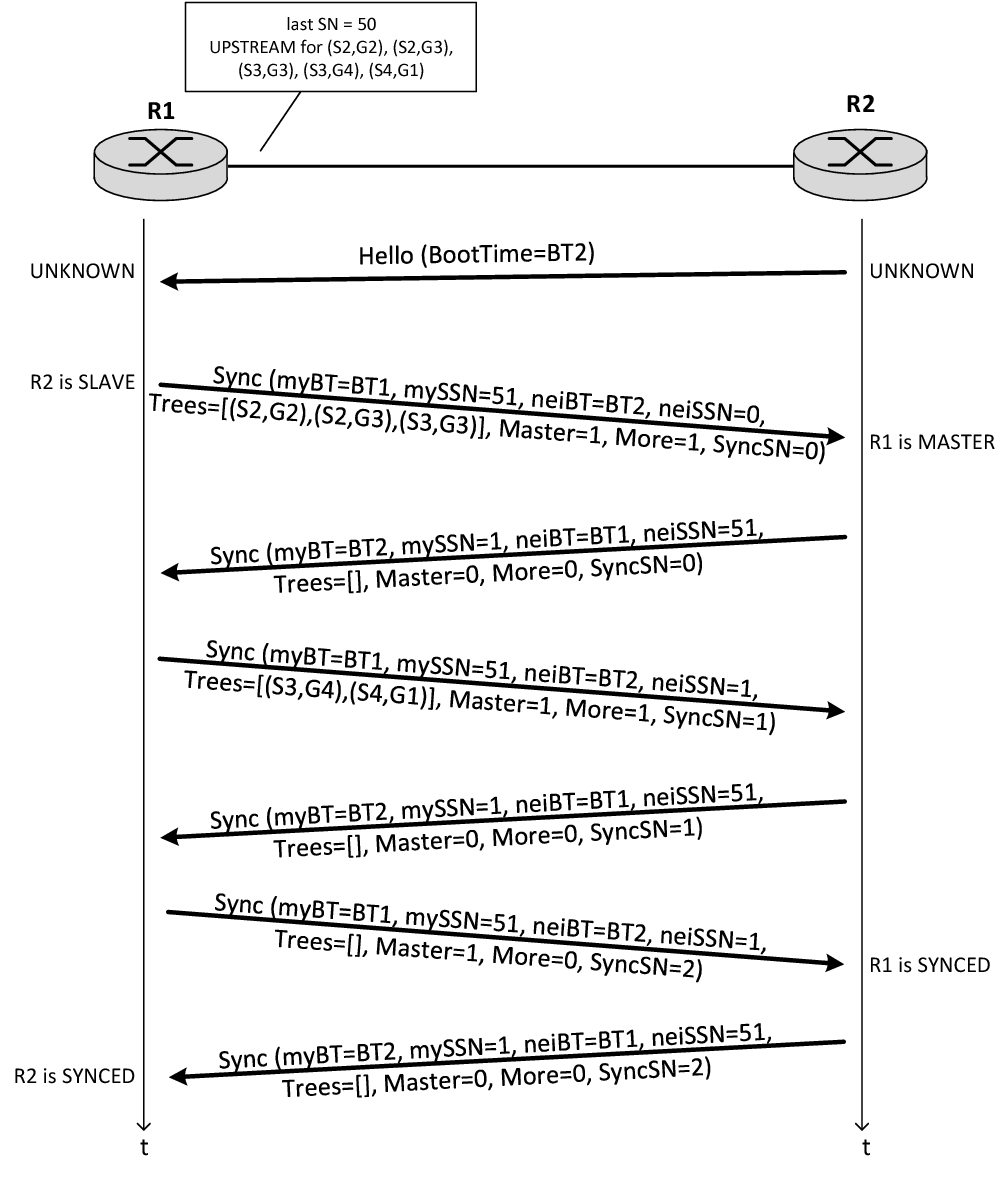}}
	\caption{Synchronization protocol example.}
	\label{fig:synchronization_example}
\end{figure}

\medskip
\noindent \textbf{Example} Figure \ref{fig:synchronization_example} illustrates the synchronization between two routers. In this example, R1 was already connected to the network and R2 is switched on. R1 considers itself UPSTREAM in relation to R2 for five trees. The initial BootTimes of R1 and R2 are denoted by BT1 and BT2, and the initial SNs are 50 for R1 and 0 for R2. Consider that each Sync message can only include information regarding three trees. In a real system the number would be much larger, and dependent on the interface's MTU and on the size of Sync header.%RV 22/8/2019

%RV 28/1/2019

When R2 is switched on, it transmits an Hello message containing its BootTime BT2. When R1 receives this message, it starts the synchronization process with R2. It assumes itself as Master, takes a snapshot of all five trees for which it considers itself UPSTREAM regarding R2, and obtains its SnapshotSN as the current SN plus one, i.e. SnapshotSN=51. It then sends a Sync message to R2 containing information on the first three trees of the snapshot, and myBT=BT1, mySSN=51, and neiBT=BT2; the neiSSN is empty since it is unknown at this point. Since R1 believes being the Master and has additional information, it sets the Master and More flags. Moreover, since this is the first message sent by the Master, SyncSN=0.%RV 22/8/2019

When R2 receives this message, it starts its synchronization process with R1, assumes itself as Slave since R1 already declared itself as Master, and sets its SnapshotSN as SnapshotSN=1. It then replies sending a Sync message with the same SyncSN of the Master, i.e. SyncSN=0, and myBT=BT2, mySSN=1, neiBT=BT1 and neiSSN=51. The Master flag is cleared, since R2 is Slave, and the More flag is also cleared, since R2 has no tree information to send to R1.%RV 22/8/2019

R1 then receives this message and learns the SSN of R2. Since the remaining information is correct, i.e. neiBT and neiSSN coincide with R1's BT and SSN, and myBT coincide with the BT already stored for R2, R1 proceeds with the synchronization process. It sends the next Sync message, with SyncSN incremented by one, i.e. SyncSN=1, and information on the remaining two trees. When R2 receives this message, it replies with a Sync message with the same SyncSN.%RV 22/8/2019

Finally, when R1 receives this message, it sends a Sync message with SyncSN=2 and the More flag cleared, since the transmission of its snapshot has now been completely acknowledged. When R2 receives this message, it terminates the synchronization process with R1 successfully  and sets R1 as UPSTREAM neighbor for the five trees included in the snapshot. It also replies with a Sync message having SyncSN=2. When R1 receives this message it also terminates the synchronization process with R2 successfully.%RV 22/8/2019

\medskip
\noindent \textbf{Neighbor states regarding the synchronization process} A router considers a neighbor to be in one of four states regarding the synchronization process:
\begin{itemize}
	\item UNKNOWN - unknown neighbor;
	\item MASTER - ongoing synchronization process and neighbor is Master;
	\item SLAVE - ongoing synchronization process and neighbor is Slave;
	\item SYNCED - router and neighbor have successfully synchronized.
\end{itemize}%RV 22/8/2018

A neighbor is initially UNKNOWN, until it is discovered through the reception of a control message. Then a synchronization is initiated and the neighbor changes to the MASTER or SLAVE states, depending on which router started the process. When the synchronization finishes successfully the neighbor changes to the SYNCED state. Finally, the neighbor returns to UNKNOWN, irrespective of current state, if it is declared dead.%RV 22/8/2019

\medskip
\noindent \textbf{Triggering the synchronization process in case of SN overflow} When a router suffers SN overflow at one interface, it obtains a higher BootTime and, therefore, needs to resynchronize with its neighbors. In this case, we require that the synchronization is only initiated by the neighbors. A neighbor knows that it must synchronize with a router that suffered SN overflow when it receives a control message from the router, e.g. an Hello message, with a BootTime higher than the current one for that router. This solution allows avoiding to attempt synchronization with neighbors that failed but for which the failure is yet to be detected.%RV 22/8/2019

\medskip
\noindent \textbf{Establishing neighborhood relationships through the synchronization process} In HPIM-DM, a neighborhood relationship between two routers is only established when the routers have synchronized. These two events, i.e. becoming neighbors and becoming synchronized, coincide. Once a neighborhood relationship is established, the neighbors need to start monitoring the liveness of each other and, at this point, they need to know the periodicity of Hello transmissions. Thus, this periodicity must be included in Sync messages. To minimize the overhead, the periodicity is only included in Sync messages with the More flag cleared, which are the last ones to be exchanged during a synchronization process.%RV 22/8/2019

\medskip
\noindent \textbf{Validity of tree information exchanged during synchronization} The tree information is only considered valid when the synchronization process finishes, since only at this point there is the guarantee that both routers are alive and have exchanged the tree information successfully. This avoids accepting tree information as valid sent by (i) a router that fails during the synchronization process and (ii) an attacker that replays Sync messages on behalf of a failed router (replay attacks will be discussed in section \ref{sec:securityissues}).%RV 22/8/2019

\subsubsection{Global versus local information of source liveness}

\begin{figure}[t!]
	\centerline{\includegraphics[scale=0.4]{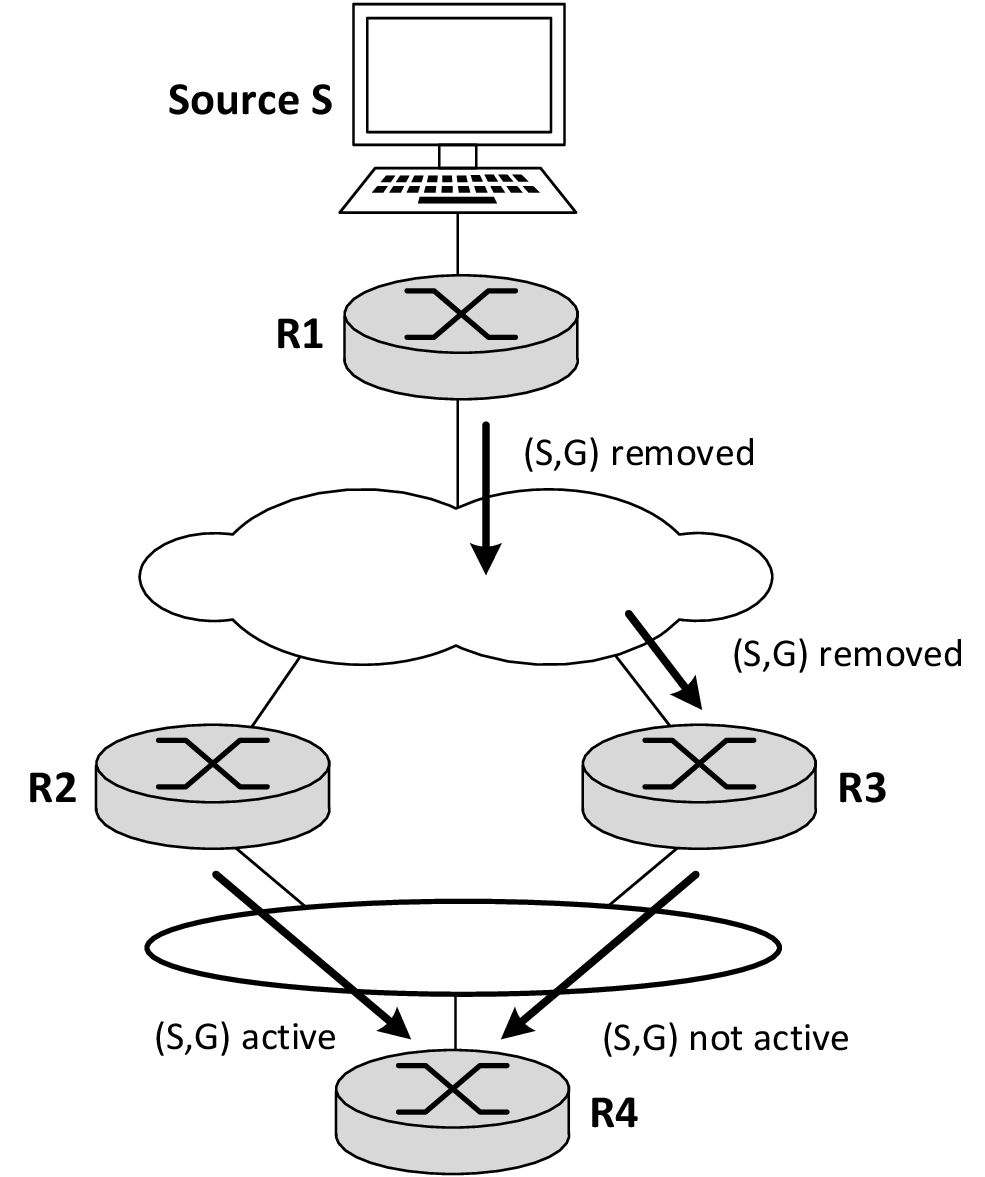}}
	\caption{Global signaling of source liveness.}
	\label{fig:globalsignaling}
\end{figure}

The synchronization process presented above aims at determining which trees are active, e.g. when a router joins the network. However, the indication at a router that a source is active is not obtained through an explicit indication sent directly by the originator. If so, a router could obtain inconsistent information from different neighbors. Take the example of Figure \ref{fig:globalsignaling}. Suppose that R4 joins the network while the (S,G) tree is being removed. R4 may obtain inconsistent information if, when the synchronization is performed, R2 still did not receive this information but R3 already did. At this point, R4 wouldn't know what to decide.%RV 22/8/2019

In HPIM-DM, the information on the liveness of sources is provided indirectly through the concept of upstream router. When a router joins the network it obtains information on whether or not the neighbor is UPSTREAM. A neighbor being UPSTREAM means that there is a chain of UPSTREAM routers between the neighbor and the originator, providing a path from which multicast data can be received and indicating that the source is active. However, being UPSTREAM is a local information (i.e. meaningful among neighbors) and not a global information such as ``source being active" (i.e. meaningful in the complete network).%RV 22/8/2019

Moreover, when information needs to be disseminated globally, i.e. from an originator to all other routers, the sequencing of the corresponding messages must be ensured on a global basis, which requires sequence numbers with global scope and controlled by the originator, and is much more complex to implement. Moreover, even if global message sequencing is ensured, one could be faced with inconsistent information in case of multiple originators.%RV 22/8/2019

This shows that the concept of upstream neighbor and local signaling of source liveness is much more suited to a hard-state protocol than the one one global signaling of source liveness.%RV 22/8/2019

In the example above, using the UPSTREAM concept, following the synchronization R4 considers R2 as UPSTREAM and R3 as NOT UPSTREAM, meaning that it is still possible to receive multicast data from the source, via R2. R2 will later understand that it lost its path to the source and communicate this information to R4 through an IamNoLongerUpstream message.%RV 22/8/2019

\subsubsection{Summary of the initial synchronization of tree information}

A router that joins the network needs to receive information from its neighbors to start receiving multicast data as soon as possible. This process is called initial synchronization and is established between pairs of neighboring routers. Specifically, each router sends to the neighbor the list of trees for which it considers being UPSTREAM regarding that neighbor. This information is transmitted in Sync messages. Interest information is not exchanged during the synchronization process. Besides the discovery of new neighbors, the initial synchronization must also be performed when a known neighbor reboots or its SN overflows, and when the bidirectional communication with a known neighbor is temporarily interrupted.%RV 22/8/2019

To ensure the consistency of the synchronization process, a snapshot of the tree information is taken when the synchronization starts, and this snapshot is completely transmitted to the neighbor even if, in the meantime, the local information changes. Moreover, we introduced a sequence number, called SnapshotSN, to distinguish among synchronization periods. When a synchronization starts, the SN of the transmitting interface is incremented by one and this value is considered the SnapshotSN. Moreover, Sync messages must carry the BootTime and SnapshotSN of both the sending router and the neighbor to which it is synchronizing. This prevents Sync messages transmitted in previous synchronization periods (and delayed) from being accepted during an ongoing synchronization.%RV 22/8/2019

The information transmitted between routers during a synchronization may not fit in a single Sync message, i.e. the snapshot may need to be fragmented. Moreover, the transmission of Sync messages needs to be protected. To address these requirements, we use a protocol similar to the one used in OSPF for the initial synchronization of the link state information. Specifically, one router is elected Master, having the role of controlling the communications process, the Master initiates a Stop-and-Wait protocol to protect and sequence the transmission of Sync messages, Sync messages include a sequence number, called SyncSN, to identify different snapshot fragments, and the Master stops the synchronization when the reception of snapshots was completed and acknowledged.%RV 22/8/2019

\subsubsection{Comparison with PIM-DM}

In PIM-DM there is no synchronization of multicast routing information between neighbors. This causes new routers to be prevented from receiving multicast data until the multicast tree is reconstructed, in case the link they attach to is pruned. PIM-DM includes the possibility of replaying State Refresh messages, which is meant to solve this problem. However, this feature is optional and, even if implemented, it may fail to provide the tree information since State Refresh messages are not protected.%RV 22/8/2019

\subsection{Message transmission reliability}\label{sec:messagereliability}%RV 22/8/2019

Control messages should be transmitted reliably between neighboring routers. For Hello messages, their periodic transmission provide a reliability mechanism. Sync messages are protected using a Stop-and-Wait protocol, as discussed in section \ref{sec:initsync}. Upstream and interest messages use an ACK-protection mechanism, which is the main subject of this section. In this case, a router sets a timer whenever it transmits a message to a neighbor and, upon receiving the message, the neighbor replies with an ACK message containing the SN of the received message. If the router does not receive the ACK within a prespecified timeout period, it retransmits the message, and this procedure is repeated until the ACK is finally received or the neighbor is declared dead. This timer is called \textit{Retransmission Timer}.%RV 3/2/2020

\medskip
\noindent \textbf{Acknowledgment of multicasted messages} Upstream messages are multicasted and interest messages are unicasted. Thus, in order to consider that an upstream message has been correctly received, an ACK must be received from all neighboring routers. ACK messages are unicasted to the router that transmitted the message being acknowledged.%RV 22/8/2019

%RV: What happens if only some but not all ACKs are received?

\medskip
\noindent \textbf{Suppressing ACK transmissions} The reception of a message needs not be acknowledged when the message conveys older information relative to previously received messages. Specifically, an ACK has only to be sent for a message received with an SN higher or equal than the SN stored for the corresponding neighbor and tree.%RV 22/8/2019

\medskip
\noindent \textbf{Suppressing pending ACKs} There are several cases where the requirement for receiving an ACK of a transmitted message can be dispensed with. This happens when a router is waiting for one or more ACKs of a previously transmitted message, and transmits a new message conveying fresher state. It can also happen in cases where the transmitted messages convey double meaning. There are then three cases:
\begin{itemize}
	\item When a new upstream message is transmitted, the pending ACKs relative to older messages (with lower SN) of the same tree can be suppressed.
	\item When a new interest message is unicasted to some neighbor, the pending ACKs relative to older messages (with lower SN) of the same tree transmitted to that neighbor can be suppressed.
	\item When a router starts a synchronization with some neighbor, the pending ACKs relative to control messages previously transmitted by the router to that neighbor (unicasted or multicasted), i.e. with an SN lower than the SnapshotSN, can be suppressed.
\end{itemize}%RV 22/8/2019

%RV: The double meaning case seems not to be included

\medskip
\noindent \textbf{Including the BootTime and SnapshotSN of the sending router and its neighbor in ACK messages} The BootTime and the SnapshotSN of the sending router and its neighbor must be included in ACK messages. This is required due to the possibility of having upstream and interest messages transmitted concurrently with a synchronization process. In fact, it may happen that an ACK acknowledges a message that placed state information in a neighbor before the last synchronization with that neighbor; in this case, the state information may be removed without notice. Including the BootTime and SnapshotSN in ACK messages allows solving this problem. Specifically, an ACK is only accepted at a router if the values of myBT, mySSN, neiBT, and neiSSN match the ones stored at the router. Rejecting non-matching ACKs forces the retransmission of the message, which may now place the correct state information at the neighbor.%RV 22/8/2019

One alternative to solve this problem would be to forbid the transmission of upstream and interest messages concurrently with a synchronization process. This is undesirable, since it would delay unnecessarily the protocol convergence.%RV 22/8/2019

\begin{figure}[t!]
	\centerline{\includegraphics[scale=0.4]{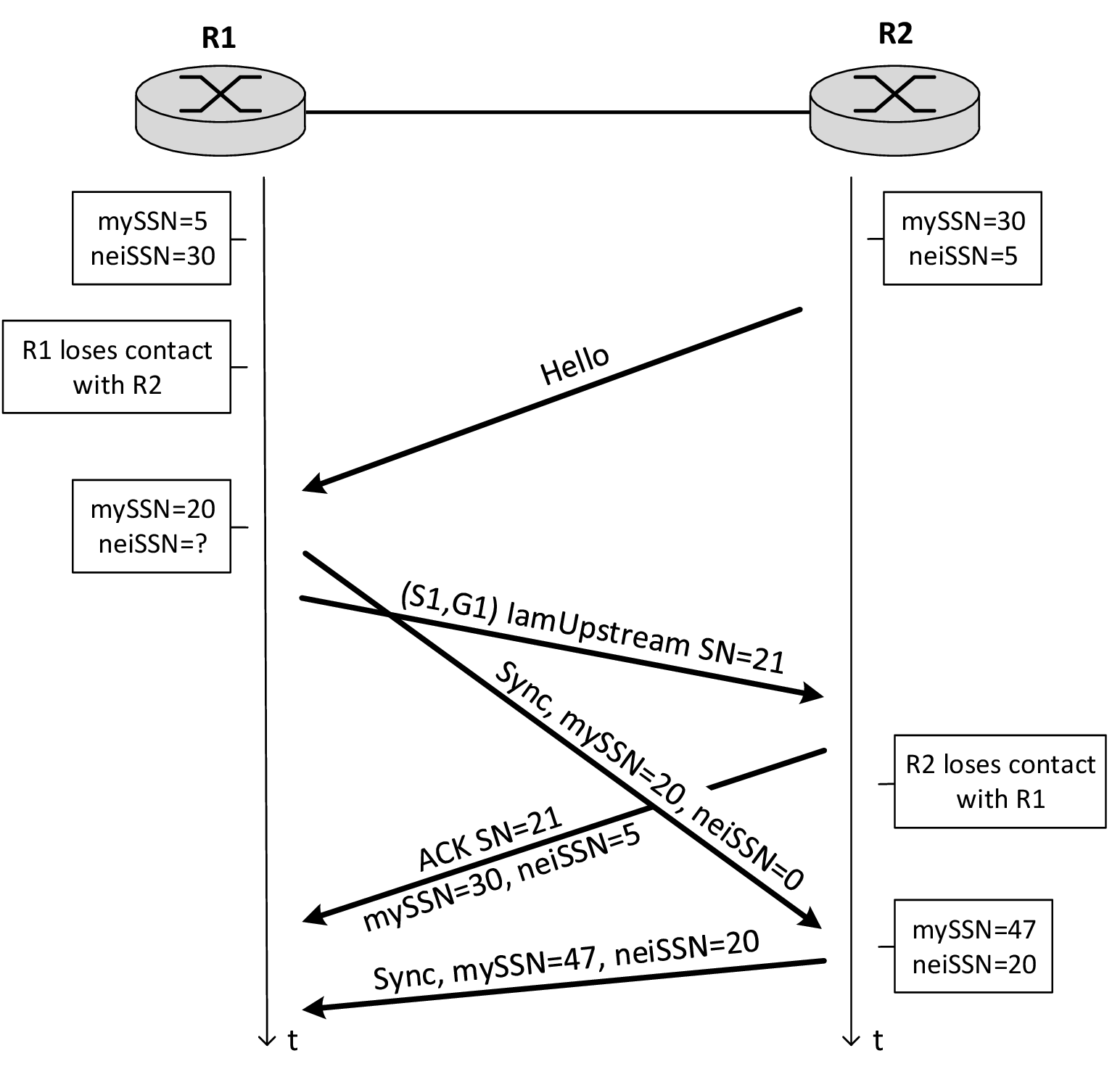}}
	\caption{Motivation for the use of all sequence numbers in ACK messages.}
	\label{fig:motivation_for_including_multiple_sn_ack_example}
\end{figure}

Figure \ref{fig:motivation_for_including_multiple_sn_ack_example} illustrates the need for including the SnapshotSN of both the sending and replying routers in ACK messages; the BootTime needs also to be included for similar reasons. Suppose that R1 and R2 have previously synchronized with each other such that, initially, the SSN of R1 is 5 and the SSN of R2 is 30. Then, R1 loses contact with R2 (e.g. because it ceases to receive Hello messages from R2 for a while) and removes its state and SSN relative to R2. Later R1 resumes receiving Hello messages from R2, and initiates a synchronization with it, by sending a Sync message with its snapshot and SSN=20. Soon after transmitting the Sync message, R1 sends an IamUpstream message relative to tree (S1,G1), which was not included in the previous snapshot. The IamUpstream message has an SN higher than the SSN, i.e. SN=21, but it arrives before the Sync at R2. At this point, R2 stores the information that router R1 is UPSTREAM for tree (S1,G1), and replies with an ACK. Suppose that this ACK does not include the SSN of the sending router and of its neighbor. When the ACK arrives at R1, R1 considers that R2 correctly stored the upstream information, and no further IamUpstream message is transmitted. Suppose now that R2 loses contact with R1 (e.g. because it ceases to receive Hello messages from R1 for a while). In this case, R2 removes its state and SSN relative to R1. If now the Sync message arrives at R2, R2 accepts R1's snapshot, which does not include the (S1,G1) tree, and completes the synchronization process with R1 by sending a Sync message with its snapshot and SSN=47. No further control messages circulate and R1 is wrongly convinced that the upstream information previous sent in its IamUpstream message is stored at R2.%RV 22/8/2019

%RV 9/10/2018

Suppose now that the ACK message carries the SSN of the sending router and of its neighbor. In this case, the ACK includes mySSN=30 and neiSSN=5 and will be rejected since the current SSN of R1 is SSN=20 and not the value indicated by R2, i.e. SSN=5. In this way, R1 detects that the IamUpstream message was received by R2 before the last synchronization between these two routers and, therefore, may have been removed at R2. Since the ACK was rejected, R1 is forced to retransmit the IamUpstream message relative to (S1,G1) which ensures that this information will be correctly stored at R2.%RV 22/8/2019

\subsubsection{Summary of reliable message transmission}

In HPIM-DM, all control messages are transmitted reliably. Hello messages are transmitted periodically (and this acts as a reliability mechanism), Sync messages are protected using a Stop-and-Wait protocol, and upstream and interest messages use an ACK-protection mechanism. ACK messages must include the BootTime and SnapshotSN of the sending router and of its neighbor, to allow the transmission of upstream and interest messages concurrently with a synchronization. The need for receiving an ACK may be suppressed in several circumstances, e.g. when a router is waiting for an ACK from an old message and transmits a new message conveying fresher state.%RV 22/8/2019

\subsubsection{Comparison with PIM-DM}

In PIM-DM, only the Hello and Graft messages are transmitted reliably. The reception of Graft messages is acknowledged using Graft-Ack messages. However, the protocol does not correlate the Graft-Ack with the Graft it is acknowledging (e.g. using sequence numbers). State refresh messages, such as Hello messages, are transmitted periodically but the period is too long to be considered a reliability mechanism. The Assert, Join and Prune messages may be lost and not delivered to the intended neighbors. This causes slow convergence since correct behavior is only established after the tree reconstruction or the reception of State Refresh messages. In HPIM-DM, all controlled messages are reliably transmitted and, therefore, the reaction to message loss is very fast.%RV 22/8/2019

\subsection{Security considerations}\label{sec:securityissues}%RV 22/8/2019

Several mechanisms were introduced in HPIM-DM to mitigate and avoid the effect of security attacks. These mechanisms will be discussed in the next sections.%RV 22/8/2019

\subsubsection{Message integrity and authentication}

In HPIM-DM, the message integrity and authentication are provided by a MAC (Message Authentication Code) included in all control messages. The source and destination IP addresses are included in the MAC construction to prevent attackers from changing these addresses.%RV 31/1/2020

\subsubsection{Replay attacks}

In a replay attack, a control message previously recorded by the attacker is retransmitted, possibly causing malfunction. This problem is similar to the out of order reception of control messages, in the sense that replayed messages carry older state. In this section, we describe the various protocol features that prevent replay attacks.%RV 31/1/2020

\begin{itemize}
	\item Upstream and interest messages include the BootTime and SN of the sending router, and ACK messages include the BootTime and SN of both the sending and destination routers. Any replayed message will have a $<BootTime,SN>$ pair older than the one currently stored and will be rejected.%RV 31/1/2020

	\item Sync messages include the SnapshotSN which, together with the BootTime, uniquely identify a synchronization period. Any replayed Sync message from a previous synchronization period will be rejected. Moreover, replayed upstream, interest, and ACK messages with an SN lower or equal than the current SnapshotSN will also be rejected.%RV 31/1/2020
	
	\item Sync messages include the BootTime and SnapshotSN of both routers that are synchronizing to each other. In this way, replaying a Sync message from a previous synchronization period will not trigger a new synchronization.%RV 31/1/2020
	
	\item Sync messages include the SyncSN to identity different snapshot fragments. Sync messages replayed during a synchronization period will have a SyncSN lower or equal than the current one and will be rejected.%RV 31/1/2020
	
	\item Replaying the initial Hello messages will not break neighborhood relationships, as in other routing protocols, since two routers only become neighbors upon synchronizing to each other. Unlike other routing protocols, Hello messages are not used to set up neighborhood relationships, but only to check their liveness.%RV 31/1/2020
	
	\item The resilience against replay attacks is achieved through the storage of the various sequence numbers, which may consume considerable memory resources. The CheckpointSN allows freeing memory resources without compromising the resilience to replay attacks.%RV 31/1/2020
	
	\item When a router fails, an attacker can extend the neighborhood relationships involving this router by replaying its Hello messages. However, the attacker can not acknowledge any subsequent control messages sent by its neighbors, and will be declared dead.%RV 31/1/2020
\end{itemize}

\medskip
\noindent \textbf{Dependence on external factors} The BootTime can be implemented either using the router's clock or by storing the last used value in the router's non-volatile memory. To avoid attacks, either the network clock synchronization system (e.g. NTP) or the non-volatile memory of routers must be properly secured. It can also happen that the BootTime overflows or the non-volatile memory fails. A simple solution in these cases is to change the secret key used for message integrity and authentication. In this way, replayed messages will be rejected due to authentication failure, even if they have a higher BootTime.%RV 31/1/2020

\subsubsection{Denial-of-Service attacks}

Denial-of-Service (DOS) attacks can be implemented by replaying messages carrying the most recent state. The receivers of those messages must necessarily acknowledge them because the stored sequence numbers are equal to the received ones (i.e. the retransmission may be due to loss of the ACK). However, the CheckpointSN helps mitigating these attacks. The CheckpointSN sent by a transmitter indicates the highest SN such that all messages with a lower or equal SN have already been acknowledged. Thus, routers can refuse replying to messages received from a transmitter with an SN lower or equal than the CheckpointSN stored for that transmitter.%RV 22/8/2019

\subsubsection{Data messages}

Attackers can also perform DOS attacks by injecting multicast data with spoofed source IP addresses on the network. The protection against this type of attack must be performed with the help of other network entities. In this section, we explain how HPIM-DM reacts to these attacks.%RV 22/8/2019

\medskip
\noindent  \textbf{Spoofed IP address belongs to the attacker subnet} If the attacker injects multicast data with spoofed IP addresses belonging to the attacker subnet, a new tree will be formed for each spoofed IP address, which can be catastrophic. This type of attack can be mitigated if routers maintain a list of authorized sources.%RV 22/8/2019

\medskip
\noindent \textbf{Spoofed IP address does not belong to the attacker subnet} If the spoofed IP address does not belong to the attacker subnet, the behavior depends on the tree state of routers regarding the spoofed IP address. If the tree is ACTIVE, i.e. if there is already a source with the same IP address as the spoofed one, the attacker data will be forwarded together with the licit data. If the tree is INACTIVE or UNSURE, the attacker data can be flooded throughout the network or not depending on how the network manager configures the initial downstream interest of interfaces. If the interest is configured as NOT DOWNSTREAM INTERESTED, the attacker data will not be forwarded beyond the subnet where it was injected. However, as discussed in section \ref{sec:dataflooding}, in this case there could be loss of licit data forwarded when the tree is being formed.%RV 22/8/2019

%One way to avoid the continuous flood of illicit data without sacrificing the initial data packets of licit sources, would be to use a lighter version of the protocol in UNSURE and INACTIVE states, similar to PIM-DM, that would force the election of the AW and the expression of interest inn these state, such that uninterested interfaces become pruned.%RV 12/2/2019

\subsection{Control message format}\label{sec:messageformat}%RV 22/8/2019

\begin{figure}[t!]
	\centerline{\includegraphics[scale=0.65]{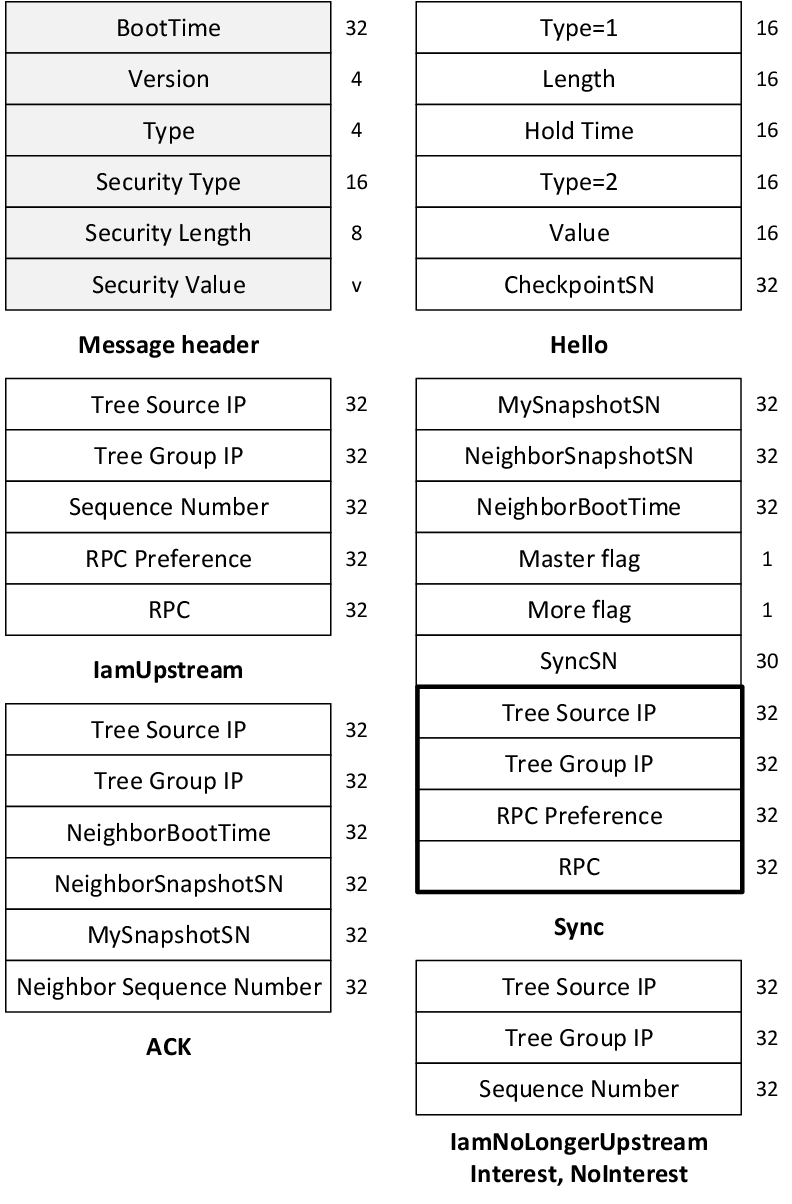}}
	\caption{Format of control messages.}
	\label{fig:message_format}
\end{figure}

This section presents the format of the control messages, which is depicted in Figure \ref{fig:message_format}. The figure indicates the bit length of the various fields, as used in our implementation. Moreover, in our implementation, we have used the IP protocol number of PIM-DM and PIM-SM (103) and the destination multicast address reserved for these protocols (224.0.0.13). In the figure, groups of fields that can repeat several times are enclosed in a bold rectangle.%RV 31/1/2020

%RV 2/2/2020

\medskip
\noindent \textbf{Header} All control messages have a common header that includes the following fields: BootTime, Version, Type, and a TLV record for security purposes. The Type field identifies the message type. The following Type values were assigned: 1 (Hello), 2 (Sync), 3 (IamUpstream), 4 (IamNoLongerUpstream), 5 (Interest), 6 (NoInterest), and 7 (ACK). The Security Type identifies the key and algorithm used in the integrity and authentication check (same as KeyID of OSPF messages \cite{RFC2328}); a value of zero indicates that no message integrity and authentication is provided. The Security Length indicates the number of bytes of the following field and the Security Value includes the MAC, and is calculated by applying the cryptographic hash function to the source and destination IP addresses, and the message contents with zeroed Security Value. We adopted a TLV format for the security information since (i) the inclusion of this type of information is optional and (ii) to accommodate the possibility of different integrity and authentication algorithms, keys and cryptographic hash lengths.%RV 3/2/2020

\medskip
\noindent \textbf{Hello} Hello messages are exclusively formed by TLV records, with Type and Length fields of two octets. We adopted a TLV format since some information to be included in Hello messages is optional. There are currently two TLVs, one for the Hold Time (Type=1) and another for the CheckpointSN (Type=2). The Hold Time is the number of seconds the receiver of the message must wait until the neighbor that sent the message is declared dead. If the Hold Time is set to zero, the information is timed out immediately (to accelerate neighbor removal). The CheckpointSN TLV carries the CheckpointSN. The Hold Time TLV is mandatory but the CheckpointSN is not. Hello messages are multicasted to the link neighbors.%RV 22/8/2019

\medskip
\noindent \textbf{Sync} Sync messages are unicasted to the neighbor synchronizing with the sending router. They include the SnapshotSN of the sending router (MySnapshotSN), the SnapshotSN of the neighboring router (NeighborSnapshotSN), the BootTime of the neighboring router (NeighborBootTime), the Master flag (M), the More Flag (m), and the sequence number that identifies the snapshot fragments (SyncSN). They also include a variable number of groups of four fields, each describing a tree of the snapshot. Each group includes the IP address of the source (Tree Source IP), the IP address of the multicast group (Tree Group IP), the preference value associated with the unicast routing protocol (RPC Preference), and the RPC (RPC).%RV 22/8/2019

\medskip
\noindent \textbf{IamUpstream} IamUpstream messages are multicasted to all link neighbors and include the message SN (Sequence Number) and, for the tree for which the sending router is declaring itself as being UPSTREAM, the IP address of the source (Tree Source IP), the IP address of the multicast group (Tree Group IP), the preference value associated with the unicast routing protocol (RPC Preference), and the RPC (RPC).%RV 22/8/2019

\medskip
\noindent \textbf{IamNoLongerUpstream, Interest, and NoInterest} IamNoLongerUpstream, Interest, and NoInterest messages include the same fields as IamUpstream messages, except the RPC Preference and RPC fields. IamNoLongerUpstream messages are multicasted to all link neighbors, and Interest and NoInterest messages are unicasted to the AW.%RV 22/8/2019

\medskip
\noindent \textbf{ACK} ACK messages acknowledge the reception of IamUpstream, IamNoLongerUpstream, Interest or NoInterest messages, and are unicasted to the transmitter of these messages. The message includes the SN of the message being acknowledged (Neighbor Sequence Number), the identification of the tree being acknowledged (Tree Source IP and the Tree Group IP) fields, and several sequence numbers for mutual acknowledgment, namely the BootTime and SnapshotSN of the neighbor (NeighborBootTime and NeighborSnapshotSN) and the Snapshot SN of the sending router (MySnapshotSN).%RV 22/8/2019

\subsection{Stored state}\label{sec:storedstate}

This section summarizes the state information that needs to be stored by a router. State information needs to be stored per router, per interface, per neighbor, or some combination of these. %This information is summarized in Figure \ref{fig:storedstate}.%RV 30/1/2020

%\begin{figure}[t!]
%	\centerline{\includegraphics[scale=0.9]{Figures/storedstate.eps}}
%	\caption{HPIM-DM timers and stored state.}
%	\label{fig:storedstate}
%\end{figure}%RV 23/8/2019

\medskip
\noindent \textbf{Per router} The per router information are the list of known trees (each tree is defined by the IP address of the source and the IP address of the multicast group) and, for each tree, the router tree state (ACTIVE, UNSURE, INACTIVE), the RPC, and the router interest state (INTERESTED, NOT INTERESTED). Recall that, in a non-originator router, the tree state is a function of the upstream state of neighbors, of their RPCs, and of the router's RPC. In an originator router, the tree state is a function of the upstream state of neighbors and of whether data is being received from the source. Finally, the interest state of a router is a function of the forwarding state of its non-root interfaces.%RV 30/1/2020

\medskip
\noindent \textbf{Per interface} The per interface information is the BootTime, the last transmitted SN (InterfaceSN), the Hello Timer, and the list of interface neighbors. This information needs not be stored in interfaces that have no neighboring router attached (passive interfaces).%RV 30/1/2020

\medskip
\noindent \textbf{Per interface and per tree} The per interface and per tree information is the interface role (root, non-root), the BestUpstreamNeighbor, the list of messages currently pending acknowledgment, and the Retransmission Timer. The three last items need not be stored in root interfaces of originator routers.%RV 30/1/2020

In addition, non-root interfaces store per tree the forwarding state (FORWARDING, PRUNED), the assert state (AW, AL), and the downstream interest state (DOWNSTREAM INTERESTED, NOT DOWNSTREAM INTERESTED). The forwarding state is a function of the assert and downstream interest states. Non-root interfaces also store information, obtained from IGMP or MLD, on which trees have local interested hosts. The downstream state of an interface is a function of the interest of neighbors and hosts. The assert state of an interface is a function of the router tree state and RPC, and of its neighbors' RPC.%RV 30/1/2020

Root interfaces of originators store the Source Active Timer. Finally, both root and non-root interfaces of non-originators store the BestUpstreamNeighbor.%RV 2/2/2020

\medskip
\noindent \textbf{Per interface and per neighbor} The per interface and per neighbor information is the synchronization state (UNKNOWN, MASTER, SLAVE, SYNCED), the neighbor BootTime, the SnapshotSN of the router and the neighbor, the CheckpointSN, the SyncSN, the Sync Retransmission Timer, and the Neighbor Liveness Timer. In addition, the snapshot sent to a neighbor during a synchronization must be stored until the end of the synchronization period. This information needs not be stored in passive interfaces.%RV 30/1/2020

\medskip
\noindent \textbf{Per interface per neighbor and per tree} The per interface per neighbor and per tree information is the upstream state (UPSTREAM, NOT UPSTREAM), the interest state of neighbors (INTERESTED, NOT INTERESTED), and the neighbor SN. This information needs not be stored in root interfaces of originator routers. The interest state only needs to be stored for non-root interfaces of ACTIVE trees. The neighbor SN only needs to be stored if it is higher than the CheckpointSN.%RV 30/1/2020

\medskip
\noindent \textbf{Default configurations} Besides the state information, there is also the need to store the default configurations, namely the timeout values of timers, the initial downstream interest (per interface and tree), and the authentication keys (per interface).%RV 30/1/2020

\subsubsection{Comparison with PIM-DM}

\begin{figure}[t!]
	\centerline{\includegraphics[scale=0.8]{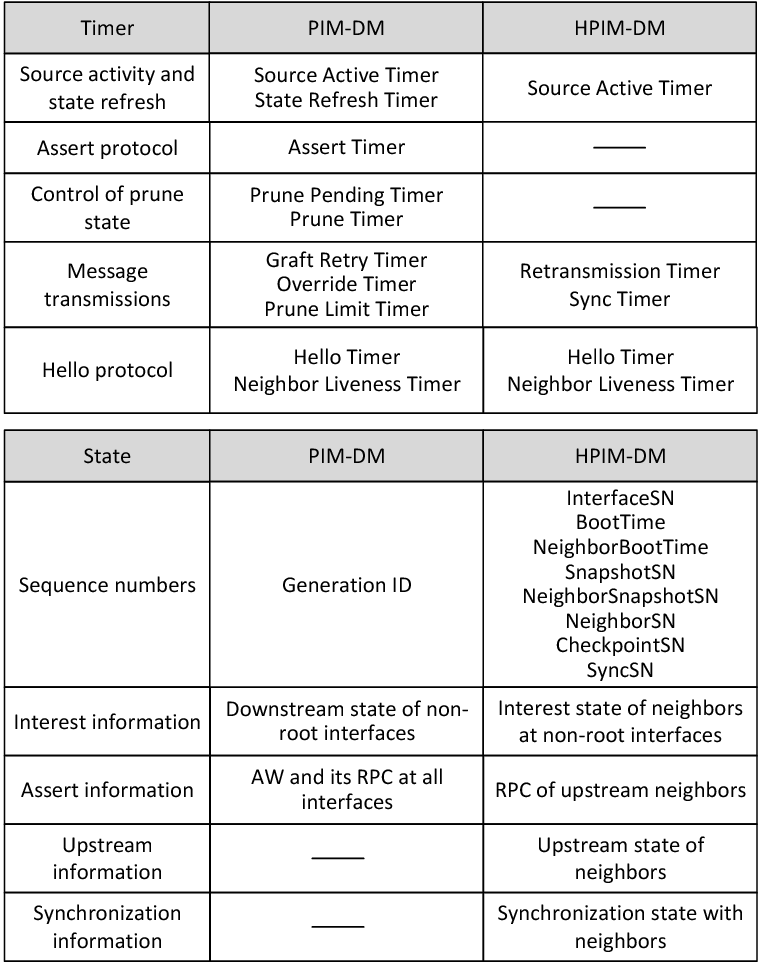}}
	\caption{PIM-DM and HPIM-DM timers and stored state.}
	\label{fig:timersstate}
\end{figure}

In this section, we compare the stored state of PIM-DM and HPIM-DM, including their timers. This is summarized in Figure \ref{fig:timersstate}.%RV 3/2/2020

PIM-DM uses more timers than HPIM-DM, since it is a soft-state protocol. The impact of each timer on the memory resources depends on its running period: some timers run for the whole period of tree activity, others run only when specific events occur, e.g. the transmission of control messages. We describe next the various timers.%RV 3/2/2020

%RV 23/8/2019

%PO: No Assert protocol do HPIM-DM: pode ser referido que pode haver um timer que regula a quantidade de tempo que um AW, que perde o assert, se mantém AW até transitar para AL -> para minimizar perda de pacotes durante troca de AW. -> isto não foi implementado
%PO: No PIM-DM em interest: Downstream Interface state of all interfaces and Upstream Interface state at the root interface
%PO: Nova linha entitulada "Tree state": o PIM-DM armazena o State Refresh state. O HPIM-DM armazena Tree state.

\begin{itemize}
	\item \textbf{Source activity and state refresh -} Both PIM-DM and HPIM-DM use the Source Active Timer to control the source activity. However, PIM-DM uses this timer only when the State Refresh option is enabled. In this case, it also uses the State Refresh Timer to regulate the transmission of State Refresh messages, which refresh the pruned state of the various routers. These timers run per originator router and per tree, whenever a tree is active.%RV 30/1/2020
	 
	\item \textbf{Assert protocol -} PIM-DM uses the Assert Timer to control the assert state, which runs per interface and per tree, whenever a tree is active. HPIM-DM has no equivalent timer.%RV 30/1/2020
	
	\item \textbf{Control of prune state -} PIM-DM uses the Prune Pending Timer to control to transition to PRUNED state when a Prune message is received; it runs for a short period of time (until it expires), per tree, but only at the AW. PIM-DM also uses the Prune Timer to control the prune state due to downstream interest. It is run per tree and non-root interface, while there is no downstream interest. There are no equivalent timers in HPIM-DM.%RV 30/1/2020
	
	\item \textbf{Message transmissions -} PIM-DM uses the Graft Retry Timer to control the reliability of Graft transmissions. With the same goal, HPIM-DM uses the Retransmission Timer for IamUpstream, IamNoLongerUpstream, Interest and NoInterest messages, and the Sync Timer for Sync messages. Moreover, PIM-DM uses the Override Timer to suppress unnecessary Join transmissions, and the Prune Limit Timer to limit the rate of Prune transmissions. These timers are run during a time period following a message transmission, e.g. until an acknowledgment is received.%RV 30/1/2020
	
	\item \textbf{Hello protocol -} Both PIM-DM and HPIM-DM use the Hello Timer to control the transmission of Hello messages, and the Neighbor Liveness Timer to control the neighbor liveness. The first timer runs per interface whenever the multicast routing protocol is active. The second timer runs per active neighbor, while the neighbor remains active.%RV 30/1/2020
	
\end{itemize}

Regarding the stored state, HPIM-DM requires more storage than PIM-DM due to its hard-state nature. We describe next the stored state. We concentrate on the state information for which storage is strictly required. Note that several state variables referred in the previous section can be computed on-the-fly from other state variables and available routing information (e.g. unicast routing tables).%RV 30/1/2020

\begin{itemize}
	\item \textbf{Sequence numbers -} HPIM-DM requires the storage of several types of sequence numbers. It must store the InterfaceSN and the BootTime per interface and, for each neighbor, the BootTime of neighbor (NeighborBootTime), and the SnapshotSN of the router and the neighbor (NeighborSnapshotSN). In addition, the SN of the neighbor (NeighborSN) must be stored for each tree and neighbor, except when this SN is lower than the CheckpointSN. Finally, it must store the SyncSN during the synchronization with a neighbor. PIM-DM does not use sequence numbers. However, it must store the Generation ID per neighbor, which is used to detect neighbor reboots.%RV 30/1/2020
	
	\item \textbf{Interest state} PIM-DM stores the downstream interest associated with each non-root interface. HPIM-DM stores the interest of each neighbor, which is only required at non-root interfaces. Thus, PIM-DM stores interest information per tree and non-root interface, and HPIM-DM stores per tree and neighbor on non-root interfaces.%RV 30/1/2020
	
	\item \textbf{Assert state} HPIM-DM stores the RPC of all upstream neighbors per tree. PIM-DM stores only the RPC of the AW at each interface and tree. However, if there is only one non-root interface at a link, no RPC information is stored, since no Assert messages are transmitted on the link and no AW election is performed; the root interfaces attached to the link determine the AW from the unicast routing table of their routers. The assert state (AW or AL) can be inferred from the RPC information.%RV 30/1/2020
	
	\item \textbf{Upstream state} HPIM-DM stores the upstream state of neighbors on each interface. PIM-DM has no similar information.%RV 30/1/2020
	
	\item \textbf{Synchronization state} HPIM-DM stores the state of the synchronization with each neighbor on each interface. PIM-DM has no similar information.%RV 30/1/2020
	
\end{itemize}

\subsection{Convergence time}

\begin{figure}[t!]
	\centerline{\includegraphics[scale=0.47]{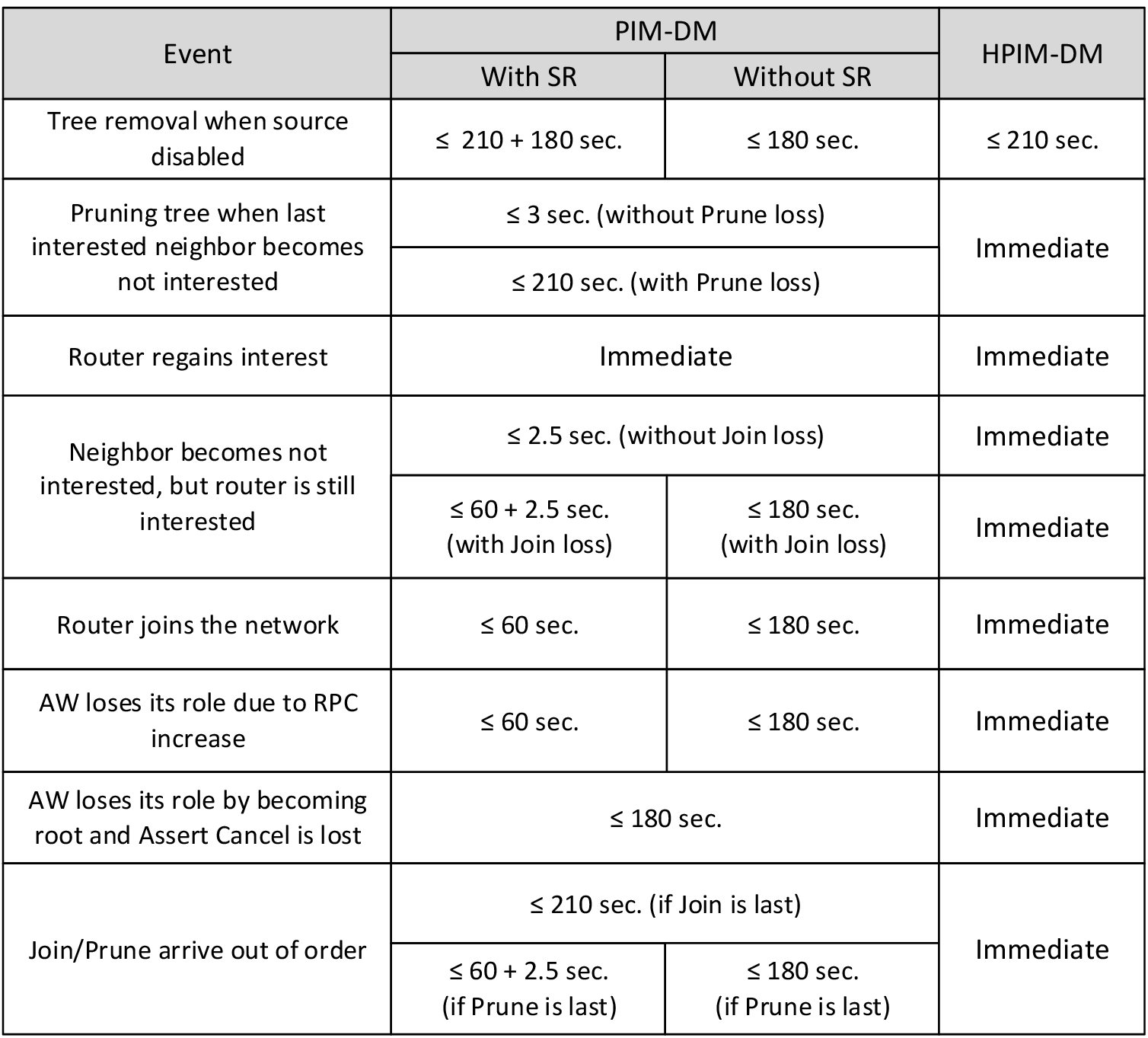}}
	\caption{PIM-DM and HPIM-DM convergence time.}
	\label{fig:convergencetime}
\end{figure}

In this section, we compare the behavior of PIM-DM and HPIM-DM in terms of convergence time. This comparison is summarized in Figure \ref{fig:convergencetime}. We have used the default timer values of both protocols. There are eight events where the convergence time may be significantly different. Except for detecting a tree removal (when the source is disabled), the reaction of HPIM-DM is immediate, in the sense that, following an event, the required actions, such as state transitions and message transmissions, are immediately triggered, without waiting for any subsequent events. In HPIM-DM, even though messages can be lost, the recovery occurs very fast due to the reliability mechanisms of message transmissions. Of course, the convergence time of PIM-DM can be improved by reducing some of its timer values, e.g. reducing the State Refresh Timer and the Prune Timer. However, this will increase the rate of control message transmissions and tree reconstructions. The convergence time of PIM-DM depends on whether State Refresh (SR) is used.%RV 30/1/2020

We describe next the various events:
\begin{itemize}
	\item \textbf{Tree removal when source disabled -} In HPIM-DM, the tree removal is performed upon expiration of the Source Active Timer (210 seconds), which triggers the flooding of IamNoLongerUpstream messages. In PIM-DM without SR, the tree removal is performed upon expiration of the Prune and Assert Timers (180 seconds). In PIM-DM with SR, two timers must expire: only after the Source Active Timer (210 seconds) expires can the originator cease transmitting State Refresh messages, and the tree is removed only when the Prune and Assert Timers (180 seconds) expire.%RV 23/8/2019
	
	\item \textbf{Pruning the tree when last interested neighbor becomes not interested -} In PIM-DM, the pruning action takes up to 3 seconds (Prune Pending Timer) if the Prune sent by the router that became not interested is not lost. Otherwise, if the Prune is lost, a new Prune will be sent when a data packet is received following the expiration of the Prune Limit Timer (210 seconds). In addition, these delays (3 seconds or 210 seconds) accumulate in the upstream links that need to be pruned as a consequence of this event.%RV 23/8/2019
	
	\item \textbf{Router regains interest -} In PIM-DM, when a router regains interest in some tree it sends a Graft message through its root interface. The reaction is immediate, since the transmission of Graft messages is protected.%RV 23/8/2019
	
	\item \textbf{Neighbor becomes not interested, but router is still interested -} In PIM-DM, the router has to reinstate its interest by sending a Join (to override the Prune). If the Join is not lost, its transmission can take up 2.5 seconds (Override Timer) due to the Join suppression mechanism. If the Join is lost, the reaction time depends on whether SR is used. If it is used, the problem is solved after the transmission of the next State Refresh message, which can take up to 60 seconds (State Refresh Timer), and the transmission of the next Join message, which can take up to 2.5 seconds (Override Timer). If SR is not used, the problem is solved when the prune state expires and data packets are reflooded throughout the network, which can take up to 180 seconds (Prune Timer).%RV 23/8/2019
	
	\item \textbf{Router joins the network -} In PIM-DM, when a new router joins the network, if SR is used, it waits up to 60 seconds (State Refresh Timer) to receive information on the active trees, which happens when the next State Refresh message is received. Otherwise, if SR is not used, it waits up to 180 seconds (Prune Timer), which happens when the prune state expires and data packets are reflooded throughout the network.%RV 23/8/2019
	
	\item \textbf{AW loses its role due to RPC increase -} In PIM-DM, if SR is used, the reelection of the new AW takes up to 60 seconds (State Refresh Timer), following the transmission of the next State Refresh message (which carries the new RPC of the current AW, causing routers with better RPC to transmit Assert messages and trigger a reelection). If SR is not used, it takes up to 180 seconds (Assert Timer), until the assert state expires.%RV 23/8/2019
	
	\item \textbf{AW loses its role by becoming root and Assert Cancel message is lost - } In PIM-DM, the correction of this problem can take up to 180 seconds (Assert Timer), until the assert state expires and a new AW election is performed.%RV 23/8/2019
	
	\item \textbf{Join/Prune arrive out of order -} In PIM-DM, if the Join is received after the Prune, meaning that the link should be pruned and it was not, it takes up to 210 seconds (Prune Limit Timer) to correct this problem, whether SR is used or not; this is the time until the next Prune message can be sent. Otherwise, if the Join is received before the Prune, meaning that the link becomes pruned and should not be, if SR is not used, it takes up to 180 seconds (Prune Timer) to correct this problem, until the prune state expires and data packets are reflooded throughout the network. If SR is used, the problem is solved when the next State Refresh message is received (carrying a flag indicating that the AW is pruned), which can take up to 60 seconds (State Refresh Time), and a Join is transmitted, which can take up to 2.5 seconds (Override Timer) due to the mechanism of suppressing Join transmissions.%RV 23/8/2019
	
\end{itemize}

%RV 4/3/2019

%% file: cc_correctness_v4.tex
\section{HPIM-DM correctness}\label{sec:correctness}

This section discusses the correctness of HPIM-DM using logical reasoning and, in some cases, model checking \cite{model_checking_primer}. The extensive tests carried out in the HPIM-DM implementation (see section \ref{sec:tests}) also contribute to the confidence on the protocol correctness.%RV 3/2/2020

The discussion on the correctness of the protocol can be subdivided in several parts, according to the protocol mechanisms: (i) initial synchronization, (ii) broadcast tree maintenance, (iii) interest maintenance, (iv) message sequencing, and (v) message transmission reliability. Note that there are dependencies among the correctness of the various parts: interest maintenance is only correct if the broadcast tree maintenance is correct; broadcast tree maintenance is only correct if the synchronization is correct; all parts are dependent on the correctness of message sequencing and reliability.%RV 24/1/2020

Model checking uses the SPIN tool \cite{SPIN, spin_book}, where the network models are written in Promela language \cite{PROMELA}. In this setting, the correctness properties of the protocol are exhaustively verified using Linear Temporal Logic (LTL). This methodology can be computationally expensive. Therefore, we use it only for addressing the most sensitive parts of the protocol and when computational resources allow it.%RV 24/1/2020

The following sections address the correction of the various protocol parts: section \ref{protocol_correctness_synchronization} covers the initial synchronization, section \ref{protocol_correctness_creation_and_removal_of_broadcast_tree} the broadcast tree maintenance, section \ref{protocol_correctness_interenst_maintenance} the interest maintenance, section \ref{protocol_correctness_message_sequencing} the message sequencing, and section \ref{protocol_correctness_message_reliability} the message transmission reliability.%RV 24/1/2020

The complete description of the Promela code used in model checking is presented in \cite{hpim_correctness}.%RV 24/1/2020

\subsection{Initial synchronization}\label{protocol_correctness_synchronization}%RV 14/2/2019

The initial synchronization of the tree information was described in section \ref{sec:initsync}. We argue about its correctness using both logical reasoning and model checking.%RV 24/1/2020

\subsubsection{Logical reasoning}

The synchronization process of HPIM-DM is based on the database description process of OSPF \cite{RFC2328}, a synchronization mechanism that is well accepted and has been in operation for many years. Our synchronization process has only two differences regarding OSPF.%RV 24/1/2020

The first difference is that, in our case, the Master is the router that initiates the synchronization process; if two routers start simultaneously and both declare themselves as Master, the interface IP address is used as a tiebreaker. In OSPF, the Master is always elected based on the router identifier. Our option does not raise any correctness concerns and speeds up the synchronization process.%RV 24/1/2020

The second difference is that the Sync messages carry additional sequence numbers in relation to OSPF, namely the BootTime and SnapshotSN of the sending router and of the neighbor to which it is synchronizing. Again, this feature does not raise any correctness concerns and offers protection against outdated Sync messages and replay attacks.%RV 24/1/2020

\subsubsection{Model checking}

We modeled the synchronization of two routers when (i) the routers exchange a random number of Sync messages and (ii) failures occur during the synchronization process. The first feature aims at testing the fragmentation process. Regarding the failures, two types were tested: (i) neighbor reboot and (ii) broken bidirectional relationship (i.e. router falsely suspects that neighbor fails).%RV 24/1/2020

Specifically, in our model, two routers connect through Promela channels where they exchange messages including the message type, the sending router identifier, and the following sequence numbers: MyBootTime, NeighborBootTime, MySnapshotSN, NeighborSnapshotSN, and SyncSN. There are only two message types: Sync and Hello. Hello messages are not sent periodically: they are only sent initially, to test the election of the Master, and after a reboot, to force resynchronization. To test the Master election a Hello message is transmitted initially by either router or by both routers simultaneously. To test the fragmentation process, the number of Sync messages is made random, between 0 and 2. The failures are modeled using a Promela channel, different from the one used to exchange Sync and Hello messages, that sends a notification failure to either router randomly. A reboot is modeled by increasing the BootTime, removing all state relative to the neighbor, and sending a Hello message with the new BootTime. A broken bidirectional relationship is modeled by removing all state relative to the neighbor and receiving a Hello from the neighbor. The Promela model monitors the Sync state of each router (i.e. UNKNOWN, MASTER, SLAVE or SYNCED), and checks the following correctness property using LTL: eventually, both routers finish the synchronization process considering its neighbor in SYNCED state and having a consistent view regarding the sequence numbers.%RV 24/1/2020

All performed tests indicated correct behavior.%RV 24/1/2020

\subsection{Broadcast tree maintenance}\label{protocol_correctness_creation_and_removal_of_broadcast_tree}

The broadcast tree maintenance was described in section \ref{sec:treemaintenance}. We argue about its correctness using both logical reasoning and model checking.%RV 24/1/2020

\subsubsection{Logical reasoning}

To maintain the broadcast tree, routers store the tree state (INACTIVE, UNSURE, ACTIVE), the upstream state of each neighbor (UPSTREAM, NOT UPSTREAM), the interface role (root, non-root), and the assert state of non-root interfaces (AW, AL). The interface role is set by the unicast routing protocol (given the source address) which we assume to be correct. The initial tree and upstream states are INACTIVE and NOT UPSTREAM; the initial assert state is AW.%RV 3/2/2020

The correctness of the broadcast maintenance process is ensured if:
\begin{itemize}
	\item Under stable conditions, a reverse shortest path tree routed at each originator and spanning all connected routers is correctly formed.%RV 3/2/2020
	\item Following a perturbation, routers always converge to a correct state. When a perturbation occurs, a router can either become disconnected from the tree (in INACTIVE state) or connected to the tree (in ACTIVE state), must elect the correct AW at each interface, and must store the correct upstream information relative to each neighbor.%RV 3/2/2020
\end{itemize}

\medskip
\noindent \textbf{Stable conditions} Under stable conditions, the correctness of the local state information (at each router) implies the correctness of the broadcast tree. Under stable conditions a router is either ACTIVE or INACTIVE regarding a tree; the UNSURE state is a transient state. If the router is INACTIVE, it is disconnected from the tree, either because a path to the source's subnet is no longer available or the source was switched off. If it is ACTIVE, this means that there is a chain of UPSTREAM routers up to the source from which multicast data can be received, and this chain corresponds to the shortest path towards the source.%RV 3/2/2020

The tree state is set according to the upstream state. Thus, the correctness of the second implies the correctness of the first. A router is ACTIVE if it has an UPSTREAM neighbor connected to its root interface that verifies the feasibility condition. The parent of a router is the neighbor that verifies the above conditions and offers the lowest RPC, i.e. the AW seen by the root interface. A router is UNSURE if it has no parent but has at least one UPSTREAM neighbor, and is INACTIVE if it has no UPSTREAM neighbor. If a neighbor fails, its upstream state is removed. If a new neighbor appears, the synchronization process sets the correct upstream state.%RV 3/2/2020

\medskip
\noindent \textbf{Role of upstream messages} The upstream state is set by control messages received from the neighbors. A router sends an IamUpstream message to a neighbor when it can serve as an upstream router for that neighbor, and sends an IamNoLongerUpstream message when it can not. Upstream messages are reliably transmitted and are processed according to its transmission order. Thus, a router having a correct tree state offers a correct upstream state to its downstream neighbors.%RV 3/2/2020

Specifically, (i) if a router becomes ACTIVE, it sends IamUpstream messages through all its non-root interfaces; (ii) if a router ceases being ACTIVE, it sends IamNoLongerUpstream messages through all its non-root interfaces; (iii) if the interface role changes (root versus non-root) and the router remains/becomes ACTIVE, it sends an IamNoLongerUpstream message through the new root interface and IamUpstream messages through the new non-root interfaces. In this way, neighbors connected to a link via non-root interfaces are considered UPSTREAM if they are ACTIVE, and neighbors connected via root interfaces are considered NOT UPSTREAM irrespective of its tree state.%RV 3/2/2020

\medskip
\noindent \textbf{Source becoming active} Starting from an originator, which is placed in ACTIVE state when the source starts sending multicast data, a chain reaction is started where (i) each ACTIVE router informs its downstream routers (if any) that it can act as an UPSTREAM neighbor, using IamUpstream messages, and (ii) routers become ACTIVE when they receive IamUpstream messages in their root interfaces respecting the feasibility condition. In this way, all connected routers (routers that have at least one path to the source) become aware of the source liveness.%RV 3/2/2020

\medskip
\noindent \textbf{Assert state} Since IamUpstream messages carry the RPC and this information is stored at routers, all routers determine the correct assert state associated with their interfaces.%RV 3/2/2020

\medskip
\noindent \textbf{Transient nature of UNSURE state} The UNSURE state is necessarily transient since a router is either connected or disconnected from the source. In the latter case, the router will necessarily change to the INACTIVE state since it will end up losing all its UPSTREAM neighbors. In the former case, the router will necessarily change to the ACTIVE state since the unicast routing protocol will necessarily set the correct interface role at all routers in the shortest path to the source, and these routers will inform their downstream neighbors that they are UPSTREAM regarding the source.%RV 3/2/2020

\medskip
\noindent \textbf{Perturbation events and corresponding actions} To demonstrate that the protocol converges in case of a perturbation, one has to ensure that (i) all perturbation events have been identified and (ii) the correct actions are taken for each identified event. The events and the corresponding actions are listed in section \ref{sec:btmaintevents} and we claim that we have been exhaustive and correct.%RV 3/2/2020

\medskip
\noindent \textbf{Reliability and ordering of message transmissions} The guarantees regarding the reliability and ordering of message transmissions play a crucial role in the correctness of the broadcast tree maintenance. In particular, upstream messages are always received and processed according to their transmission order. This ensures that control messages always place the correct upstream state at the receiving routers.%RV 3/2/2020

\medskip
\noindent \textbf{Synchronization mechanism} The synchronization mechanism ensures that a router joining the network gets all state information needed to connect to the tree. Moreover, the synchronization mechanism has been devised such that fresher state information concurrent with a synchronization period is never lost.%RV 3/2/2020

\medskip
\noindent \textbf{Cold start} In a cold start situation, i.e. all routers and sources started at the same time, routers start synchronizing with each other. In this case, they may not obtain state information on all active trees as part of the synchronization with a neighbor. However, if not obtained as part of this process, they will certainly obtain it later or concurrently with the synchronization, through the reception of IamUpstream messages. In fact, if a router does not report, in a Sync message, a tree that has recently become active and for which the router is UPSTREAM, this is because the router is still not ACTIVE regarding that tree, and therefore is yet to receive an IamUpstream message from its parent on the tree. Once it receives the message, the router sends IamUpstream messages to all its downstream neighbors reporting that the tree is now ACTIVE, and these messages are received and processed by the neighbors even if received concurrently with a synchronization.%RV 24/1/2020

\subsubsection{Model checking}

\begin{figure}[t!]
	\centerline{\includegraphics[scale=0.35]{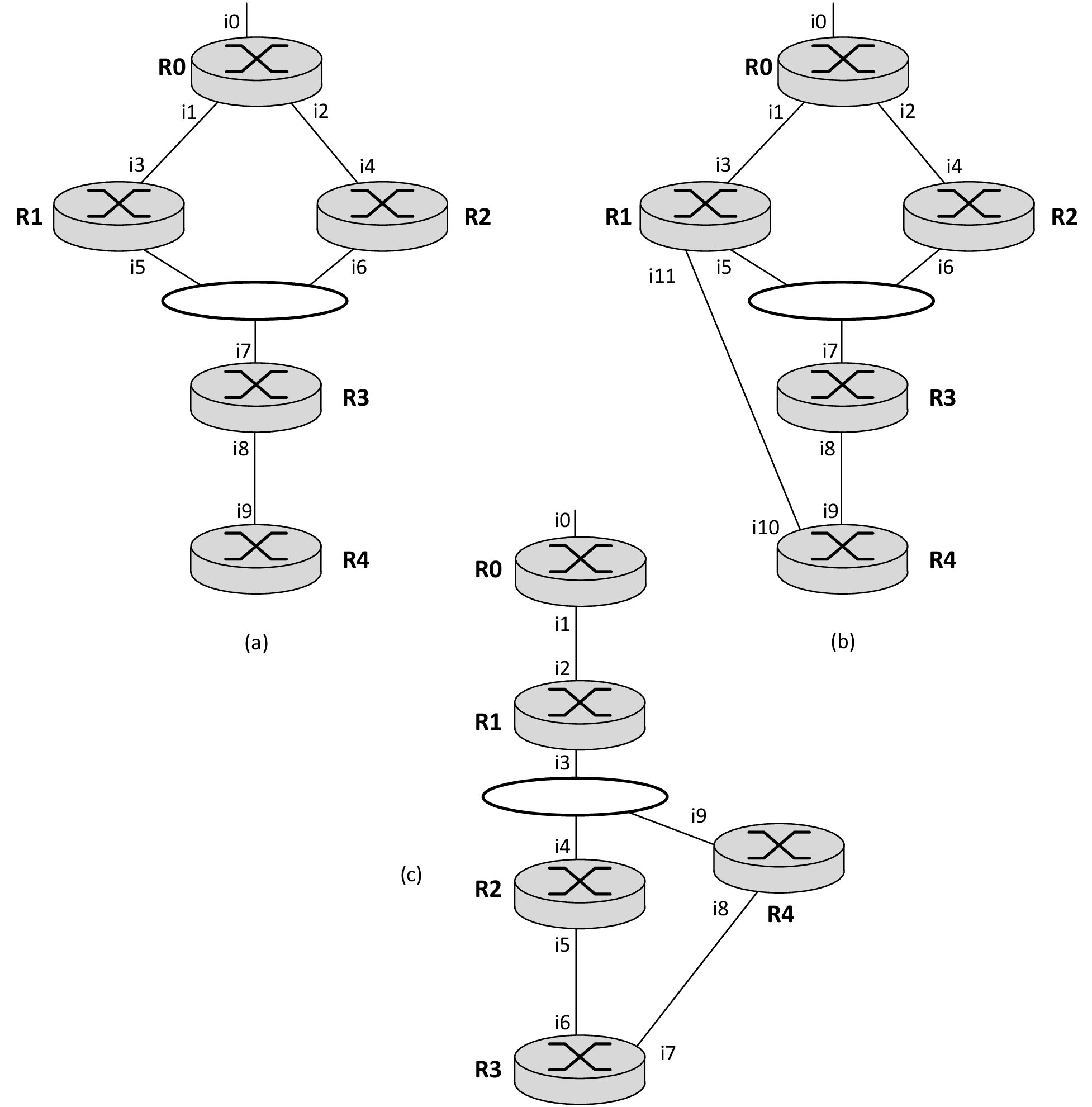}}
	\caption{Network topologies used in model checking; (a) topology 1, (b) topology 2, and (c) topology 3.}
	\label{fig:promela_tests}
\end{figure}

We modeled in Promela the formation of the broadcast tree concurrent with router failures and interface role changes, for the three network topologies of Figure \ref{fig:promela_tests}.%RV 24/1/2020

%RV 15/2/2019

%RV: Do we need to have interface i0 in the model?
%PO: Estive a ver o código e à primeira vista parece-me que não. No entanto se não apresentarmos uma interface root no originator parece um pouco contraditório a explicação do teste porque isto na realidade corresponderia a ter um originator sem interface root (que na realidade nunca iria transitar para um estado ACTIVO).
%RV: Yes, but this is just a model. We don't need i0 for a credible model.

The router model includes the tree state (INACTIVE, UNSURE, ACTIVE), the interface role (root, non-root), the RPC, and the upstream state of neighbors (UPSTREAM, NOT UPSTREAM). Routers are connected through Promela channels to exchange IamUpstream and IamNoLongerUpstream messages. The model does not include the possibility of losing messages and receiving messages out of order. The AW election was also not modeled. A router failure is modeled by removing the upstream state relative to the failed router on all its neighbors, and triggering the appropriate actions (state changes and upstream message transmissions). An interface role change is simply modeled by modifying two interface role state variables at a router (one changes from root to non-root and the other from non-root to root) and triggering the appropriate actions (state changes and upstream message transmissions). The RPC of each router corresponds to the minimum-hop count towards the source. The feasibility condition was checked in order to determine the tree state of a router. The originator is router R0 in all topologies. All tests start with the routers in the INACTIVE state, and by manually placing the originator in the ACTIVE state. This triggers a sequence of events leading to the formation of the tree. The correctness properties that were checked using LTL were defined in terms of the tree states of routers.%RV 24/1/2020

Topology 1 was designed such that (i) there is more than one path to the originator and (ii) there is at least one router failure that separates the network in two parts. We performed five tests in this topology: (1) tree formation without router failures, (2) tree formation concurrent with failure of router R1, (3) tree formation concurrent with failure of router R2, (4) tree formation concurrent with failure of router R3, and (5) tree formation concurrent with failure of router R0. Tests 1, 2, and 3 checked whether all routers except the failed one reached the ACTIVE state. Test 4 checked whether routers R0, R1, and R2 reached the ACTIVE state while R4 reached the INACTIVE state. Finally, test 5 checked if all routers except router R0 reached the INACTIVE state.%RV 24/1/2020

Topology 2 is equal to topology 1 except for the link introduced between R1 and R4. This aims at testing the reaction to a router failure involving a change in the root interface of one router. Topology 2 was used for one test: (6) tree formation concurrent with failure of router R3 and interface role change at router R4. Specifically, we changed the root interface from i9 to i10, which simulates the reaction to the failure of router R3. This test checked whether all routers except the failed one reached the ACTIVE state.%RV 24/1/2020

Topology 3 was used to verify the formation and removal of a tree in the presence of network loops. The root interfaces were selected such that a loop is formed between routers R2, R3, and R4. Specifically, the root interfaces were i0, i2, i4, i6, and i8. The RPCs were selected consistently with the root interfaces. Specifically, we selected R0-RPC=0, R1-RPC=20, R2-RPC=30, R3-RPC=40, and R4-RPC=50. We performed two tests: (7) tree formation in the presence of network loops, and (8) tree removal in the presence of network loops. The latter test was performed by forcing a failure of the originator, similar to the one of test (5). In test 7 we checked if all routers reached the ACTIVE state. In test 8 we checked if all routers reached the INACTIVE state.%RV 24/1/2020

All tests performed in the three topologies indicated correct behavior.%RV 24/1/2020

\subsection{Interest maintenance}\label{protocol_correctness_interenst_maintenance}

The interest maintenance was described in section \ref{sec:interestmaintenance}. Its correctness is only discussed using logical reasoning. We assume that the state information related to the broadcast maintenance tree is correct, namely the upstream state of neighbors and the link AWs.%RV 24/1/2020

The interest information is only stored in ACTIVE state and needs only to be correct at the AW. Moreover, only the NOT UPSTREAM interfaces need to transmit interest information to the AW. UPSTREAM interfaces are not interested in receiving multicast data by definition. Since all interfaces store the upstream state of their neighbors in the ACTIVE and UNSURE states, the AW knows which neighbors are UPSTREAM and NOT UPSTREAM, and infers that UPSTREAM neighbors are not interested.%RV 24/1/2020

Interest messages are unicasted to the AW, to save in the amount of acknowledgments. However, during a transient period when the AW is changing, an interface that sends an interest message may not have the correct perception on who the AW is. This raises correctness concerns. To address this problem we require that all non-root interfaces, either AL or AW, store the latest interest information received from their neighbors, when ACTIVE.%RV 24/1/2020

When an interest message is transmitted, there are four cases regarding the message receiver:%RV 24/1/2020
\begin{enumerate}
	\item \textbf{The message is transmitted to the correct AW, which believes it is AW and is ACTIVE -} In this case the interest state is correctly stored at the AW.%RV 24/1/2020
	\item \textbf{The message is transmitted to the correct AW, which still believes it is AL and is ACTIVE -} In this case, the interest state is stored but, since the receiving interface still considers itself AL, this state will not influence forwarding decisions. When the interface later understands it is the AW, the correct interest state is already available.%RV 24/1/2020
	\item \textbf{The message is transmitted to the wrong AW -} This case occurs when the receiver fails or makes a transition from AW to AL, but the transmitter did not perceive this before sending the interest message. In this case, the transmitter will eventually change its view on the AW (if any) and send it the interest message. Thus, the correct interest state is placed at the AW.%RV 24/1/2020
	\item \textbf{The message is transmitted to the correct AW which lost the interest information -} This case happens when the AW changes to a non ACTIVE state and then returns to ACTIVE state without other routers noticing it. When the AW changes to a non ACTIVE state it must send an IamNoLongerUpstream message, which would provide information that the router is no longer AW. However, this message may not be received, if it is initially lost and, in the meantime, the router changes to ACTIVE. In this case, the router sends an IamUpstream message and cancels the transmission of the previous IamNoLongerUpstream. To ensure that the correct interest state gets placed at the AW in this case, we force the transmission of interest messages when an IamUpstream message is received from the current AW.%RV 24/1/2020
\end{enumerate}

\begin{figure}[t!]
	\centerline{\includegraphics[scale=0.7]{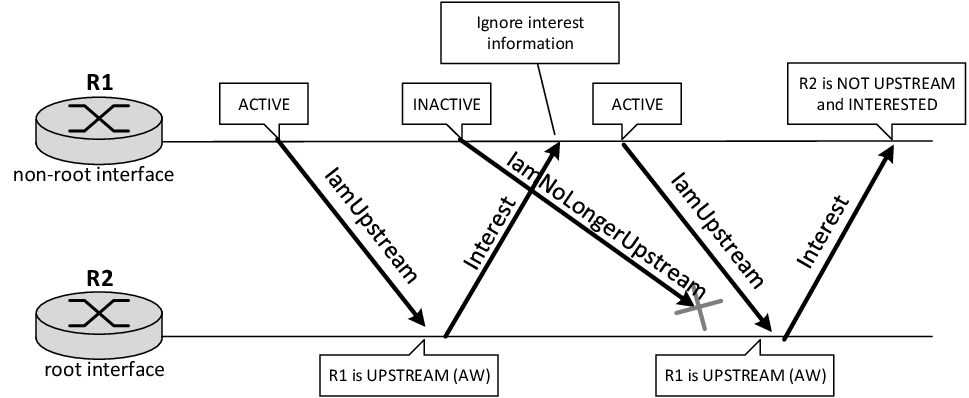}}
	\caption{Loss of interest information by the AW.}
	\label{fig:interest_correction_example3}
\end{figure}

We illustrate the last case through the example of Figure \ref{fig:interest_correction_example3}. Router R1 becomes ACTIVE and sends an IamUpstream message. When R2 receives this message, it elects R1 as the link AW and sends it an interest message. Then, R1 becomes INACTIVE (e.g. because its root interface was temporarily switched off) and sends an IamNoLongerUpstream message. When R1 receives the interest message, it is INACTIVE and, therefore, ignores the interest information sent by R2. Suppose that the first IamNoLongerUpstream message is lost and, in the meantime, R1 becomes ACTIVE again. In this case, R1 cancels the transmission of the previous IamNoLongerUpstream and sends an IamUpstream. R2 kept believing that R1 was AW, but R1 did not receive its interest information. To correct this problem, R2 sends its interest information when it receives the IamUpstream message sent by R1 (which R2 always believed being the AW).%RV 24/1/2020

%RV 14/2/2019

\subsection{Message sequencing}\label{protocol_correctness_message_sequencing}

The message sequencing mechanism was described in section \ref{sec:messagesequencing}. Upstream and interest messages are sequenced using linear Sequence Numbers. The possibility of SN restart due to SN overflow or router reboot is addressed through the BootTime sequence number. The BootTime correctness requires that either routers be clock synchronized or have non-volatile memory. The possibility of BootTime overflow or non-volatile memory failure is addressed by changing the security key.%RV 24/1/2020

The message sequencing includes several optimizations to reduce the amount of stored sequencing information, which do not compromise correctness.%RV 24/1/2020

The first optimization is to sequence messages per transmitting interface, irrespective of tree and message type, and to store the SNs of received messages per neighbor and per tree, irrespective of message type. This optimization is made possible by enlarging the semantic scope of upstream and interest messages. Specifically, IamUpstream messages are also interpreted as NoInterest, and Interest or NoInterest messages are also interpreted as IamNoLongerUpstream. The first double meaning follows directly from the fact that UPSTREAM interfaces are never interested in receiving multicast traffic. The second double meaning follows from the fact that, according to the first double meaning, interest messages are only transmitted by NOT UPSTREAM interfaces. Thus, both double meanings are valid.%RV 24/1/2020

These two double meanings allow suppressing interest messages when an IamUpstream is transmitted and suppressing IamNoLongerUpstream messages when interest messages are transmitted. We suppress the interest messages when an IamUpstream is transmitted, but do not suppress the IamNoLongerUpstream, since interest messages are unicasted to the AW and, therefore, cannot convey upstream information to all neighbors. The latter case maintains its correctness: IamNoLongerUpstream messages are transmitted before interest message; however, if they arrive after (and, therefore, get discarded) the double meaning of interest messages ensures that the correct upstream state is delivered to all intended receivers.%RV 24/1/2020

In transmitted messages, the SN space is shared not only among message types (upstream and interest), but also among different trees. This poses no problem since the receivers store the SNs per tree and, in this way, only discard messages on a per tree basis. Thus, if a transmitter sends two consecutive messages from different trees, both conveying fresher state, they will both be accepted even if they arrive out of order.%RV 24/1/2020

The second optimization is the use of the CheckpointSN, which is optional. The CheckpointSN is sent by a transmitter and corresponds to the highest SN that has been acknowledged so far, such that all SNs lower than it have also been acknowledged. When this information is received, the receivers replace the SN of trees that have an SN lower or equal than the CheckpointSN by the CheckpointSN itself. This saves memory resources (one SN instead of many), and is correct since, for the trees that made the replacement, any message received with an SN lower or equal than the CheckpointSN is necessarily outdated (and should be discarded).%RV 24/1/2020

\subsection{Message transmission reliability}\label{protocol_correctness_message_reliability}%RV 18/2/2019

The message transmission reliability was discussed in section \ref{sec:messagereliability}. Hello messages are periodically transmitted. Sync messages are transmitted using a Stop-and-Wait protocol which has been proven to be correct if both the transmitted messages and their ACKs include sequence numbers, which is the case.%RV 24/1/2020

Upstream and interest messages use ACK-protection, which is a widely used method of ensuring message transmission reliability. Several optimizations were introduced to suppress ACK transmissions and pending ACKs. However, this does not compromise the protocol correctness. When a new message conveying fresher state relative to some tree is transmitted, older messages no longer need to be acknowledged. Thus, regarding these messages, transmitters must cancel the wait for pending ACKs and receivers must suppress ACK replies. Actually, ACK replies are suppressed whenever a message with a higher SN was already acknowledged.%RV 3/2/2020

Moreover, once a synchronization with a neighbor is initiated, all upstream and interest messages sent previously to that neighbor become outdated. Therefore, the router must cancel the wait for pending ACKs relative to messages transmitted with an SN lower than the SnapshotSN.%RV 24/1/2020

Finally, it may happen that an upstream message concurrent with a synchronization and conveying state fresher than the synchronization snapshot does not get its state stored at the neighbor. This may happen if the state is received, acknowledged, and then lost, before the actual start of the synchronization. This problem is addressed by including the SnapshotSN and BootTime of both the sending router and its neighbor in ACK messages. In this way, messages are only considered to have been correctly received if they were received after the initiation of the last synchronization period.%RV 24/1/2020

%RV: Can we use model checking to evaluate the concurrency between upstream messages and synchronization? We may need only two routers with two parallel channels to simulate messages being received out of order, and include ACKs and router failures.
%PO: Poderia ser feito no entanto não percebo o porquê de ter dois canais. Isto poderia ser feito apenas com um canal no entanto o receptor lia uma mensagem aleatoria da queue das mensagens que se encontram armazenadas no canal (mensagens ainda não lidas). Isto implica incluir informação das árvores na sincronização (atualmente não é feito), incluir SNs em mensagens upstream e incluir ACKs. Eu acho que isto iria requerer demasiados recursos uma vez que o Promela teria que testar todas as possibilidades de chegada fora de ordem de mensagens (Sync, IamUpstream e ACK) concorrentemente com a simulação de reinicio da relação de vizinhança em todos os instantes possiveis.

%% file: cc_implementation_v4.tex
\section{Implementation}\label{sec:implementation}%RV 26/2/2019

HPIM-DM was implemented in Python, for platforms with the Linux operating system. In this way, we benefited from the easy-to-use API related to multicast. The Linux kernel allows a user-level process to manipulate entries in the multicast routing table, namely to define, for each multicast stream, the interface that should receive data packets (root interface), and the interfaces that should forward data packets (non-root interfaces in FORWARDING state). We selected Python against other lower level and faster languages such as C, since it provides several abstractions that ease the development and improve the readability, and since our goal was only to develop a proof-of-concept. We implemented the IPv4 version of HPIM-DM. In addition to HPIM-DM, we developed the router side of IGMPv2 since, contrarily to the host side, it is not implemented in the Linux kernel. The implementation is publicly available in GitHub \cite{pedro_repositorio_git_new_protocol}. As part of this project, we also implemented PIM-DM in Python \cite{pedro_repositorio_git_pim_dm}.%RV 3/2/2020

\begin{figure}
	\centering
	\includegraphics[scale=0.6,angle=90]{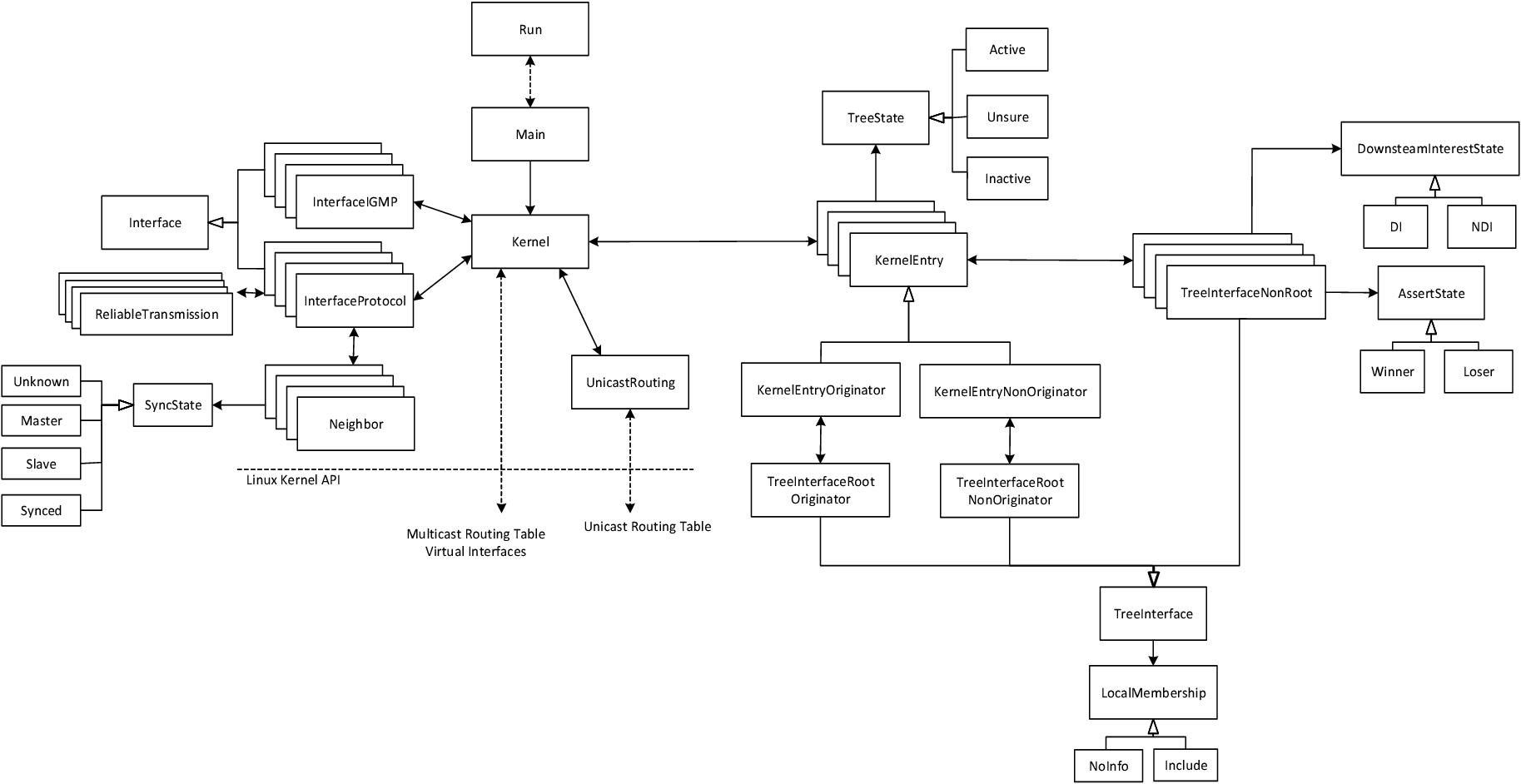}
	\caption{HPIM-DM implementation diagram.}
	\label{fig:new_protocol_digram1}
\end{figure}

The implementation of our protocol followed an objected oriented approach. The diagram of Figure \ref{fig:new_protocol_digram1} represents the software architecture; rectangles represent classes, lines with a closed arrowhead represent associations and lines with an open arrowhead represent inheritance. In what follows, we briefly explain each class.%RV 24/1/2020

A detailed description of the implementation can be found in \cite{python_implementations}.%RV 24/1/2020

%Pedro: Isto está ainda um pouco desligado da descrição do protocolo. Tente incluir mais informação, nomedamente que explique o dinamismo do protocolo: quem notifica quem no caso de alterações. Indique também onde são implementados os diversos timers e onde é feita a eleição do AW. Na descrição também não é feita qualquer referência ao interesse e às mensagens de interesse. Indique também onde são armazenados os vários números de sequência. Descreva sumariamente as threads que são usadas em cada classe, e as threads genéricas. Indique também quais são as máquinas de estado, dizendo simplemente qual a função (sem repetir coisas que já estejam ditas na explicação do protocolo), e indicando a relação com a classe.

\begin{itemize}
	\item \textbf{Run} - Run is used to start/stop the protocol process and for user interactions. This module is daemonized, in order to have the protocol running in the background.%RV 3/2/2020
	
	We have implemented a set of commands in the Run module, to allow the interaction of the user with the protocol process. There are commands for starting/stopping/restarting the process, listing the interfaces enabled for HPIM-DM and IGMP, listing neighbors and their state information, listing sequence numbers, displaying the multicast routing table (output of command ip mroute show), enabling/disabling HPIM-DM and IGMP at interfaces, reading the state of all state machines, presenting detailed information on running events (e.g. state transitions and message receptions), and enabling tests (i.e. sending to a server event logs in real time).%RV 24/1/2020
	
	%PO: no Run indicar que também há comandos para associar e desassociar segurança em interfaces (verificação de integridade e autenticação em mensagens de controlo).
	
	\item \textbf{Main} - Main is a service used to interact with the protocol. This module receives commands from Run and execute them in Kernel. If the command returns an output, the service returns it back to Run in a human readable format (table format).%RV 24/1/2020
	
	\item \textbf{Interface} - Interface is an abstract class that has code related to InterfaceIGMP and InterfaceProtocol, e.g. related to the reception and transmission of control messages (through a socket).%RV 25/1/2020
	
	The reception of control messages is performed through a thread, running in the background, that is constantly waiting for packets from the socket. If a packet is received, the corresponding control message is processed according to its type.%RV 25/1/2020
	
	\item \textbf{InterfaceIGMP} - InterfaceIGMP interfaces with the IGMPv2 implementation. Thus, this module is responsible for monitoring the membership information of all hosts directly connected to physical interfaces.%RV 25/1/2020
	
	\item \textbf{InterfaceProtocol} - InterfaceProtocol represents a physical interface where the protocol was enabled. It sends and receives control messages via a raw socket, including the periodic transmission of Hello messages. It also stores the SN and BootTime used to transmit control messages across the interface. This class uses the Hello Timer, in order to control the periodic transmission of Hello messages. It is associated with the ReliableTransmission and Neighbor classes.%RV 3/2/2020
	
	\item \textbf{ReliableTransmission} - ReliableTransmission is responsible for guaranteeing the reliable transmission of control messages of a given tree. It stores the transmitted messages for which an acknowledgment is pending and a list of neighbors that already acknowledged the message, and uses the Retransmission Timer to process the acknowledgments and retransmissions.%RV 25/1/2020
	
	%PO: Na minha implementação quando uma nova mensagem multicast é transmitida, todas as mensagens enviadas anteriormente (relativamente à mesma árvore) deixam de ser monitorizadas. Quando uma nova mensagem unicast é transmitida e se uma mensagem multicast enviada anteriormente ainda não tiver sido entregue de forma fiável, deixo de monitorizar a mensagem multicast para o destino da mensagem unicast (porque a mensagem unicast terá um SN maior que a mensagem multicast, logo esse vizinho apenas necessita de fazer ACK à ultima mensagem - devido ao duplo significado das mensagens de interesse).
	
	\item \textbf{Neighbor} - Neighbor represents a neighbor detected by InterfaceProtocol. It stores neighbor state, such as the upstream, synchronization, and interest state for each tree, along with the received sequence numbers; it also monitors the neighbor liveness using the Neighbor Liveness Timer.%RV 3/2/2020
	
	The class stores all sequence numbers relative to each neighborhood relationship, i.e. the BootTime and SnapshotSN of the router and of its neighbor and the CheckpointSN of the neighbor. It also stores the SNs received from the neighbor relative to each tree (the ones higher than the CheckpointSN), in a Python dictionary with an (S,G) key.%RV 3/2/2020
	
	The upstream and interest states are also stored in Python dictionaries. The upstream dictionary is (S,G) keyed and includes the RPC if the neighbor is UPSTREAM and is empty otherwise. The interest dictionary is also (S,G) keyed and has a Boolean value that is true if the neighbor is INTERESTED and false otherwise; if this dictionary is empty for some tree, the default interest is assumed.%RV 3/2/2020
	
	When an upstream/interest message is received from a neighbor the received state is stored in the corresponding dictionary if the received SN is greater than the ones stored for that neighbor and tree and greater than the CheckpointSN.%RV 3/2/2020
	
	The Neighbor class is also associated with the synchronization states. In particular, it implements the Sync Timer to control the reliability of Sync message transmissions.%RV 25/1/2020
	
	\item \textbf{SyncState, Unknown, Master, Slave and Synced} - These classes implement the synchronization state machine.%RV 25/1/2020

	%PO: Indicar que SyncState com base no estado actual da sincronização e progresso da mesma obtém informação do que transmitir a partir da class Neighbor (snapshot e SNs), armazena estado das mensagens de controlo recebidas em Neighbor (upstream, interesse (NO INTEREST) e SyncSN) e altera timers de Neighbor (SyncTimer e NeighborLivenessTimer).
	
	\item \textbf{Kernel} - Kernel serves as an abstraction regarding all interactions with the Linux kernel, namely to create/remove virtual interfaces (the ones that will receive/transmit multicast data), set/remove/change entries in the multicast routing table, and be notified about the arrival of multicast data from new trees. It also stores each KernelEntry, and the interfaces added by the user. The exchange of information with the Linux kernel is performed by an mroute socket. A thread is continuously running to receive the kernel notifications.%RV 25/1/2020
	
	\item \textbf{KernelEntry, KernelEntryOriginator, KernelEntryNonOriginator} - KernelEntry represents a tree that is present in the multicast routing table. It can be of two types: KernelEntryOriginator if the router is originator for this tree, or KernelEntryNonOriginator otherwise. This distinction is required since the tree states are calculated differently in each case. This class is associated with a TreeState, representing the state of a tree, and is associated with multiple TreeInterfaces. One of those interfaces must be a root interface, represented by classes TreeInterfaceRootOriginator and TreeInterfaceRootNonOriginator, and the remaining interfaces must be non-root, represented by the TreeInterfaceNonRoot class. Whenever there is a change in the tree state, KernelEntry notifies all interfaces (TreeInterfaceRootOriginator/TreeInterfaceRootNonOriginator and TreeInterfaceNonRoot) of this change.%RV 25/1/2020
	
	Each KernelEntry also stores, for each interface, if there is neighbor interest and what is the RPC of the BestUpstreamNeighbor. This information is updated upon the reception of control messages and the removal/addition of neighbors. By having this information cached it is possible to determine at all times the tree state, whether an interface is AW or AL, and whether there are neighbors interested.%RV 25/1/2020
	
	\item \textbf{TreeState, Active, Inactive and Unsure} - These classes represent the various tree states.%RV 25/1/2020
	
	\item \textbf{TreeInterfaceRootOriginator} - This class represents the root interface of the originator. When an object from this class is created, a socket to receive multicast data from this tree and interface is opened, and a thread is continuously running to receive this data. Moreover, the object monitors the liveness of the source using the Source Active Timer.%RV 25/1/2020
		
	\item \textbf{TreeInterfaceRootNonOriginator} - This class implements the state machines of the root interface of a non-originator to determine when the interface should transmit control messages. Interest messages are transmitted to the BestUpstreamNeighbor (stored in KernelEntry).%RV 25/1/2020
	
	\item \textbf{TreeInterfaceNonRoot} - This class implements the state machines of a non-root interface to determine when the interface should transmit control messages. The class determines if the interface is AW or AL based on the tree state and on the RPCs of the router and of the BestUpstreamNeighbor; it also determines if the interface is DOWNSTREAM INTERESTED or NOT DOWNSTREAM INTERESTED based on the interest of its neighbors and hosts (through the IGMP state machine - Local Membership class).%RV 25/1/2020
		
	\item \textbf{AssertState, Winner and Loser} - These classes represent the various assert states.%RV 25/1/2020

	\item \textbf{DownstreamInterestState, DI and NDI} - These classes represent the various interface downstream interest states.%RV 3/2/2020
	
	\item \textbf{Local Membership} - This class defines if an interface has members interested in receiving multicast data traffic, and depends on the IGMP state machine.%RV 3/2/2020
	
	\item \textbf{UnicastRouting} - Obtains the unicast routing table, and notifies about its changes, to determine the root interface and RPC of a specific tree.%RV 25/1/2020
	
\end{itemize}

To give an example, suppose that an upstream message is received. The message is received by InterfaceProtocol. Then, Neighbor is invoked to compare the SN and BootTime of the received message with the ones stored for the neighbor, to decide if the message is discarded, accepted, or triggers a synchronization (in case of higher BootTime).%RV 3/2/2020

If the message is accepted without synchronization triggering, the received SN, upstream state, and interest state are stored in Neighbor, and InterfaceProtocol is invoked to recalculate the BestUpstreamNeighbor and the global interest of routers associated with this receiving interface. The outcome is sent to Kernel and the corresponding KernelEntry recalculates the router tree state, and the assert state and downstream interest state of the receiving interface. Finally, this may trigger message transmissions (involving InterfaceProtocol and ReliableTransmission) and changes in the FORWARDING state of non-root interfaces (involving Kernel).%RV 3/2/2020

If the message triggers a synchronization, the Kernel is invoked to take the tree snapshot, and InterfaceProtocol is invoked to determine the SnapshotSN and transmit the first Sync message. Moreover, the tree snapshot and the neighbor BootTime are stored in Neighbor, and the Neighbor Liveness Timer and Sync Retransmission Timer are set.%RV 25/1/2020

%% file: cc_tests_v1.tex
\section{Implementation tests}\label{sec:tests}

The HPIM-DM implementation described in section \ref{sec:implementation} was exhaustively tested using Netkit-NG \cite{netkit}. Netkit-NG is a network emulator where network elements (routers/hosts) are implemented through virtual machines running Linux. Regarding the unicast routing protocol we used OSPF \cite{RFC2328}, implemented through the Quagga routing suite \cite{quagga}. In Netkit-NG, the messages exchanged between routers can be captured using \textit{tcpdump} \cite{tcpdump}, to be further analyzed using \textit{Wireshark} \cite{Wireshark}. We developed a Wireshark dissector for HPIM-DM control messages using the Lua language.%RV 3/2/2020

%RV: Give reference to Lua

To perform the tests, we log several types of events at each router and send them to a central node, which processes the information and decides if the test was successful or not. The remote logging was accomplished by the \textit{logging} module of Python. The logged information was: for IGMP (i) expired timers, (ii) state transitions, (ii) reception of control messages; for HPIM-DM, (i) reception of control messages, and state transitions related to (ii) tree, (iii) downstream interest, (iv) assert, (v) interface role, (vi) upstream state, (vii) synchronization state.%RV 25/1/2020

The central node filters the log information relevant for each test and checks if the state transitions are according to what is expected. Moreover, the control messages are manually inspected using Wireshark.%RV 25/1/2020

We have performed a total of 17 tests to HPIM-DM, divided into six categories. All tests were successful. The detailed description of the tests can be found in \cite{python_tests}. We also tested our IGMP implementation.%RV 25/1/2020

We have tested (i) the establishment and maintenance of neighborhood relationships, (ii) the formation and maintenance of the broadcast tree, (iii) the AW election, (iv) the interest maintenance and forwarding decisions, (v) the discovery of trees through the synchronization process and (vi) the tree maintenance in the presence of loops.%RV 25/1/2020

\subsection{Establishment and maintenance of neighborhood relationships}

These tests were performed using two routers connected through a single link. We focused on the correct establishment and maintenance of neighborhood relationships in the presence of reboots and failures, without having any tree involved in the synchronization process. We checked the evolution of the synchronization states, from UNKNOWN to SYNCED, and analyzed the Hello and Sync messages, namely the sequence numbers (BootTime, SnapshotSN, and SyncSN), and the flags (Master and More).%RV 25/1/2020

The following tests were performed:
\begin{itemize}
	\item Test 1: Establishment of neighborhood relationship between two unknown neighbors, without any existing trees
	\item Test 2: Reestablishment of neighborhood relationship after a known neighbor reboots
	\item Test 3: Neighborhood relationship break after known neighbor fails
\end{itemize}%RV 25/1/2020

Test 2 was implemented by restarting the HPIM-DM process at one router and test 3 by stopping the process at one router.%RV 25/1/2020

\subsection{Formation and maintenance of the broadcast tree}

\begin{figure}[!t]
	\centering
	\includegraphics[scale=0.65]{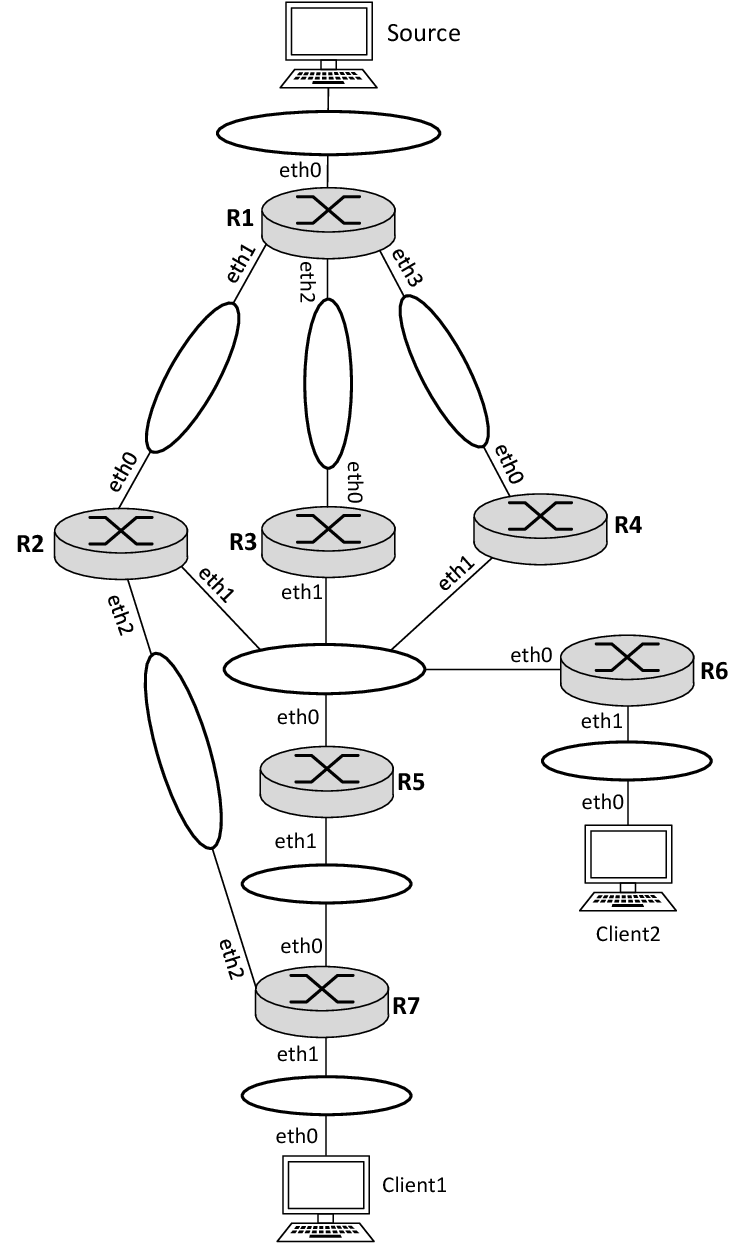}
	\caption{First network used for testing the HPIM-DM implementation.}
	\label{fig:implementationtest1}
\end{figure}

These tests targeted the initial formation of the tree and its reconfiguration in the case of neighbor reboots and failures. They were performed using the network of Figure \ref{fig:implementationtest1}. We selected this topology because it offers multiple paths and has a shared link connecting multiple non-root interfaces.%RV 25/1/2020

The unicast routing protocol was initially configured with all interfaces having a cost of 10. In this way, the root interface was eth0 for all routers, except router R7 where it was eth2. The routers were configured with initial downstream interest in all interfaces, to allow the initial flooding of multicast data.%RV 25/1/2020

As part of these tests we checked the evolution of the tree states, the interface roles, and the RPC values of all routers. We also analyzed the upstream messages exchanged between routers.%RV 25/1/2020

The following tests were performed:
\begin{itemize}
	\item Test 4: Initial formation of the broadcast tree
	\item Test 5: Broadcast tree reconfiguration in reaction to RPC change causing interface role change
	\item Test 6: Broadcast tree reconfiguration in reaction to router failure, detected by both the unicast routing protocol and the multicast routing protocol
\end{itemize}%RV 25/1/2020

Test 4 started by switching on the source. For test 5, we changed the cost of interface eth2 of R7 to 100, which forced this interface to become non-root and interface eth0 to become root. In this test, we payed special attention to the transmission of upstream messages through the new root and non-root interfaces of R7. For test 6, we switched off router R5.%RV 25/1/2020

%RV 6/3/2019

\subsection{Election of AW}

These tests focused on the shared link connecting routers R2 through R6 of Figure \ref{fig:implementationtest1}, and verified if the correct AW was elected at the link, in the presence of router failures and interface role changes. We checked the evolution of the assert states at non-root interfaces and of the BestUpstreamNeighbor at root interfaces. Initially, the interfaces of R2, R3 and R4 were non-root interfaces and the ones of R5 and R6 were root interfaces.%RV 25/1/2020

The following tests were performed:
\begin{itemize}
	\item Test 7: Initial AW election
	\item Test 8: Reelection of AW when AW fails	
	\item Test 9: Reelection of AW when AW becomes root
\end{itemize}%RV 25/1/2020

Test 7 checked if R4 became AW following the formation of the tree. In test 8 we switched off R4, and checked if R3 became the new AW. In test 9, performed in the conditions of test 8, the AW interface became root by changing the cost of interface eth0 of R3 to 100, and we checked if R2 became AW.%RV 25/1/2020

\subsection{Interest maintenance and forwarding decisions}

These tests addressed the forwarding state of non-root interfaces according to the interest of downstream devices. We focused on the shared link connecting routers R2 through R6, as in the tests of previous section. The cost of all interfaces was set to 10, except the cost of interface eth2 of R7 which was set to 100. In this way, the interest at the shared link is directly controlled by the clients: Client1 controls the interest of R5 and Client2 controls the interest of R6.%RV 25/1/2020

We checked the evolution of the downstream interest and forwarding states of the non-root interfaces, and the transmission of interest messages.%RV 25/1/2020

The following tests were performed:
\begin{itemize}
	\item Test 10: Initial forwarding state
	\item Test 11: Forwarding state when one root interface becomes not interested, leaving other root interfaces interested	
	\item Test 12: Forwarding state when one root interface becomes not interested, leaving no other root interfaces interested	
	\item Test 13: Forwarding state when one root interface becomes interested, while no others are interested
\end{itemize}%RV 25/1/2020

In test 10 both clients are interested. In test 11, the interest of Client2 was disabled, making R6 NOT INTERESTED. In test 12, the interest of Client1 was also disabled, making R5 NOT INTERESTED. Finally, in test 13, the interest of Client2 was again enabled, making R6 INTERESTED again.%RV 25/1/2020

\subsection{Tree discovery through synchronization}

In this test we analyzed the synchronization process at the shared link when one interface reboots. Before the test, we started 20 different trees by having the source send multicast packets to 20 different groups. In order to test the fragmentation process, we configured the routers to include only information about five trees in each Sync message.%RV 27/2/2019

We checked the evolution of upstream and tree states, the exchange of Sync messages, in particular which trees were reported in each Sync message, and the exchange of interest messages following the synchronization.%RV 25/1/2020

The following tests were performed:
\begin{itemize}
	\item Test 14: Synchronization after reboot of non-root interface
	\item Test 15: Synchronization after reboot of root interface
\end{itemize}%RV 25/1/2020

In test 14, we rebooted the non-root interface of R4, which triggered synchronizations with all its neighbors connected to the shared link. In test 15, we rebooted the root interface of R5, which forced the router to lose its parent, and regain the tree information through the synchronization with the routers connected to the shared link.%RV 25/1/2020

\subsection{Tree maintenance in the presence of network loops}

\begin{figure}[t!]
	\centering
	\includegraphics[scale=0.65]{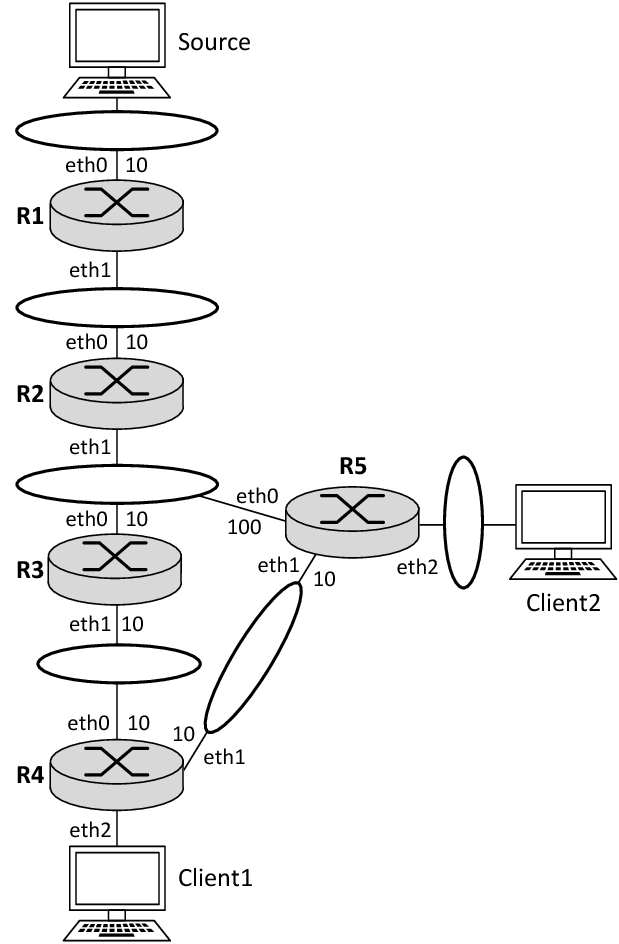}
	\caption{Second network used for testing the HPIM-DM implementation.}
	\label{fig:implementationtest2}
\end{figure}

These tests checked the tree formation and the tree removal in the presence of network loops. We resorted to the network of Figure \ref{fig:implementationtest2}. The loop is composed by R3, R4 and R5: non-root interface of R5 connected to root interface of R3; non-root interface of R3 connected to root interface of R4; non-root interface of R4 connected to root interface of R5.%RV 25/1/2020

We checked the evolution of the tree states and the upstream messages.%RV 25/1/2020

The following test were performed:
\begin{itemize}
	\item Test 16: Tree formation in the presence of network loops
	\item Test 17: Tree removal in the presence of network loops
\end{itemize}%RV 25/1/2020

%RV 6/3/2019

%RV: R1 seems unnecessary. Remove?

Tests 16 and 17 were simply implemented by starting and stopping the source.%RV 25/1/2020

%% file: cc_conclusions.tex
\section{Conclusions}\label{sec:conclusions}

This paper presented a novel multicast routing protocol, called HPIM-DM (Hard-state Protocol Independent Multicast - Dense Mode), which may be seen as a hard-state version of PIM-DM. HPIM-DM overcomes several issues of PIM-DM, that translate into poor convergence and make PIM-DM unsuitable for high-speed networks. These improvements were made possible by introducing (i) the concept of upstream neighbors, i.e. neighbors that can deliver multicast traffic coming from the source, (ii) mechanisms that ensure the reliable transmission and sequencing of control messages, and (iii) a synchronization process enabling a router joining the network to obtain immediately information on the active multicast trees. In this way, there is no need for transmitting periodically control messages (for state update) and the protocol reacts immediately to any event susceptible of changing the configuration of the multicast trees. Moreover, the protocol was optimized for resilience against replay attacks. The correctness of HPIM-DM was assessed using logical reasoning and model checking. Moreover, we made a full implementation of HPIM-DM in Python, and performed extensive tests on this implementation, which also validated the protocol correctness.%RV 25/1/2020

Many of the solutions found for HPIM-DM can be used to improve PIM-SM (PIM - Sparse Mode), which suffers from convergence problems similar to PIM-DM. This is left for future work.%RV 25/1/2020